\documentclass[twoside, 10pt]{article}

\usepackage{jmlr2e}
\usepackage{caption}
\usepackage{color}
\usepackage{hyperref}
\hypersetup{
    colorlinks,
    citecolor=blue,
    filecolor=black,
    linkcolor=blue,
    urlcolor=blue
}
\PassOptionsToPackage{hyphens}{url}\usepackage{hyperref}
\usepackage{breakurl}

\newcommand{\todo}[1]{}
\renewcommand{\todo}[1]{{\color{red} TODO: {#1}}}

\usepackage{multicol}

\begin{document}

\title{Towards an integration of deep learning and neuroscience}

\author{\name Adam H. Marblestone \email amarbles@media.mit.edu \\
       \addr MIT Media Lab\\
       Cambridge, MA 02139, USA
       \AND
       \name Greg Wayne \email gregwayne@google.com \\
       \addr Google Deepmind\\
       London, EC4A 3TW, UK
       \AND
       \name Konrad P. Kording \email 
       koerding@gmail.com\\
       \addr Rehabilitation Institute of Chicago\\
Northwestern University\\
       Chicago, IL 60611, USA
       }

\editor{TBN}

\maketitle

\begin{abstract}
Neuroscience has focused on the detailed implementation of computation, studying neural codes, dynamics and circuits. In machine learning, however, artificial neural networks tend to eschew precisely designed codes, dynamics or circuits in favor of brute force optimization of a cost function, often using simple and relatively uniform initial architectures. Two recent developments have emerged within machine learning that create an opportunity to connect these seemingly divergent perspectives. First, structured architectures are used, including dedicated systems for attention, recursion and various forms of short- and long-term memory storage. Second, cost functions and training procedures have become more complex and are varied across layers and over time. Here we think about the brain in terms of these ideas. We hypothesize that (1) the brain optimizes cost functions, (2) these cost functions are diverse and differ across brain locations and over development, and (3) optimization operates within a pre-structured architecture matched to the computational problems posed by behavior. Such a heterogeneously optimized system, enabled by a series of interacting cost functions, serves to make learning data-efficient and precisely targeted to the needs of the organism. We suggest directions by which neuroscience could seek to refine and test these hypotheses.
\end{abstract}


\tableofcontents
\newpage

\begin{figure}[]
\includegraphics[width=1.1\textwidth]{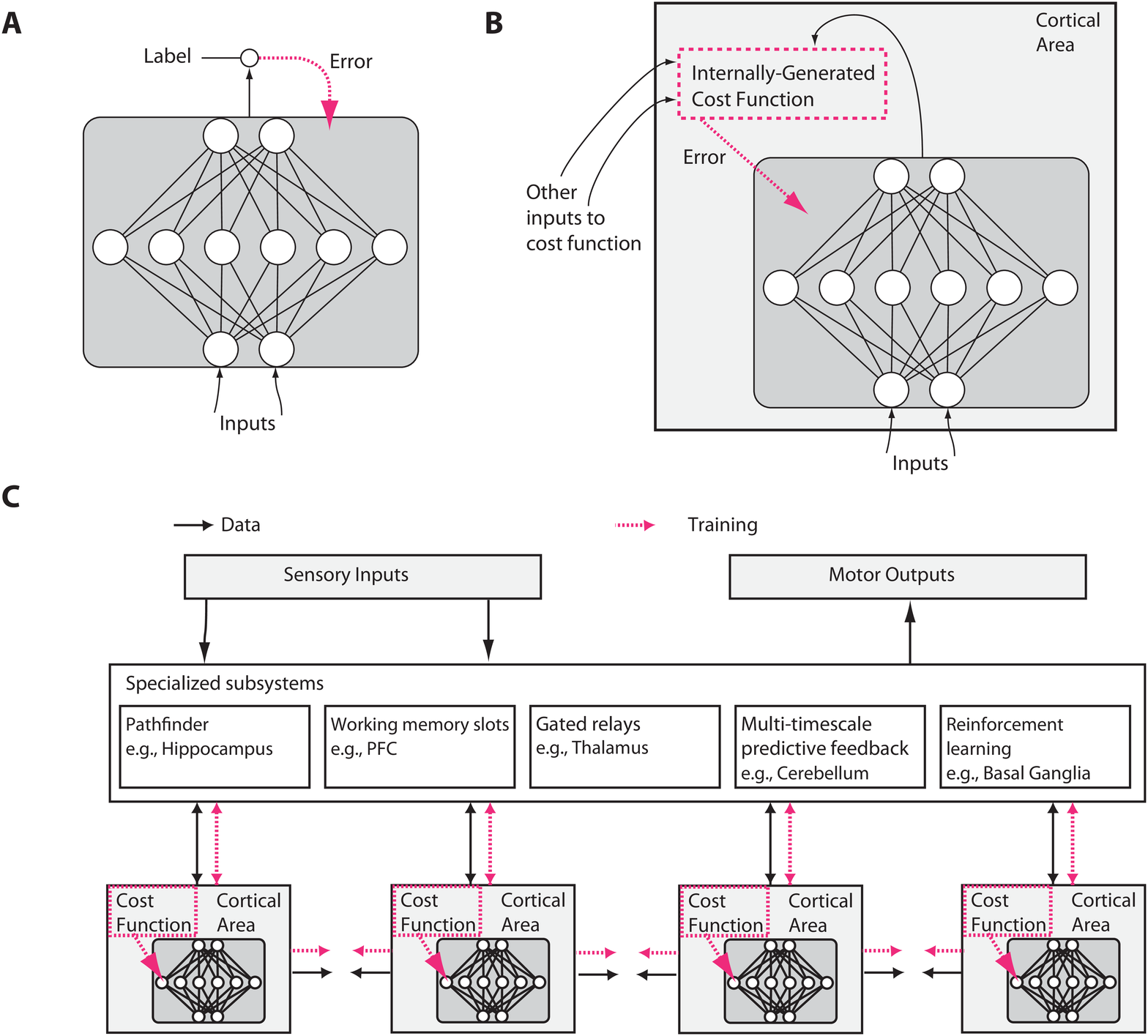}
\captionsetup{labelformat=empty}
\caption{\textbf{Fig 1: Putative differences between conventional and brain-like neural network designs.}\\\textbf{A)} In conventional deep learning, supervised training is based on externally-supplied, labeled data.\\\textbf{B)} In the brain, supervised training of networks can still occur via gradient descent on an error signal, but this error signal must arise from internally generated cost functions. These cost functions are themselves computed by neural modules specified by both genetics and learning. Internally generated cost functions create heuristics that are used to bootstrap more complex learning. For example, an area which recognizes faces might first be trained to detect faces using simple heuristics, like the presence of two dots above a line, and then further trained to discriminate salient facial expressions using representations arising from unsupervised learning and error signals from other brain areas related to social reward processing.\\\textbf{C)} Internally generated cost functions and error-driven training of cortical deep networks form part of a larger architecture containing several specialized systems. Although the trainable cortical areas are schematized as feedforward neural networks here, LSTMs or other types of recurrent networks may be a more accurate analogy, and many neuronal properties such as spiking, dendritic computation, neuromodulation, adaptation, timing-dependent plasticity, direct electrical connections, transient synaptic dynamics, spontaneous activity, and others, will influence what and how such networks learn.}
\label{Fig1}
\end{figure}

\begin{multicols}{2}{
\raggedcolumns

\section{Introduction}

Machine learning and neuroscience speak different languages today. Brain science has discovered a dazzling array of brain areas, cell types, molecules, cellular states, and mechanisms for computation and information storage. Machine learning, in contrast, has largely focused on instantiations of a single principle: function optimization. It has found that simple optimization objectives, like minimizing classification error, can lead to the formation of rich internal representations and powerful algorithmic capabilities in multilayer and recurrent networks~\citep{LeCun2015, schmidhuber2015deep}. Here we seek to connect these perspectives. 

The artificial neural networks now prominent in machine learning were, of course, originally inspired by neuroscience~\citep{McCulloch1943}. While neuroscience has continued to play a role~\citep{Cox2014}, many of the major developments were guided by insights into the mathematics of efficient optimization, rather than neuroscientific findings~\citep{Sutskever2013}. The field has advanced from simple linear systems~\citep{Minsky1972}, to nonlinear networks~\citep{Haykin1994}, to deep and recurrent networks~\citep{schmidhuber2015deep, LeCun2015}. Backpropagation of error~\citep{werbos1974beyond, Werbos1982, Rumelhart1986} enabled neural networks to be trained efficiently, by providing an efficient means to compute the gradient with respect to the weights of a multi-layer network. Methods of training have improved to include momentum terms, better weight initializations, conjugate gradients and so forth, evolving to the current breed of networks optimized using batch-wise stochastic gradient descent. These developments have little obvious connection to neuroscience. 

We will argue here, however, that neuroscience and machine learning are, once again, ripe for convergence. Three aspects of machine learning are particularly important in the context of this paper. First, machine learning has focused on the optimization of cost functions (\textbf{Figure \ref{Fig1}A}). 

Second, recent work in machine learning has started to introduce complex cost functions, those that are not uniform across layers and time, and those that arise from interactions between different parts of a network. For example, introducing the objective of temporal coherence for lower layers (non-uniform cost function over space) improves feature learning~\citep{Sermanet2013}, cost function schedules (non-uniform cost function over time) improve\footnote{Hyper-parameter optimization shows that complicated schedules of training, which differ across parts of the network, lead to optimal performance~\citep{Maclaurin2015}.} generalization~\citep{Saxe2013,Goodfellow2014a, Gulcehre2016} and adversarial networks -- an example of a cost function arising from internal interactions -- allow gradient-based training of generative models~\citep{Goodfellow2014}. Networks that are easier to train are being used to provide ``hints'' to help bootstrap the training of more powerful networks~\citep{romero2014fitnets}. 

Third, machine learning has also begun to diversify the architectures that are subject to optimization. It has introduced simple memory cells with multiple persistent states~\citep{Hochreiter1997, Chung2014}, more complex elementary units such as ``capsules'' and other structures~\citep{Hinton2011,Livni2013, Delalleau2011, Tang2012}, content addressable~\citep{Weston2014, Graves2014} and location addressable memories~\citep{Graves2014}, as well as pointers~\citep{Kurach2015} and hard-coded arithmetic operations~\citep{Neelakantan2015}. 

These three ideas have, so far, not received much attention in neuroscience. We thus formulate these ideas as three hypotheses about the brain, examine evidence for them, and sketch how experiments could test them. But first, let us state the hypotheses more precisely. 

\subsection{\textit{Hypothesis 1 --} The brain optimizes cost functions.}

The central hypothesis for linking the two fields is that biological systems, like many machine-learning systems, are able to optimize cost functions. The idea of cost functions means that neurons in a brain area can somehow change their properties, e.g., the properties of their synapses, so that they get better at doing whatever the cost function defines as their role. Human behavior sometimes approaches optimality in a domain, e.g., during movement~\citep{Kording2007}, which suggests that the brain may have learned optimal strategies. Subjects minimize energy consumption of their movement system~\citep{Taylor2011}, and minimize risk and damage to their body, while maximizing financial and movement gains. Computationally, we now know that optimization of trajectories gives rise to elegant solutions for very complex motor tasks~\citep{mordatch2012discovery, todorov2002optimal, harris1998signal}. We suggest that cost function optimization occurs much more generally in shaping the internal representations and processes used by the brain. We also suggest that this requires the brain to have mechanisms for efficient credit assignment in multilayer and recurrent networks.

\subsection{\textit{Hypothesis 2 --} Cost functions are diverse across areas and change over development.}

A second realization is that cost functions need not be global. Neurons in different brain areas may optimize different things, e.g., the mean squared error of movements, surprise in a visual stimulus, or the allocation of attention. Importantly, such a cost function could be locally generated. For example, neurons could locally evaluate the quality of their statistical model of their inputs (\textbf{Figure \ref{Fig1}B}). Alternatively, cost functions for one area could be generated by another area. Moreover, cost functions may change over time, e.g., guiding young humans to understanding simple visual contrasts early on, and faces a bit later. This could allow the developing brain to bootstrap more complex knowledge based on simpler knowledge. Cost functions in the brain are likely to be complex and to be arranged to vary across areas and over development.

\subsection{\textit{Hypothesis 3 --} Specialized systems allow efficiently solving key computational problems.}

A third realization is that structure matters. The patterns of information flow seem fundamentally different across brain areas, suggesting that they solve distinct computational problems. Some brain areas are highly recurrent, perhaps making them predestined for short-term memory storage~\citep{Wang2012}. Some areas contain cell types that can switch between qualitatively different states of activation, such as a persistent firing mode versus a transient firing mode, in response to particular neurotransmitters~\citep{Hasselmo2006}. Other areas, like the thalamus appear to have the information from other areas flowing through them, perhaps allowing them to determine information routing~\citep{Sherman2005}. Areas like the basal ganglia are involved in reinforcement learning and gating of discrete decisions~\citep{Sejnowski2014, doya1999computations}. As every programmer knows, specialized algorithms matter for efficient solutions to computational problems, and the brain is likely to make good use of such specialization (\textbf{Figure \ref{Fig1}C}).

These ideas are inspired by recent advances in machine learning, but we also propose that the brain has major differences from any of today's machine learning techniques. In particular, the world gives us a relatively limited amount of information that we could use for supervised learning~\citep{fodor2002understanding}. There is a huge amount of information available for unsupervised learning, but there is no reason to assume that a \emph{generic} unsupervised algorithm, no matter how powerful, would learn the precise things that humans need to know, in the order that they need to know it. The evolutionary challenge of making unsupervised learning solve the ``right'' problems is, therefore, to find a sequence of cost functions that will deterministically build circuits and behaviors according to prescribed developmental stages, so that in the end a relatively small amount of information suffices to produce the right behavior. For example, a developing duck imprints~\citep{tinbergen1965behavior} a template of its parent, and then uses that template to generate goal-targets that help it develop other skills like foraging. 

Generalizing from this and from other studies~\citep{Ullman2012, Minsky1977}, we propose that many of the brain's cost functions arise from such an internal bootstrapping process. Indeed, we propose that biological development and reinforcement learning can, in effect, program the emergence of a sequence of cost functions that precisely anticipates the future needs faced by the brain's internal subsystems, as well as by the organism as a whole. This type of developmentally programmed bootstrapping generates an internal infrastructure of cost functions which is diverse and complex, while simplifying the learning problems faced by the brain's internal processes. Beyond simple tasks like familial imprinting, this type of bootstrapping could extend to higher cognition, e.g., internally generated cost functions could train a developing brain to properly access its memory or to organize its actions in ways that will prove to be useful later on. The potential bootstrapping mechanisms that we will consider operate in the context of unsupervised and reinforcement learning, and go well beyond the types of curriculum learning ideas used in today's machine learning~\citep{bengio2009curriculum}.

In the rest of this paper, we will elaborate on these hypotheses. First, we will argue that both local and multi-layer optimization is, perhaps surprisingly, compatible with what we know about the brain. Second, we will argue that cost functions differ across brain areas and change over time and describe how cost functions interacting in an orchestrated way could allow bootstrapping of complex function. Third, we will list a broad set of specialized problems that need to be solved by neural computation, and the brain areas that have structure that seems to be matched to a particular computational problem. We then discuss some implications of the above hypotheses for research approaches in neuroscience and machine learning, and sketch a set of experiments to test these hypotheses. Finally, we discuss this architecture from the perspective of evolution.

} 
\end{multicols}

\begin{multicols}{2}{
\raggedcolumns

\section{The brain can optimize cost functions}

Much of machine learning is based on efficiently optimizing functions, and, as we will detail below, the ability to use backpropagation of error~\citep{werbos1974beyond,Rumelhart1986} to calculate gradients of arbitrary parametrized functions has been a key breakthrough. In \textbf{Hypothesis 1}, we claim that the brain is also, at least in part, an optimization machine. But what exactly does it mean to say that the brain can optimize cost functions? After all, many processes can be viewed as optimizations. For example, the laws of physics are often viewed as minimizing an action functional, while evolution optimizes the fitness of replicators over a long timescale. To be clear, our main claims are: that \textbf{a)} the brain has powerful mechanisms for credit assignment during learning that allow it to optimize global functions in multi-layer networks by adjusting the properties of each neuron to contribute to the global outcome, and that \textbf{b)} the brain has mechanisms to specify exactly which cost functions it subjects its networks to, i.e., that the cost functions are highly tunable, shaped by evolution and matched to the animal's ethological needs. Thus, the brain uses cost functions as a key driving force of its development, much as modern machine learning systems do.

To understand the basis of these claims, we must now delve into the details of how the brain might efficiently perform credit assignment throughout large, multi-layered networks, in order to optimize complex functions. We argue that the brain uses several different types of optimization to solve distinct problems. In some structures, it may use genetic pre-specification of circuits for problems that require only limited learning based on data, or it may exploit local optimization to avoid the need to assign credit through many layers of neurons. It may also use a host of proposed circuit structures that would allow it to actually perform, in effect, backpropagation of errors through a multi-layer network, using biologically realistic mechanisms -- a feat that had once been widely believed to be biologically implausible~\citep{Crick1989, Stork1989}. Potential such mechanisms include circuits that literally backpropagate error derivatives in the manner of conventional backpropagation, as well as circuits that provide other efficient means of approximating the effects of backpropagation, i.e., of rapidly computing the approximate gradient of a cost function relative to any given connection weight in the network. Lastly, the brain may use algorithms that exploit specific aspects of neurophysiology -- such as spike timing dependent plasticity, dendritic computation, local excitatory-inhibitory networks, or other properties -- as well as the integrated nature of higher-level brain systems. Such mechanisms promise to allow learning capabilities that go even beyond those of current backpropagation networks.

\subsection{Local self-organization and optimization without multi-layer credit assignment}

Not all learning requires a general-purpose optimization mechanism like gradient descent\footnote{Of course, some circuits may also be heavily genetically pre-specified to minimize the burden on learning. For instance, particular cell adhesion molecules~\citep{Hattori2007} expressed on particular parts of particular neurons defined by a genetic cell type~\citep{Zeisel2015}, and the detailed shapes and placements of neuronal arbors, may constrain connectivity in some cases, though in other cases local connectivity is thought to be only weakly constrained~\citep{Kalisman2005}. Genetics is sufficient to specify complex circuits involving hundreds of neurons, such as central pattern generators~\citep{Yuste2005} which create complex self-stabilizing oscillations, or the entire nervous systems of small worms. Genetically guided wiring should not be thought of as fixed ``hard-wiring'' but rather as a programmatic construction process that can also accept external inputs and interact with learning mechanisms~\citep{marcus2004birth}.}. Many theories of cortex~\citep{George2009, Hoerzer2014, Kappel2014} emphasize potential self-organizing and unsupervised learning properties that may obviate the need for multi-layer backpropagation as such. Hebbian plasticity, which adjusts weights according to correlations in pre-synaptic and post-synaptic activity, is well established\footnote{Hebbian plasticity even has a well-understood biological basis in the form of the NMDA receptors, which are activated by the simultaneous occurrence of chemical transmitter delivered from the pre-synaptic neuron, and voltage depolarization of the post-synaptic neuron.}. Various versions of Hebbian plasticity~\citep{Miller1994} can give rise to different forms of correlation and competition between neurons, leading to the self-organized formation of ocular dominance columns, self-organizing maps and orientation columns~\citep{Ferster2003, Miller1989}. Often these types of local self-organization can also be viewed as optimizing a cost function: for example, certain forms of Hebbian plasticity can be viewed as extracting the principal components of the input, which minimizes a reconstruction error. 

To generate complex temporal patterns, the brain may also implement other forms of learning that do not require any equivalent of full backpropagation through a multilayer network. For example, ``liquid-''~\citep{Maass2002} or ``echo-state machines''~\citep{Jaeger2004} are randomly connected recurrent networks that form a basis set of random filters, which can be harnessed for learning with tunable readout weights. Variants exhibiting chaotic, spontaneous dynamics can even be trained by feeding back readouts into the network and suppressing the chaotic activity~\citep{Sussillo2009}. Learning only the readout layer makes the optimization problem much simpler (indeed, equivalent to regression for supervised learning). Additionally, echo state networks can be trained by reinforcement learning as well as supervised learning~\citep{bush2007echo}.

\subsection{Biological implementation of optimization}

We argue that the above mechanisms of local self-organization are insufficient to account for the brain’s powerful learning performance. To elaborate on the need for an efficient means of gradient computation in the brain, we will first place backpropagation into it's computational context~\citep{Hinton1989, Baldi2015}. Then we will explain how the brain could plausibly implement approximations of gradient descent.

\subsubsection{The need for efficient gradient descent in multi-layer networks}

The simplest mechanism to perform cost function optimization is sometimes known as the ``twiddle'' algorithm or, more technically, as ``serial perturbation''. This mechanism works by perturbing (i.e., ``twiddling''), with a small increment, a single weight in the network, and verifying improvement by measuring whether the cost function has decreased compared to the network's performance with the weight unperturbed. If improvement is noticeable, the perturbation is used as a direction of change to the weight; otherwise, the weight is changed in the opposite direction (or not changed at all). Serial perturbation is therefore a method of ``coordinate descent'' on the cost, but it is slow and requires global coordination: each synapse in turn is perturbed while others remain fixed. 

Weight perturbation (or parallel perturbation) perturbs all of the weights in the network at once. It is able to optimize small networks to perform tasks but generally suffers from high variance. That is, the measurement of the gradient direction is noisy and changes drastically from perturbation to perturbation because a weight's influence on the cost is masked by the changes of all other weights, and there is only one scalar feedback signal indicating the change in the cost\footnote{The variance can be mitigated by averaging out many perturbations before making a change to the baseline value of the weights, but this would take significant time for a network of non-trivial size as the variance of weight perturbation's estimates scales in proportion to the number of synapses in the network.}. Weight perturbation is dramatically inefficient for large networks. In fact, parallel and serial perturbation learn at approximately the same rate if the time measure counts the number of times the network propagates information from input to output~\citep{Werfel2005}. 

Some efficiency gain can be achieved by perturbing neural activities instead of synaptic weights, acknowledging the fact that any long-range effect of a synapse is mediated through a neuron. Like weight perturbation and unlike serial perturbation, minimal global coordination is needed: each neuron only needs to receive a feedback signal indicating the global cost. The variance of node perturbation's gradient estimate is far smaller than that of weight perturbation under the assumptions that either all neurons or all weights, respectively, are perturbed and that they are perturbed at the same frequency. In this case, node perturbation's variance is proportional to the number of cells in the network, not the number of synapses.  

All of these approaches are slow either due to the time needed for serial iteration over all weights or the time needed for averaging over low signal-to-noise ratio gradient estimates. To their credit however, none of these approaches requires more than knowledge of local activities and the single global cost signal. Real neural circuits in the brain have mechanisms (e.g., diffusible neuromodulators) that appear to code the signals relevant to implementing those algorithms. In many cases, for example in reinforcement learning, the cost function, which is computed based on interaction with an unknown environment, cannot be differentiated directly, and an agent has no choice but to deploy clever twiddling to explore at some level of the system~\citep{Williams1992}.

Backpropagation, in contrast, works by computing the sensitivity of the cost function to each weight based on the layered structure of the system. The derivatives of the cost function with respect to the last layer can be used to compute the derivatives of the cost function with respect to the penultimate layer, and so on, all the way down to the earliest layers\footnote{If the error derivatives of the cost function with respect to the last layer of unit activities are unknown, then they can be replaced with node-perturbation-like correlations, as is common in reinforcement learning.}. Backpropagation can be computed rapidly, and for a single input-output pattern, it exhibits no variance in its gradient estimate. The backpropagated gradient has no more noise for a  large system than for a  small system, so  deep and wide architectures with great computational power can be trained efficiently.

\subsubsection{Biologically plausible approximations of gradient descent}

To permit biological learning with efficiency approaching that of machine learning methods, some provision for more sophisticated gradient propagation may be suspected. Contrary to what was once a common assumption, there are now many proposed ``biologically plausible'' mechanisms by which a neural circuit could implement optimization algorithms that, like backpropagation, can efficiently make use of the gradient. These include Generalized Recirculation~\citep{OReilly1996}, Contrastive Hebbian Learning~\citep{Xie2003}, random feedback weights together with synaptic homeostasis~\citep{Lillicrap2014, Liao2015}, spike timing dependent plasticity (STDP) with iterative inference and target propagation~\citep{Scellier2016, Bengio2015}, complex neurons with backpropagating action-potentials~\citep{Kording2000}, and others~\citep{Balduzzi2014a}. While these mechanisms differ in detail, they all invoke feedback connections that carry error phasically. Learning occurs by comparing a prediction with a target, and the prediction error is used to drive top-down changes in bottom-up activity.

As an example, consider O'Reilly's temporally eXtended Contrastive Attractor Learning (XCAL) algorithm~\citep{OReilly2012,OReilly2014a}. Suppose we have a multilayer neural network with an input layer, an output layer, and a set of hidden layers in between. O’Reilly showed that the same functionality as backpropagation can be implemented by a bidirectional network with the same weights but symmetric connections. After computing the outputs using the forward connections only, we set the output neurons to the values they should have. The dynamics of the network then cause the hidden layers' activities to evolve toward a stable attractor state linking input to output. The XCAL algorithm performs a type of local modified Hebbian learning at each synapse in the network during this process~\citep{OReilly2012}. The XCAL Hebbian learning rule compares the local synaptic activity (pre x post) during the early phase of this settling (before the attractor state is reached) to the final phase (once the attractor state has been reached), and adjusts the weights in a way that should make the early phase reflect the later phase more closely. These contrastive Hebbian learning methods even work when the connection weights are not precisely symmetric~\citep{OReilly1996}. XCAL has been implemented in biologically plausible conductance-based neurons and basically implements the backpropagation of error approach. 

Approximations to backpropagation could also be enabled by the millisecond-scale timing of of neural activities~\citep{OReilly2014a}. Spike timing dependent plasticity (STDP)~\citep{Markram1997}, for example, is a feature of some neurons in which the sign of the synaptic weight change depends on the precise millisecond-scale relative timing of pre-synaptic and post-synaptic spikes. This is conventionally interpreted as Hebbian plasticity that measures the potential for a causal relationship between the pre-synaptic and post-synaptic spikes: a pre-synaptic spike could have contributed to causing a post-synaptic spike only if it occurs shortly beforehand\footnote{Interestingly, STDP is not a unitary phenomenon, but rather a diverse collection of different rules with different timescales and temporal asymmetries~\citep{Sjostrom2010, Mishra2016}. Effects include STDP with the inverse temporal asymmetry, symmetric STDP, STDP with different temporal window sizes. STDP is also frequency dependent, which can be explained by rules that depend on triplets rather than pairs of spikes~\citep{pfister2006triplets}. While STDP is often included explicitly in models, biophysical derivations of STDP from various underlying phenomena are also being attempted, some of which involve the post-synaptic voltage~\citep{clopath2010voltage} or a local dendritic voltage~\citep{Urbanczik2014}. Meanwhile, other theories suggest that STDP may enable the use of precise timing codes based on temporal coincidence of inputs, the generation and unsupervised learning of temporal sequences~\citep{Fiete2010, Abbott1996}, enhancements to distal reward processing in reinforcement learning~\citep{Izhikevich2007}, stabilization of neural responses~\citep{Kempter2001}, or many other higher-level properties~\citep{Nessler2013, Kappel2014}.}. To enable a backpropagation mechanism, Hinton has suggested an alternative interpretation: that neurons could encode the types of error derivatives needed for backpropagation in the temporal derivatives of their firing rates~\citep{Hinton2007, Hinton2016talk}. STDP then corresponds to a learning rule that is sensitive to these error derivatives~\citep{Bengio2015b, Xie2000}. In other words, in an appropriate network context, STDP learning could give rise to a biological implementation of backpropagation\footnote{Hinton has suggested~\citep{Hinton2007, Hinton2016talk} that this could take place in the context of autoencoders and recirculation~\citep{Hinton1988}. Bengio and colleagues have proposed~\citep{Scellier2016, Bengio2015a, Bengio2014} another context in which the connection between STDP and plasticity rules that depend on the temporal derivative of the post-synaptic firing rate can be exploited for biologically plausible multilayer credit assignment. This setting relies on clamping of outputs and stochastic relaxation in energy-based models~\citep{Ackley1985}, which leads to a continuous network dynamics~\citep{Hopfield1984} in which hidden units are perturbed towards target values~\citep{Bengio2015a}, loosely similar to that which occurs in XCAL. This dynamics then allows the STDP-based rule to correspond to gradient descent on the energy function with respect to the weights~\citep{Scellier2016}. This scheme requires symmetric weights, but in an autoencoder context, Bengio notes that these can arise spontaneously~\citep{Arora2015}.}. 

Another possible mechanism, by which biological neural networks could approximate backpropagation, is ``feedback alignment''~\citep{Lillicrap2014, Liao2015}. There, the feedback pathway in backpropagation, by which error derivatives at a layer are computed from error derivatives at the subsequent layer, is replaced by a set of random feedback connections, with no dependence on the forward weights. Subject to the existence of a synaptic normalization mechanism and approximate sign-concordance between the feedforward and feedback connections~\citep{Liao2015}, this mechanism of computing error derivatives works nearly as well as backpropagation on a variety of tasks. In effect, the forward weights are able to adapt to bring the network into a regime in which the random backwards weights actually carry the information that is useful for approximating the gradient. This is a remarkable and surprising finding, and is indicative of the fact that our understanding of gradient descent optimization, and specifically of the mechanisms by which backpropagation itself functions, are still incomplete. In neuroscience, meanwhile, we find feedback connections almost wherever we find feed-forward connections, and their role is the subject of diverse theories~\citep{Callaway2004, Maass2007}. It should be noted that feedback alignment as such does not specify exactly how neurons represent and make use of the error signals; it only relaxes a constraint on the transport of the error signals. Thus, feedback alignment is more a primitive that can be used in fully biological (approximate) implementations of backpropagation, than a fully biological implementation in its own right. As such, it may be possible to incorporate it into several of the other schemes discussed here.

The above ``biological'' implementations of backpropagation still lack some key aspects of biological realism. For example, in the brain, neurons tend to be either excitatory or inhibitory but not both, whereas in artificial neural networks a single neuron may send both excitatory and inhibitory signals to its downstream neurons. Fortunately, this constraint is unlikely to limit the functions that can be learned~\citep{Parisien2008, Tripp2016}. Other biological considerations, however, need to be looked at in more detail: the highly recurrent nature of biological neural networks, which show rich dynamics in time, and the fact that most neurons in mammalian brains communicate via spikes. We now consider these two issues in turn.

\paragraph{Temporal credit assignment:} The biological implementations of backpropagation proposed above, while applicable to feedforward networks, do not give a natural implementation of ``backpropagation through time'' (BPTT)~\citep{Werbos1990} for recurrent networks, which is widely used in machine learning for training recurrent networks on sequential processing tasks. BPTT ``unfolds'' a recurrent network across multiple discrete time steps and then runs backpropagation on the unfolded network to assign credit to particular units at particular time steps\footnote{Even BPTT has arguably not been completely successful in recurrent networks. The problems of vanishing and exploding gradients led to long short term memory networks with gated memory units. An alternative is to use optimization methods that go beyond first order derivatives~\citep{Martens2011}. This suggests the need for specialized systems and structures in the brain to mitigate problems of temporal credit assignment.}. While the network unfolding procedure of BPTT itself does not seem biologically plausible, to our intuition, it is unclear to what extent temporal credit assignment is truly needed~\citep{Ollivier2015} for learning particular temporally extended tasks. 

If the system is given access to appropriate memory stores and representations~\citep{Buonomano1995, gershman2012successor, gershman2014time} of temporal context, this could potentially mitigate the need for temporal credit assignment as such -- in effect, memory systems could ``spatialize'' the problem of temporal credit assignment\footnote{Interestingly, the hippocampus seems to ``time stamp'' memories by encoding them into ensembles with cellular compositions and activity patterns that change gradually as a function of time on the scale of days~\citep{rubin2015hippocampal, cai2016shared}, and may use ``time cells'' to mark temporal positions within episodes on a timescale of seconds~\citep{kraus2013hippocampal}.}. For example, memory networks~\citep{Weston2014} store everything by default up to a certain buffer size, eliminating the need to perform credit assignment over the write-to-memory events, such that the network only needs to perform credit assignment over the read-from-memory events. In another example, certain network architectures that are superficially very deep, but which possess particular types of ``skip connections'', can actually be seen as ensembles of comparatively shallow networks~\citep{Veit2016}; applied in the time domain, this could limit the need to propagate errors far backwards in time. Other, similar specializations or higher-levels of structure could, potentially, further ease the burden on credit assignment.

Can generic recurrent networks perform temporal credit assignment in in a way that is more biologically plausible than BPTT? Indeed, new discoveries are being made about the capacity for supervised learning in continuous-time recurrent networks with more realistic synapses and neural integration properties. In internal FORCE learning~\citep{Sussillo2009}, internally generated random fluctuations inside a chaotic recurrent network are adjusted to provide feedback signals that drive weight changes internal to the network while the outputs are clamped to desired patterns. This is made possible by a learning procedure that rapidly adjusts the network output to a state where it is close to the clamped values, and exerts continuous control to keep this difference small throughout the learning process\footnote{Control theory concepts also appear to be useful for simplifying optimization problems in certain other settings~\citep{todorov2009efficient}.}. This procedure is able to control and exploit the chaotic dynamical patterns that are spontaneously generated by the network. Werbos has proposed in his ``error critic'' that an online approximation to BPTT can be achieved by learning to predict the backward-through-time gradient signal (costate) in a manner analogous to the prediction of value functions in reinforcement learning~\citep{si2004handbook}. Broadly, we are only beginning to understand how neural activity can itself represent the time variable~\citep{Finnerty2015}, and how recurrent networks can learn to generate trajectories of population activity over time~\citep{Liu2009}. Moreover, as we discuss below, a number of cortical models also propose means, other than BPTT, by which networks could be trained on sequential prediction tasks, even in an online fashion~\citep{Cui2015, OReilly2014a}. A broad range of ideas can be used to approximate BPTT in more realistic ways.

\paragraph{Spiking networks:} It has been difficult to apply gradient descent learning directly to spiking neural networks\footnote{Analogs of weight perturbation and node perturbation are known for spiking networks~\citep{Seung2003, Fiete2006}. \citet{Seung2003} also discusses implications of gradient based learning algorithms for neuroscience, echoing some of our considerations here.}\footnote{A related, but more general, question is how to learn over many layers of non-differentiable structures. One option is to perform updates via finite-sized rather than infinitesimal steps, e.g., via target-propagation~\citep{Bengio2014}.}. A number of optimization procedures have been used to generate, indirectly, spiking networks which can perform complex tasks, by performing optimization on a continuous representation of the network dynamics and embedding variables into high-dimensional spaces with many spiking neurons representing each variable~\citep{Abbott, DePasquale2016, Komer2016, Thalmeier2015}. The use of recurrent connections with multiple timescales can remove the need for backpropagation in the direct training of spiking recurrent networks~\citep{Bourdoukan2015}. Fast connections maintain the network in a state where slow connections have local access to a global error signal. While the biological realism of these methods is still unknown, they all allow connection weights to be learned in spiking networks.

These and other novel learning procedures illustrate the fact that we are only beginning to understand the connections between the temporal dynamics of biologically realistic networks, and mechanisms of temporal and spatial credit assignment. Nevertheless, we argue here that existing evidence suggests that biologically plausible neural networks can solve these problems -- in other words, it is possible to efficiently optimize complex functions of temporal history in the context of spiking networks of biologically realistic neurons. In any case, there is little doubt that spiking recurrent networks using realistic population coding schemes can, with an appropriate choice of connection weights, compute complicated, cognitively relevant functions\footnote{Eliasmith and others have shown~\citep{Eliasmith2013, Eliasmith2004, Eliasmith2012} that complex functions and control systems can be compiled onto such networks, using nonlinear encoding and linear decoding of high-dimensional vectors.}. The question is how the developing brain efficiently learns such complex functions.

\subsection{Alternative mechanisms for learning}

The brain has mechanisms and structures that could support learning mechanisms different from typical gradient-based optimization algorithms.

\subsubsection{Exploiting biological neural mechanisms}

The complex physiology of individual biological neurons may not only help explain how some form of efficient gradient descent could be implemented within the brain, but also could provide mechanisms for learning that go beyond backpropagation. This suggests that the brain may have discovered mechanisms of credit assignment quite different from those dreamt up by machine learning. 

One such biological primitive is dendritic computation, which could impact prospects for learning algorithms in several ways. First, real neurons are highly nonlinear, with the dendrites of each \emph{single} neuron implementing\footnote{Dendritic computation may also have other functions, e.g., competitive interactions between dendrites in a single neuron could also allow neurons to contribute to multiple different ensembles~\citep{Legenstein2011a}.} something computationally similar to a three-layer neural network~\citep{Mel1992}. Second, when a neuron spikes, its action potential propagates back from the soma into the dendritic tree. However, it propagates more strongly into the branches of the dendritic tree that have been active~\citep{Williams2000}, potentially simplifying the problem of credit assignment~\citep{Kording2000}. Third, neurons can have multiple somewhat independent dendritic compartments, as well as a somewhat independent somatic compartment, which means that the neuron should be thought of as storing more than one variable. Thus, there is the possibility for a neuron to store both its activation itself, and the error derivative of a cost function with respect to its activation, as required in backpropagation, and biological implementations of backpropagation based on this principle have been proposed~\citep{Kording2001}\footnote{Interestingly, in the model of~\citep{Kording2001}, single spikes are used to transmit activations and burst spikes are used to transmit error information. In other models, including the dendritic voltage in a plasticity rule leads to simple error-driven and predictive learning~\citep{Urbanczik2014}. Single neurons with active dendrites and many synapses may embody learning rules of greater complexity, such as the storage and recall of temporal patterns~\citep{Hawkins2015}.}. Overall, the implications of dendritic computation for credit assignment in deep networks are only beginning to be considered. 

Beyond dendritic computation, diverse mechanisms~\citep{Marblestone2014a} like retrograde (post-synaptic to pre-synaptic) signals using cannabinoids~\citep{Wilson2001}, or rapidly-diffusing gases such as nitric oxide~\citep{Arancio1996}, are among many that could enable learning rules that go beyond backpropagation. Harris has suggested~\citep{Harris2008, Lewis2014} how slow, retroaxonal (i.e., from the outgoing synapses back to the parent cell body) transport of molecules like neurotrophins could allow neural networks to implement an analog of an exchangeable currency in economics, allowing networks to self-organize to efficiently provide information to downstream ``consumer'' neurons that are trained via faster and more direct error signals. The existence of these diverse mechanisms may call into question traditional, intuitive notions of ``biological plausibility'' for learning algorithms.

Another biological primitive is neuromodulation. The same neuron or circuit can exhibit different input-output responses and plasticity depending on a global circuit state, as reflected by the concentrations of various \emph{neuromodulators} like dopamine, serotonin, norepinephrine, acetylcholine, and hundreds of different neuropeptides such as opiods~\citep{Bargmann2013, Bargmann2012}. These modulators interact in complex and cell-type-specific ways to influence circuit function. Interactions with glial cells also play a role in neural signaling and neuromodulation, leading to the concept of ``tripartite'' synapses that include a glial contribution~\citep{Perea2009}. Modulation could have many implications for learning. First, modulators can be used to gate synaptic plasticity on and off selectively in different areas and at different times, allowing precise, rapidly updated orchestration of where and when cost functions are applied. Furthermore, it has been argued that a single neural circuit can be thought of as multiple overlapping circuits with modulation switching between them~\citep{Bargmann2013, Bargmann2012}. In a learning context, this could potentially allow sharing of synaptic weight information between overlapping circuits. \citet{Dayan2012} discusses further computational aspects of neuromodulation. Overall, neuromodulation seems to expand the range of possible algorithms that could be used for optimization.

\subsubsection{Learning in the cortical sheet}

A number of models attempt to explain cortical learning on the basis of specific architectural features of the 6-layered cortical sheet. These models generally agree that a primary function of the cortex is some form of unsupervised learning via prediction~\citep{OReilly2014a}. Some cortical learning models are explicit attempts to map cortical structure onto the framework of message-passing algorithms for Bayesian inference~\citep{George2009, Dean2005, Lee2003}, while others start with particular aspects of cortical neurophysiology and seek to explain those in terms of a learning function. For example, the nonlinear and dynamical properties of cortical pyramidal neurons -- the principal excitatory neuron type in cortex -- are of particular interest here, especially because these neurons have multiple dendritic zones that are targeted by different kinds of projections, which may allow the pyramidal neuron to make comparisons of top-down and bottom-up inputs\footnote{This idea has been used by Hawkins and colleagues to suggest mechanisms for continuous online sequence learning~\citep{Hawkins2015, Cui2015} and by Larkum and colleagues for comparison of top-down and bottom-up signals~\citep{Larkum2013}. The Larkum model focuses on the layer 5 (L5) pyramidal neuron type. The cell body of this neuron lies in L5 but which extends its ``apical'' dendritic tree all the way up to a tuft at the top of the cortex in layer 1 (L1), which is a primary target of feedback projections. In the model, interactions between local spiking in these different dendritic zones, which are targeted by different kinds of projections, are crucial to the learning function. The model of Hawkins~\citep{Hawkins2015, Cui2015} also focused on the unique dendritic structure of the L5 pyramidal neuron, and distinguishes internal states of the neuron, which impact its responsiveness to other inputs, from activation states, which directly translate into spike rates. Three integration zones in each neuron, and dendritic NMDA spikes~\citep{Palmer2014} acting as local coincidence detectors, allow temporal patterns of dendritic input to impact the cell’s internal state. Intra-column inhibition is also used in this model. Other cortical models pay less attention to the details of dendritic computation, but still provide detailed interpretations of the inter-laminar projection patterns of the neocortex. For example, in~\citep{OReilly2014a}, an architecture is presented for continuous learning based on prediction of the next input. Time is discretized into 100 millisecond bins via an alpha oscillation, and the deep vs. shallow layers maintain different information during these time bins, with deep layers maintaining a record of the previous time step, and shallow layers representing the current state. The stored information in the deep layers leads to a prediction of the current state, which is then compared with the actual current state. Periodic bursting locked to the oscillation provides a kind of clock that causes the current state to be shifted into the deep layers for maintenance during the subsequent time step, and recurrent loops with the thalamus allow this representation to remain stable for sufficiently long to be used to generate the prediction.}. 

Other aspects of the laminar cortical architecture could be crucial to how the brain implements learning. Local inhibitory neurons targeting particular dendritic compartments of the L5 pyramidal could be used to exert precise control over when and how the relevant feedback signals and associative mechanisms are utilized. Notably, local inhibitory networks could also give rise to competition~\citep{Petrov2010} between different representations in the cortex, perhaps allowing one cortical column to suppress others nearby, or perhaps even to send more sophisticated messages to gate the state transitions of its neighbors~\citep{Bach2015a}. Moreover, recurrent connectivity with the thalamus, structured bursts of spiking, and cortical oscillations (not to mention other mechanisms like neuromodulation) could control the storage of information over time, to facilitate learning based on temporal prediction. These concepts begin to suggest preliminary, exploratory models for how the detailed anatomy and physiology of the cortex could be interpreted within a machine-learning framework that goes beyond backpropagation. But these are early days: we still lack detailed structural/molecular and functional maps of even a single local cortical microcircuit.

\subsubsection{One-shot learning}

Human learning is often one-shot: it can take just a single exposure to a stimulus to never forget it, as well as to generalize from it to new examples. One way of allowing networks to have such properties is what is described by I-theory, in the context of learning invariant representations for object recognition~\citep{Anselmi2015}. Instead of training via gradient descent, image templates are stored in the weights of simple-complex cell networks while objects undergo transformations, similar to the use of stored templates in HMAX~\citep{Serre2007}. The theories then aim to show that you can invariantly and discriminatively represent objects using a single sample, even of a new class~\citep{Anselmi2015}. 

Additionally, the nervous system may have a way of replaying reality over and over, allowing to move an item from episodic memory into a long-term memory in the neural network~\citep{Ji2007}. This solution effectively uses many iterations of weight updating to fully learn a single item, even if one has only been exposed to it once. 

Finally, higher-level systems in the brain may be able to implement Bayesian learning of sequential programs, which is a powerful means of one-shot learning~\citep{Lake2015}. This type of cognition likely relies on an interaction between multiple brain areas such as the prefrontal cortex and basal ganglia. Computer models, and neural network based models in particular~\citep{Rezende2016}, have not yet reached fully human-like performance in this area, despite significant recent advances~\citep{Lake2015}. 

These potential substrates of one-shot learning rely on mechanisms other than simple gradient descent. It should be noted, though, that recent architectural advances, including specialized spatial attention and feedback mechanisms~\citep{Rezende2016}, as well as specialized memory mechanism~\citep{Santoro2016}, do allow some types of one-shot generalization to be driven by backpropagation-based learning.

\subsubsection{Active learning}

Human learning is often active and deliberate. It seems likely that, in human learning, actions are chosen so as to generate interesting training examples, and sometimes also to test specific hypotheses. Such ideas of active learning and ``child as scientist'' go back to Piaget and have been elaborated more recently~\citep{gopnik2000scientist}. We want our learning to be based on maximally informative samples, and active querying of the environment (or of internal subsystems) provides a way route to this.

At some level of organization, of course, it would seem useful for a learning system to develop explicit representations of its uncertainty, since this can be used to guide the system to actively seek the information that would reduce its uncertainty most quickly. Moreover, there are population coding mechanisms that could support explicit probabilistic computations~\citep{Ma2006, zemel1997combining, gershman2016complex, eliasmith2011normalization, rao2004bayesian, sahani2003doubly}. Yet it is unclear to what extent and at what levels the brain uses an explicitly probabilistic framework, or to what extent probabilistic computations are emergent from other learning processes~\citep{Orhan2016}. 

Standard gradient descent does not incorporate any such adaptive sampling mechanism, e.g., it does not deliberately sample data so as to maximally reduce its uncertainty. Interestingly, however, stochastic gradient descent can be used to generate a system that samples adaptively~\citep{bouchard2015accelerating, alain2015variance}. In other words, a system can learn, by gradient descent, how to choose its own input data samples in order to learn most quickly from them by gradient descent.

Ideally, the learner learns to choose actions that will lead to the largest improvements in its prediction or data compression performance~\citep{Schmidhuber2010}. In~\citep{Schmidhuber2010}, this is done in the framework of reinforcement learning, and incorporates a mechanisms for the system to measure its own rate of learning. In other words, it is possible to reinforcement-learn a policy for selecting the most interesting inputs to drive learning. Adaptive sampling methods are also known in reinforcement learning that can achieve optimal Bayesian exploration of Markov Decision Process environments~\citep{guez2012efficient, sun2011planning}. 

These approaches achieve optimality in an arbitrary, abstract environment. But of course, evolution may also encode its implicit knowledge of the organism's natural environment, the behavioral goals of the organism, and the developmental stages and processes which occur inside the organism, as priors or heuristics which would further constrain the types of adaptive sampling that are optimal in practice. For example, simple heuristics like seeking certain perceptual signatures of novelty, or more complex heuristics like monitoring situations that other people seem to find interesting, might be good ways to bias sampling of the environment so as to learn more quickly. Other such heuristics might be used to give internal brain systems the types of training data that will be most useful to those particular systems at any given developmental stage.

We are only beginning to understand how active learning might be implemented in the brain. We speculate that multiple mechanisms, specialized to different brain systems and spatio-temporal scales, could be involved. The above examples suggest that at least some such mechanisms could be understood from the perspective of optimizing cost functions.

\subsection{Differing biological requirements for supervised and reinforcement learning}

We have described how the brain could implement learning mechanisms of comparable power to backpropagation. But in many cases, the system may be more limited by the available training signals than by the optimization process itself. In machine learning, one distinguishes supervised learning, reinforcement learning and unsupervised learning, and the training data limitation manifests differently in each case. 

Both supervised and reinforcement learning require some form of teaching signal, but the nature of the teaching signal in supervised learning is different from that in reinforcement learning. In supervised learning, the trainer provides the entire vector of errors for the output layer and these are back-propagated to compute the gradient: a locally optimal direction in which to update all of the weights of a potentially multi-layer and/or recurrent network. In reinforcement learning, however, the trainer provides a scalar evaluation signal, but this is not sufficient to derive a low-variance gradient. Hence, some form of trial and error twiddling must be used to discover how to increase the evaluation signal. Consequently, reinforcement learning is generally much less efficient than supervised learning. 

Reinforcement learning in shallow networks is simple to implement biologically. For reinforcement learning of a deep network to be biologically plausible, however, we need a more powerful learning mechanism, since we are learning based on a more limited evaluation signal than in the supervised case: we do not have the full target pattern to train towards. Nevertheless, approximations of gradient descent can be achieved in this case, and there are cases in which the scalar evaluation signal of reinforcement learning can be used to efficiently update a multi-layer network by gradient descent. The ``attention-gated reinforcement learning'' (AGREL) networks of~\citep{Stanisor2013, Brosch2015, Roelfsema2005}, and variants like KickBack~\citep{Balduzzi2014}, give a way to compute an approximation to the full gradient in a reinforcement learning context using a feedback-based attention mechanism for credit assignment within the multi-layer network. The feedback pathway, together with a diffusible reward signal, together gate plasticity. For networks with more than three layers, this gives rise to a model based on columns containing parallel feedforward and feedback pathways. The process is still not as efficient as backpropagation, but it seems that this form of feedback can make reinforcement learning in multi-layer networks more efficient than a naive node perturbation or weight perturbation approach. 

The machine-learning field has recently been tackling the question of credit assignment in deep reinforcement learning. Deep Q-learning~\citep{Mnih2015} demonstrates reinforcement learning in a deep network, wherein most of the network is trained via backpropagation. In regular Q learning, we define a function Q, which estimates the best possible sum of future rewards (the return) if we are in a given state and take a given action. In deep Q learning, this function is approximated by a neural network that, in effect, estimates action-dependent returns in a given state. The network is trained using backpropagation of local errors in Q estimation, using the fact that the return decomposes into the current reward plus the discounted estimate of future return at the next moment. During training, as the agent acts in the environment, a series of loss functions is generated at each step, defining target patterns that can be used as the supervision signal for backpropagation. As Q is a highly nonlinear function of the state, tricks are sometimes needed to make deep Q learning efficient and stable, including experience replay and a particular type of mini-batch training. It is also necessary to store the outputs from the previous iteration (or clone the entire network) in evaluating the loss function for the subsequent iteration\footnote{Many other reinforcement learning algorithms, including REINFORCE~\citep{Williams1992}, can be implemented as fully online algorithms using ``eligibility traces'', which accumulate the sensitivity of action distributions to parameters in a temporally local manner~\citep{sutton1998reinforcement}.}.

This process for generating learning targets provides a kind of bridge between reinforcement learning and efficient backpropagation-based gradient descent learning\footnote{\citet{zaremba2015reinforcement} also bridges reinforcement learning and backpropagation learning in the same system, in the context of a neural network controlling discrete interfaces, and illustrates some of the challenges of this approach: compared to an end-to-end backpropagation-trained Neural Turing Machine~\citep{Graves2014}, reinforcement based training allows training of only relatively simple algorithmic tasks. Special measures need to be taken to make reinforcement efficient, including limiting the number of possible actions, subtracting a baseline reward, and training the network using a curriculum schedule.}. Importantly, only temporally local information is needed making the approach relatively compatible with what we know about the nervous system. 

Even given these advances, a key remaining issue in reinforcement learning is the problem of long timescales, e.g., learning the many small steps needed to navigate from London to Chicago. Many of the formal guarantees of reinforcement learning~\citep{williams1993tight}, for example, suggest that the difference between an optimal policy and the learned policy becomes increasingly loose as the discount factor shifts to take into account reward at longer timescales. Although the degree of optimality of human behavior is unknown, people routinely engage in adaptive behaviors that can take hours or longer to carry out, by using specialized processes like \emph{prospective memory} to ``remember to remember'' relevant variables at the right times, permitting extremely long timescales of coherent action. Machine learning has not yet developed methods to deal with such a wide range of timescales and scopes of hierarchical action. Below we discuss ideas of hierarchical reinforcement learning that may make use of callable procedures and sub-routines, rather than operating explicitly in a time domain.

As we will discuss below, some form of deep reinforcement learning may be used by the brain for purposes beyond optimizing global rewards, including the training of local networks based on diverse internally generated cost functions. Scalar reinforcement-like signals are easy to compute, and easy to deliver to other areas, making them attractive mechanistically. If the brain does employ internally computed scalar reward-like signals as a basis for cost functions, it seems likely that it will have found an efficient means of reinforcement-based training of deep networks, but it is an open question whether an analog of deep Q networks, AGREL, or some other mechanism entirely, is used in the brain for this purpose. Moreover, as we will discuss further below, it is possible that reinforcement-type learning is made more efficient in the context of specialized brain systems like short term memories, replay mechanisms, and hierarchically organized control systems. These specialized systems could reduce reliance on a need for powerful credit assignment mechanisms for reinforcement learning. Finally, if the brain uses a diversity of scalar reward-like signals to implement different cost functions, then it may need to mediate delivery of those signals via a comparable diversity of molecular substrates. The great diversity of neuromodulatory signals, e.g., neuropeptides, in the brain~\citep{Bargmann2012, Bargmann2013} makes such diversity quite plausible, and moreover, the brain may have found other, as yet unknown, mechanisms of diversifying reward-like signaling pathways and enabling them to act independently of one another.

\section{The cost functions are diverse across brain areas and time}

In the last section, we argued that the brain can optimize functions. This raises the question of what functions it optimizes. Of course, in the brain, a cost function will itself be created (explicitly or implicitly) by a neural network shaped by the genome. Thus, the cost function used to train a given sub-network in the brain is a key innate property that can be built into the system by evolution. It may be much cheaper in biological terms to specify a cost function that allows the rapid learning of the solution to a problem than to specify the solution itself. 

In \textbf{Hypothesis 2}, we proposed that the brain optimizes not a single ``end-to-end'' cost function, but rather a diversity of internally generated cost functions specific to particular functions. To understand how and why the brain may use a diversity of cost functions, it is important to distinguish the differing types of cost functions that would be needed for supervised, unsupervised and reinforcement learning. We can also seek to identify types of cost functions that the brain may need to generate from a functional perspective, and how each may be implemented as supervised, unsupervised, reinforcement-based or hybrid systems.

\subsection{How cost functions may be represented and applied}

What additional circuitry is required to actually impose a cost function on an optimizing network? In the most familiar case, supervised learning may rely on computing a vector of errors at the output of a network, which will rely on some comparator circuitry to compute the difference between the network outputs and the target values. This difference could then be backpropagated to earlier layers. An alternative way to impose a cost function is to ``clamp'' the output of the network, forcing it to occupy a desired target state. Such clamping is actually assumed in some of the putative biological implementations of backpropagation described above, such as XCAL and target propagation. Alternatively, as described above, scalar reinforcement signals are attractive as internally-computed cost functions, but using them in deep networks requires special mechanisms for credit assignment.

In unsupervised learning, cost functions may not take the form of externally supplied training or error signals, but rather can be built into the dynamics inherent to the network itself, i.e., there may be no need for a \emph{separate} circuit to compute and impose a cost function on the network. Indeed, beginning with Hopfield's definition of an energy function for learning in certain classes of symmetric network~\citep{Hopfield1982}, researchers have discovered networks with inherent learning dynamics that implicitly optimizes certain objectives, such as statistical reconstruction of the input (e.g., via stochastic relaxation in Boltzmann machines~\citep{Ackley1985}), or the achievement of certain properties like temporally stable or sparse representations. 

Alternatively, explicit cost functions could be computed, delivered to a network, and used for unsupervised learning, following a variety of principles being discovered in machine learning (e.g.,~\citep{Radford2015, Lotter2015}), which typically find a way to encode the cost function into the error derivatives which are backpropagated. For example, prediction errors naturally give rise to error signals for unsupervised learning, as do reconstruction errors in autoencoders, and these error signals can also be augmented with additional penalty or regularization terms that enforce objectives like sparsity or continuity, as described below. In the next sections, we elaborate on these and other means of specifying and delivering cost functions in different learning contexts. 

\subsection{Cost functions for unsupervised learning}

There are many objectives that can be optimized in an unsupervised context, to accomplish different kinds of functions or guide a network to form particular kinds of representations.

\subsubsection{Matching the statistics of the input data using generative models}

In one common form of unsupervised learning, higher brain areas attempt to produce samples that are statistically similar to those actually seen in lower layers. For example, the wake-sleep algorithm~\citep{Hinton1995} requires the sleep mode to sample potential data points whose distribution should then match the observed distribution. Unsupervised pre-training of deep networks is an instance of this~\citep{Erhan2009}, typically making use of a stacked auto-encoder framework. Similarly, in target propagation~\citep{Bengio2014}, a top-down circuit, together with lateral information, has to produce data that directs the local learning of a bottom-up circuit and vice-versa. Ladder autoencoders make use of lateral connections and local noise injection to introduce an unsupervised cost function, based on internal reconstructions, that can be readily combined with supervised cost functions defined on the network’s top layer outputs~\citep{Valpola2015}. Compositional generative models generate a scene from discrete combinations of template parts and their transformations~\citep{Wang2014}, in effect performing a rendering of a scene based on its structural description. Hinton and colleagues have also proposed cortical ``capsules''~\citep{Tang2013, Tang2012, Hinton2011} for compositional inverse rendering. The network can thus implement a statistical goal that embodies some understanding of the way that the world produces samples.

Learning rules for generative models have historically involved local message passing of a form quite different from backpropagation, e.g., in a multi-stage process that first learns one layer at a time and then fine-tunes via the wake-sleep algorithm~\citep{Hinton2006}. Message-passing implementations of probabilistic inference have also been proposed as an explanation and generalization of deep convolutional networks~\citep{Patel2015, Chen2014}. Various mappings of such processes onto neural circuitry have been attempted~\citep{Sountsov2015, George2009, Lee2011}. Feedback connections tend to terminate in distinct layers of cortex relative to the feedforward ones~\citep{Callaway2004} making the idea of separate but interacting networks for recognition and generation potentially attractive. Interestingly, such sub-networks might even be part of the same neuron and map onto ``apical'' versus ``basal'' parts of the dendritic tree~\citep{Kording2001, Urbanczik2014}. 

Generative models can also be trained via backpropagation. Recent advances have shown how to perform variational approximations to Bayesian inference inside backpropagation-based neural networks~\citep{Kingma2013}, and how to exploit this to create generative models~\citep{Eslami2016, Gregor2015, Goodfellow2014, Radford2015}. Through either explicitly statistical or gradient descent based learning, the brain can thus obtain a probabilistic model that simulates features of the world. 

\subsubsection{Cost functions that approximate properties of the world}

A perceiving system should exploit statistical regularities in the world that are not present in an arbitrary dataset or input distribution. For example, objects are sparse: there are far fewer objects than there are potential places in the world, and of all possible objects there is only a small subset visible at any given time. As such, we know that the output of an object recognition system must have sparse activations. Building the assumption of sparseness into simulated systems replicates a number of representational properties of the early visual system~\citep{Olshausen1997, Rozell2008}, and indeed the original paper on sparse coding obtained sparsity by gradient descent optimization of a cost function~\citep{Olshausen1996}. A range of unsupervised machine learning techniques, such as the sparse autoencoders~\citep{Le2011} used to discover cats in YouTube videos, build sparseness into neural networks. Building in such spatio-temporal sparseness priors should serve as an ``inductive bias''~\citep{mitchell1980need} that can accelerate learning.

But we know much more about the regularities of objects. As young babies, we already know~\citep{Bremner2015} that objects tend to persist over time. The emergence or disappearance of an object from a region of space is a rare event. Moreover, object locations and configurations tend to be coherent in time. We can formulate this prior knowledge as a cost function, for example by penalizing representations which are not temporally continuous. This idea of continuity is used in a great number of artificial neural networks and related models~\citep{Mobahi2009, Wiskott2002, Foldiak2008}. Imposing continuity within certain models gives rise to aspects of the visual system including complex cells~\citep{Kording2004}, specific properties of visual invariance~\citep{Isik2012}, and even other representational properties such as the existence of place cells~\citep{Wyss2006, Franzius2007}. Unsupervised learning mechanisms that maximize temporal coherence or slowness are increasingly used in machine learning\footnote{Temporal continuity is exploited in~\citet{Poggio2015}, which analyzes many properties of deep convolutional networks with respect to their biological plausibility, including their apparent need for large amounts of supervised training data, and concludes that the environment may in fact provide a sufficient number of ``implicitly'', though not explicitly, labeled examples to train a deep convolutional network for object recognition. Implicit labeling of object identity, in this case, arises from temporal continuity: successive frames of a video are likely to have the same objects in similar places and orientations. This allows the brain to derive an invariant signature of object identity which is independent of transformations like translations and rotations, but which does not yet associate the object with a specific name or label. Once such an invariant signature is established, however, it becomes basically trivial to associate the signature with a label for classification~\citep{Anselmi2015}. \citet{Poggio2015} also suggests specific means, in the context of I-theory~\citep{Anselmi2015}, by which this training could occur via the storage of image templates using Hebbian mechanisms among simple and complex cells in the visual cortex. Thus, in this model, the brain has used its implicit knowledge of the temporal continuity of object motion to provide a kind of minimal labeling that is sufficient to bootstrap object recognition. Although not formulated as a cost function, this shows how usefully the assumption of temporal continuity could be exploited by the brain.}.

We also know that objects tend to undergo predictable sequences of transformations, and it is possible to build this assumption into unsupervised neural learning systems~\citep{George2009}. The minimization of prediction error explains a number of properties of the nervous system~\citep{Friston2007, Huang2011}, and biologically plausible theories are available for how cortex could learn using prediction errors by exploiting temporal differences~\citep{OReilly2014a} or top-down feedback~\citep{George2009}. In one implementation, a system can simply predict the next input delivered to the system and can then use the difference between the actual next input and the predicted next input as a full vectorial error signal for supervised gradient descent. Thus, rather than optimization of prediction error being implicitly implemented by the network dynamics, the prediction error is used as an explicit cost function in the manner of supervised learning, leading to error derivatives which can be back-propagated. Then, no special learning rules beyond simple backpropagation are needed. This approach has recently been advanced within machine learning~\citep{Lotter2015, lotter2016deep}. Recently, combining such prediction-based learning with a specific gating mechanism has been shown to lead to unsupervised learning of disentangled representations~\citep{Whitney2016}. Neural networks can also be designed to learn to invert spatial transformations~\citep{Jaderberg2015a}. Statistically describing transformations or sequences is thus an unsupervised way of learning representations.

Furthermore, there are multiple modalities of input to the brain. Each sensory modality is primarily connected to one part of the brain\footnote{Although, some multi-sensory integration appears to occur even in the early sensory cortices~\citep{murray2012cortical}.}. But higher levels of cortex in each modality are heavily connected to the other modalities. This can enable forms of self-supervised learning: with a developing visual understanding of the world we can predict its sounds, and then test those predictions with the auditory input, and vice versa. The same is true about multiple parts of the same modality: if we understand the left half of the visual field, it tells us an awful lot about the right. Maximizing mutual information is a natural way of improving learning~\citep{becker1992self, Mohamed2015}, and there are many other ways in which multiple modalities or processing streams could mutually train one another. Relatedly, we can use observations of one part of a visual scene to predict the contents of other parts~\citep{Oord2016, noroozi2016unsupervised}, and optimize a cost function that reflects the discrepancy. This way, each modality effectively produces training signals for the others.

\subsection{Cost functions for supervised learning}

In what cases might the brain use supervised learning, given that it requires the system to ``already know'' the exact target pattern to train towards? One possibility is that the brain can store records of states that led to good outcomes. For example, if a baby reaches for a target and misses, and then tries again and successfully hits the target, then the difference in the neural representations of these two tries reflects the direction in which the system should change. The brain could potentially use a comparator circuit -- a non-trivial task since neural activations are always positive, although different neuron types can be excitatory vs. inhibitory -- to directly compute this vectorial difference in the neural population codes and then apply this difference vector as an error signal.

Another possibility is that the brain uses supervised learning to implement a form of ``chunking'', i.e., a consolidation of something the brain already knows how to do: routines that are initially learned as multi-step, deliberative procedures could be compiled down to more rapid and automatic functions by using supervised learning to train a network to mimic the overall input-output behavior of the original multi-step process. Such a process is assumed to occur in cognitive models like ACT-R~\citep{Servan-Schreiber1990}, and methods for compressing the knowledge in neural networks into smaller networks are also being developed~\citep{Ba2014}. Thus supervised learning can be used to train a network to do in ``one step'' what would otherwise require long-range routing and sequential recruitment of multiple systems.

\subsection{Repurposing reinforcement learning for diverse internal cost functions}

Certain generalized forms of reinforcement learning may be ubiquitous throughout the brain. Such reinforcement signals may be repurposed to optimize diverse internal cost functions. These internal cost functions could be specified at least in part by genetics. 

Some brain systems such as in the striatum appear to learn via some form of temporal difference reinforcement learning~\citep{Tesauro1995, Foster2000}. This is reinforcement learning based on a global value function~\citep{OReilly2014} that predicts total future reward or utility for the agent. Reward-driven signaling is not restricted to the striatum, and is present even in primary visual cortex~\citep{Chubykin2013, Stanisor2013}. Remarkably, the reward signaling in primary visual cortex is mediated in part by glial cells~\citep{Takata2011}, rather than neurons, and involves the neurotransmitter acetylcholine~\citep{Chubykin2013, Hangya2015}. On the other hand, some studies have suggested that visual cortex learns the basics of invariant object recognition in the absence of reward~\citep{Li2012}, perhaps using reinforcement only for more refined perceptual learning~\citep{Roelfsema2010}. 

But beyond these well-known global reward signals, we argue that the basic mechanisms of reinforcement learning may be widely re-purposed to train local networks using a variety of internally generated error signals. These internally generated signals may allow a learning system to go beyond what can be learned via standard unsupervised methods, effectively guiding or steering the system to learn specific features or computations~\citep{Ullman2012}.

\subsubsection{Cost functions for bootstrapping learning in the human environment}

Special, internally-generated signals are needed specifically for learning problems where standard unsupervised methods -- based purely on matching the statistics of the world, or on optimizing simple mathematical objectives like temporal continuity or sparsity -- will fail to discover properties of the world which are statistically weak in an objective sense but nevertheless have special significance to the organism~\citep{Ullman2012}. Indigo bunting birds, for example, learn a template for the constellations of the night sky long before ever leaving the nest to engage in navigation-dependent tasks~\citep{emlen1967migratory}. This memory template is directly used to determine the direction of flight during migratory periods, a process that is modulated hormonally so that winter and summer flights are reversed. Learning is therefore a multi-phase process in which navigational cues are memorized prior to the acquisition of motor control.

In humans, we suspect that similar multi-stage bootstrapping processes are arranged to occur. Humans have innate specializations for social learning. We need to be able to read their expressions as indicated with hands and faces. Hands are important because they allow us to learn about the set of actions that can be produced by agents~\citep{Ullman2012}. Faces are important because they give us insight into what others are thinking. People have intentions and personalities that differ from one another, and their feelings are important. How could we hack together cost functions, built on simple genetically specifiable mechanisms, to make it easier for a learning system to discover such behaviorally relevant variables?

Some preliminary studies are beginning to suggest specific mechanisms and heuristics that humans may be using to bootstrap more sophisticated knowledge. In a groundbreaking study, \citet{Ullman2012} asked how could we explain hands, to a system that does not already know about them, in a cheap way, without the need for labeled training examples? Hands are common in our visual space and have special roles in the scene: they move objects, collect objects, and caress babies. Building these biases into an area specialized to detect hands could guide the right kind of learning, by providing a downstream learning system with many likely positive examples of hands on the basis of innately-stored, heuristic signatures about how hands tend to look or behave~\citep{Ullman2012}. Indeed, an internally supervised learning algorithm containing specialized, hard-coded biases to detect hands, on the basis of their typical motion properties, can be used to bootstrap the training of an image recognition module that learns to recognize hands based on their appearance. Thus, a simple, hard-coded module bootstraps the training of a much more complex algorithm for visual recognition of hands. 

\citet{Ullman2012} then further exploits a combination of hand and face detection to bootstrap a predictor for gaze direction, based on the heuristic that faces tend to be looking towards hands. Of course, given a hand detector, it also becomes much easier to train a system for reaching, crawling, and so forth. Efforts are underway in psychology to determine whether the heuristics discovered to be useful computationally are, in fact, being used by human children during learning~\citep{fausey2016faces, yu2013joint}.

Ullman refers to such primitive, inbuilt detectors as innate ``proto-concepts''~\citep{Ullman2012}. Their broader claim is that such pre-specification of mutual supervision signals can make learning the relevant features of the world far easier, by giving an otherwise unsupervised learner the right kinds of hints or heuristic biases at the right times.  Here we call these approximate, heuristic cost functions ``bootstrap cost functions''. The purpose of the bootstrap cost functions is to reduce the amount of data required to learn a specific feature or task, but at the same time to avoid a need for fully unsupervised learning.

Could the neural circuitry for such a bootstrap hand-detector be pre-specified genetically? The precedent from other organisms is strong: for example, it is famously known that the frog retina contains circuitry sufficient to implement a kind of ``bug detector''~\citep{Lettvin1959}. Ullman's hand detector, in fact, operates via a simple local optical flow calculation to detect ``mover'' events. This type of simple, local calculation could potentially be implemented in genetically-specified and/or spontaneously self-organized neural circuitry in the retina or early dorsal visual areas~\citep{biilthoff1989parallel}, perhaps similarly to the frog's ``bug detector''. 

How could we explain faces without any training data? Faces tend to have two dark dots in their upper half, a line in the lower half and tend to be symmetric about a vertical axis. Indeed, we know that babies are very much attracted to things with these generic features of upright faces starting from birth, and that they will acquire face-specific cortical areas\footnote{In the visual system, it is still unknown why a clustered spatial pattern of representational categories arises, e.g., a physically localized ``area'' that seems to correspond to representations of faces~\citep{kanwisher1997fusiform}, another area for representations of visual word forms~\citep{mccandliss2003visual}, and so on. It is also unknown why this spatial pattern seems to be largely reproducible across individuals. Some theories are based on bottom-up correlation-based clustering or neuronal competition mechanisms, which generate category-selective regions as a byproduct. Other theories suggest a computational reason for this organization, in the context of I-theory~\citep{Anselmi2015}, involving the limited ability to generalize transformation-invariances learned for one class of objects to other classes~\citep{leibo2015invariance}. Areas for abstract culture-dependent concepts, like the visual word form area, suggest that the decomposition cannot be ``purely genetic''. But it is conceivable that these areas could at least in part reflect different local cost functions.} in their first few years of life if not earlier~\citep{mckone2009cognitive}. It is easy to define a local rule that produces a kind of crude face detector (e.g., detecting two dots on top of a horizontal line), and indeed some evidence suggests that the brain can rapidly detect faces without even a single feed-forward pass through the ventral visual stream~\citep{Crouzet2011}. The crude detection of human faces used together with statistical learning should be analogous to semi-supervised learning~\citep{sukhbaatar2014training} and could allow identifying faces with high certainty.

Humans have areas devoted to emotional processing, and the brain seems to embody prior knowledge about the structure of emotions: emotions should have specific types of strong couplings to various other higher-level variables, should be expressed through the face, and so on. This prior knowledge, encoded into brain structure via evolution, could allow learning signals to come from the right places and to appear developmentally at the right times. What about agency? It makes sense to describe, when dealing with high-level thinking, other beings as optimizers of their own goal functions. It appears that heuristically specified notions of goals and agency are infused into human psychological development from early infancy and that notions of agency are used to bootstrap heuristics for ethical evaluation~\citep{Skerry2014, Hamlin2007}. Algorithms for establishing more complex, innately-important social relationships such as joint attention are under study~\citep{Gao2014}, building upon more primitive proto-concepts like face detectors and Ullman's hand detectors~\citep{Ullman2012}. The brain can thus use innate detectors to create cost functions and training procedures to train the next stages of learning.

It is intuitive to ask whether this type of bootstrapping poses a kind of ``chicken and egg'' problem: if the brain already has an inbuilt heuristic hand detector, how can it be used to train a detector that performs any better than those heuristics? After all, isn't a trained system only as good as its training data? The work of \citet{Ullman2012} illustrates why this is not the case. First, the ``innate detector'' can be used to train a downstream detector that operates based on different cues: for example, based on the spatial and body context of the hand, rather than its motion. Second, once multiple such pathways of detection come into existence, they can be used to improve each other. In \citet{Ullman2012}, appearance, body context, and mover motion are all used to bootstrap off of one another, creating a detector that is better than any of it training heuristics. In effect, the innate detectors are used not as supervision signals per se, but rather to guide or steer the learning process, enabling it to discover features that would otherwise be difficult. If such affordances can be found in other domains, it seems likely that the brain would make extensive use of them to ensure that developing animals learn the precise patterns of perception and behavior needed to ensure their later survival and reproduction.

Thus, generalizing previous ideas \citep{Ullman2012,Poggio2015}, we suggest that the brain uses optimization with respect to internally generated heuristic detection signals to bootstrap learning of biologically relevant features which would otherwise be missed by an unsupervised learner. In one possible implementation, such bootstrapping may occur via reinforcement learning, using the outputs of the innate detectors as local reinforcement signals, and perhaps using mechanisms similar to~\citep{Stanisor2013, Rombouts2015, Brosch2015, Roelfsema2005} to perform reinforcement learning through a multi-layer network. It is also possible that the brain could use such internally generated heuristic detectors in other ways, for example to bias the inputs delivered to an unsupervised learning network towards entities of interest to humans (Joscha Bach, personal communication), or to directly train simple classifiers~\citep{Ullman2012}.

\subsubsection{Cost functions for story generation and understanding}

It has been widely noticed in cognitive science and AI that the generation and understanding of stories are crucial to human cognition. Researchers such as Winston have framed story understanding as the key to human-like intelligence~\citep{winston2011strong}. Stories consist of a linear sequence of episodes, in which one episode refers to another through cause and effect relationships, with these relationships often involving the implicit goals of agents. Many other cognitive faculties, such as conceptual grounding of language, could conceivably emerge from an underlying internal representation in terms of stories.

Perhaps the ultimate series of bootstrap cost functions would be those which would direct the brain to utilize its learning networks and specialized systems so as to construct representations that are specifically useful as components of stories, to spontaneously chain these representations together, and to update them through experience and communication. How could such cost functions arise? One possibility is that they are bootstrapped through imitation and communication, where a child learns to mimic the story-telling behavior of others. Another possibility is that useful representations and primitives for stories emerge spontaneously from mechanisms for learning state and action chunking in hierarchical reinforcement learning and planning. Yet another is that stories emerge from learned patterns of saliency-directed memory storage and recall (e.g.,~\citep{xiong2016dynamic}). In addition, priors that direct the developing child's brain to learn about and attend to social agency seem to be important for stories. These systems will be discussed in more detail below.

\section{Optimization occurs in the context of specialized structures}

Optimization of initially unstructured ``blank slate'' networks is not sufficient to generate complex cognition in the brain, we argue, even given a diversity of powerful genetically-specified cost functions and local learning rules, as we have posited above. Instead, in \textbf{Hypothesis 3}, we suggest that specialized, pre-structured architectures are needed for at least two purposes. 

First, pre-structured architectures are needed to allow the brain to find efficient solutions to certain types of problems. When we write computer code, there are a broad range of algorithms and data structures employed for different purposes: we may use dynamic programming to solve planning problems, trees to efficiently implement nearest neighbor search, or stacks to implement recursion. Having the right kind of algorithm and data structure in place to solve a problem allows it to be solved efficiently, robustly and with a minimum amount of learning or optimization needed. This observation is concordant with the increasing use of pre-specialized architectures and specialized computational components in machine learning~\citep{Graves2014, Weston2014, Neelakantan2015}. In particular, to enable the learning of efficient computational solutions, the brain may need pre-specialized systems for planning and executing sequential multi-step processes, for accessing memories, and for forming and manipulating compositional and recursive structures.

Second, the training of optimization modules may need to be coordinated in a complex and dynamic fashion, including delivering the right training signals and activating the right learning rules in the right places and at the right times. To allow this, the brain may need specialized systems for storing and routing data, and for flexibly routing training signals such as target patterns, training data, reinforcement signals, attention signals, and modulatory signals. These mechanisms may need to be at least partially in place in advance of learning.

Looking at the brain, we indeed seem to find highly conserved structures, e.g., cortex, where it is theorized that a similar type of learning and/or computation is happening in multiple places~\citep{Douglas2004, braitenberg1991anatomy}. But we also see a large number of specialized structures, including thalamus, hippocampus, basal ganglia and cerebellum~\citep{Solari2011}. Some of these structures evolutionarily pre-date~\citep{Lee2015} the cortex, and hence the cortex may have evolved to work in the context of such specialized mechanisms. For example, the cortex may have evolved as a trainable module for which the training is orchestrated by these older structures.

Even within the cortex itself, microcircuitry within different areas may be specialized: tinkered variations on a common ancestral microcircuit scaffold could potentially allow different cortical areas, such as sensory areas vs. prefrontal areas, to be configured to adopt a number of qualitatively distinct computational and learning configurations~\citep{Marcus2014, Marcus2014a, Yuste2005}, even while sharing a common gross physical layout and communication interface. Within cortex, over forty distinct cell types – differing in such aspects as dendritic organization, distribution throughout the six cortical layers, connectivity pattern, gene expression, and electrophysiological properties – have already been found~\citep{Zeisel2015, Markram2015}. Central pattern generator circuits provide an example of the kinds of architectures that can be pre-wired into neural microcircuitry, and may have evolutionary relationships with cortical circuits~\citep{Yuste2005}. Thus, while the precise degree of architectural specificity of particular cortical regions is still under debate~\citep{Marcus2014,Marcus2014a}, various mechanism could offer pre-specified heterogeneity.

In this section, we explore the kinds of computational problems for which specialized structures may be useful, and attempt to map these to putative elements within the brain. Our preliminary sketch of a functional decomposition can be viewed as a summary of suggestions for specialized functions that have been made throughout the computational neuroscience literature, and is influenced strongly by the models of O'Reilly, Eliasmith, Grossberg, Marcus, Hayworth and others~\citep{OReilly2006a, Eliasmith2012, Marcus2001, Grossberg2013, Hayworth2012}. The correspondence between these models and actual neural circuitry is, of course, still the subject of extensive debate. 

Many of the computational and neural concepts sketched here are preliminary and will need to be made more rigorous through future study. Our knowledge of the functions of particular brain areas, and thus our proposed mappings of certain computations onto neuroanatomy, also remains tentative. Finally, it is still far from established which processes in the brain emerge from optimization of cost functions, which emerge from other forms of self-organization, which are pre-structured through genetics and development, and which rely on an interplay of all these mechanisms. Our discussion here should therefore be viewed as a sketch of potential directions for further study.

\subsection{Structured forms of memory}

One of the central elements of computation is memory. Importantly, multiple different kinds of memory are needed~\citep{squire2004memory}. For example, we need memory that is stored for a long period of time and that can be retrieved in a number of ways, such as in situations similar to the time when the memory was first stored (content addressable memory). We also need memory that we can keep for a short period of time and that we can rapidly rewrite (working memory). Lastly, we need the kind of implicit memory that we cannot explicitly recall, similar to the kind of memory that is classically learned using gradient descent on errors, i.e., sculpted into the weight matrix of a neural network.

\subsubsection{Content addressable memories}

Content addressable memories\footnote{Attractor models of memory in neuroscience tend to have the property that only one memory can be accessed at a time. Yet recent machine learning systems have constructed differentiable addressable memory~\citep{Graves2014} and gating~\citep{Whitney2016} systems by allowing weighted superpositions of memory registers or gates to be queried. It is unclear whether the brain uses such mechanisms.} are classic models in neuroscience~\citep{Hopfield1982}. Most simply, they allow us to recognize a situation similar to one that we have seen before, and to ``fill in'' stored patterns based on partial or noisy information, but they may also be put to use as sub-components of many other functions. Recent research has shown that including such memories allows deep networks to learn to solve problems that previously were out of reach, even of LSTM networks that already have a simpler form of local memory and are already capable of learning long-term dependencies~\citep{Weston2014, Graves2014}. Hippocampal area CA3 may act as an auto-associative memory\footnote{Computational analogies have also been drawn between associative memory storage and object recognition~\citep{Leibo2015}, suggesting the possibility of closely related computations occurring in parts of neocortex and hippocampus. Indeed both areas have some apparent anatomical similarities such as the presence of pyramidal neurons, and it has been suggested that the hippocampus can be thought of as the top of the cortical hierarchy~\citep{Hawkins2007}, responsible for handling and remembering information that could not be fully explained by lower levels of the hierarchy. These connections are still tentative.} capable of content-addressable pattern completion, with pattern separation occurring in the dentate gyrus~\citep{Rolls2013}. Such systems could permit the retrieval of complete memories from partial cues, enabling networks to perform operations similar to database retrieval or to instantiate lookup tables of historical stimulus-response mappings, among numerous other possibilities.

\subsubsection{Working memory buffers}

Cognitive science has long characterized properties of the working memory. It is somewhat limited, with the old idea being that it can represent ``seven plus or minus two'' elements~\citep{Miller1956}. There are many models of working memory~\citep{Wang2012, Singh2006, OReilly2006, Buschman2014, Warden2007}, some of which attribute it to persistent, self-reinforcing patterns of neural activation~\citep{goldman2003robust} in the recurrent networks of the prefrontal cortex. Prefrontal working memory appears to be made up of multiple functionally distinct subsystems~\citep{Markowitz2015}. Neural models of working memory can store not only scalar variables~\citep{SebastianSeung1998}, but also high-dimensional vectors~\citep{Eliasmith2004, Eliasmith2012} or sequences of vectors~\citep{ChooandEliasmith2010}. Working memory buffers seem crucial for human-like cognition, e.g., reasoning, as they allow short-term storage while also -- in conjunction with other mechanisms -- enabling generalization of operations across anything that can fill the buffer.

\subsubsection{Storing state in association with saliency}

Saliency, or interestingness, measures can be used to tag the importance of a memory~\citep{GonzalezAndino2012}. This can allow removal of the boring data from the training set, allowing a mechanism that is more like optimal experimentation. Moreover, saliency can guide memory replay or sampling from generative models, to generate more training data drawn from a distribution useful for learning~\citep{Mnih2015, Ji2007}.  Conceivably, hippocampal replay could allow a batch-like training process, similar to how most machine learning systems are trained, rather than requiring all training to occur in an online fashion. Plasticity mechanisms in memory systems which are gated by saliency are starting to be uncovered in neuroscience~\citep{dudman2007role}. Importantly, the notions of ``saliency'' computed by the brain could be quite intricate and multi-faceted, potentially leading to complex schemes by which specific kinds of memories would be tagged for later context-dependent retrieval. As a hypothetical example, representations of both timing and importance associated with memories could perhaps allow retrieval only of important memories that happened within a certain window of time~\citep{macdonald2011hippocampal, kraus2013hippocampal,rubin2015hippocampal}. Storing and retrieving information selectively based on specific properties of the information itself, or of ``tags'' appended to that information, is a powerful computational primitive that could enable learning of more complex tasks.

\subsection{Structured routing systems}

To use its information flexibly, the brain needs structured systems for routing data. Such systems need to address multiple temporal and spatial scales, and multiple modalities of control. Thus, there are several different kinds of information routing systems in the brain which operate by different mechanisms and under different constraints.

\subsubsection{Attention}

If we can focus on one thing at a time, we may be able to allocate more computational resources to processing it, make better use of scarce data to learn about it, and more easily store and retrieve it from memory\footnote{Attention also arguably solves certain types of perceptual binding problem~\citep{Reynolds1999}.}. Notably in this context, attention allows improvements in learning: if we can focus on just a single object, instead of an entire scene, we can learn about it more easily using limited data. Formal accounts in a Bayesian framework talk about attention reducing the sample complexity of learning~\citep{Chikkerur2010}. Likewise, in models, the processes of applying attention, and of effectively making use of incoming attentional signals to appropriately modulate local circuit activity, can themselves be learned by optimizing cost functions~\citep{mnih2014recurrent, jaramillo2004normative}. The right kinds of attention make processing and learning more efficient, and also allow for a kind of programmatic control over multi-step perceptual tasks.

How does the brain determine where to allocate attention, and how is the attentional signal physically mediated? Answering this question is still an active area of neuroscience. Higher-level cortical areas may be specialized in allocating  attention. The problem is made complex by the fact that there seem to be many different types of attention -- such as object-based, feature-based and spatial attention in vision -- that may be mediated by interactions between different brain areas. The frontal eye fields (area FEF), for example, are important in visual attention, specifically for controlling saccades of the eyes to attended locations. Area FEF contains ``retinotopic'' spatial maps whose activation determines the saccade targets in the visual field. Other prefrontral areas such as the dorsolateral prefrontal cortex and inferior frontal junction are also involved in maintaining representations that specify the targets of certain types of attention. Certain forms of attention may require a complex interaction between brain areas, e.g., to determine targets of attention based on higher-level properties that are represented across multiple areas, like the identity and spatial location of a specific face~\citep{Baldauf2014}.

There are many proposed neural mechanisms of attention, including the idea that synchrony plays a role~\citep{Baldauf2014}, perhaps by creating resonances that facilitate the transfer of information between synchronously oscillating neural populations in different areas. Other proposed mechanisms include specific circuits for attention-dependent signal routing~\citep{anderson1987shifter, Olshausen1993}. Various forms of attention also have specific neurophysiological signatures, such as enhancements in synchrony among neural spikes and with the ambient local field potential, changes in the sharpness of neural tuning curves, and other properties. These diverse effects and signatures of attention may be consequences of underlying pathways that wire up to particular elements of cortical microcircuits to mediate different attentional effects.

\subsubsection{Buffers}

One possibility is that the brain uses distinct groups of neurons, which we can call ``buffers'', to store distinct variables, such as the subject or object in a sentence~\citep{Frankland2015}. Having memory buffers allows the abstraction of a variable. As is ubiquitously used in computer science, this comes with the ability to generalize operations across any variable that could meaningfully fill the buffer and makes computation flexible.

Once we establish that the brain has a number of memory buffers, we need ways for those buffers to interact. We need to be able to take a buffer, do a computation on its contents and store the output into another buffer. But if the representations in each of two groups of neurons are learned, and hence are coded differently, how can the brain ``copy and paste'' information between these groups of neurons? Malsburg argued that such a system of separate buffers is impossible because the neural pattern for ``chair'' in buffer 1 has nothing in common with the neural pattern for ``chair'' in buffer 2 -- any learning that occurs for the contents of buffer 1 would not automatically be transferable to buffer 2. Various mechanisms have been proposed to allow such transferability, which focus on ways in which all buffers could be trained jointly and then later separated so that they can work independently when they need to\footnote{One idea for achieving such transferability is that of a partitionable~\citep{Hayworth2012} or annexable~\citep{Bostrom1996} network. These models posit that a large associative memory network links all the different buffers. This large associative memory network has a number of stable attractor states. These are called ``global'' attractor states since they link across all the buffers. Forcing a given buffer into an activity pattern resembling that of its corresponding ``piece'' of an attractor state will cause the entire global network to enter that global attractor state. During training, all of the connections between buffers are turned on, so that their learned contents, though not identical, are kept in correspondence by being part of the same attractor. Later, the connections between specific buffers can be turned off to allow them to store different information. Copy and paste is then implemented by turning on the connections between a source buffer and a destination buffer~\citep{Hayworth2012}. Copying between a source and destination buffer can also be implemented, i.e., learned, in a deep learning system using methods similar to the addressing mechanisms of the Neural Turing Machine~\citep{Graves2014}.}.

\subsubsection{Discrete gating of information flow between buffers}

Dense connectivity is only achieved locally, but it would be desirable to have a way for any two cortical units to talk to one another, if needed, regardless of their distance from one another, and without introducing crosstalk\footnote{Micro-stimulation experiments, in which an animal learns to behaviorally report stimulation of electrode channels located in diverse cortical regions, suggest that many areas can be routed or otherwise linked to behavioral ``outputs''~\citep{histed2013insights}, although the mechanisms behind this -- e.g., whether this stimulation gives rise to a high-level percept that the animal then uses to make a decision -- are unclear. Likewise, it is possible to reinforcement-train an animal to control the activity of individual neurons~\citep{fetz1969operant, fetz2007volitional}.}. It is therefore critical to be able to dynamically turn on and off the transfer of information between different source and destination regions, in much the manner of a switchboard. Together with attention, such dedicated routing systems can make sure that a brain area receives exactly the information it needs. Such a discrete routing system is, of course, central to cognitive architectures like ACT-R~\citep{Anderson2007}. The key feature of ACT-R is the ability to evaluate the IF clauses of tens of thousands of symbolic rules (called ``productions''), in parallel, approximately every 50 milliseconds. Each rule requires equality comparisons between the contents of many constant and variable memory buffers, and the execution of a rule leads to the conditional routing of information from one buffer to another. 

What controls which long-range routing operations occur when, i.e., where is the switchboad and what controls it? Several models, including ACT-R, have attributed such parallel rule-based control of routing to the action selection circuitry~\citep{Gurney2001, TerrenceC.StewartXuanChoo2010} of the basal ganglia (BG)~\citep{Stocco2010, OReilly2006}, and its interaction with working memory buffers in the prefrontal cortex. In conventional models of thalamo-cortico-striatal loops, competing actions of the direct and indirect pathways through the basal ganglia can inhibit or disinhibit an area of motor cortex, thereby gating a motor action\footnote{Conventionally, models of the basal ganglia involve all or none gating of an action, but recent evidence suggests that the basal ganglia may also have continuous, analog outputs~\citep{yttri2016opponent}.}. Models like~\citep{Stocco2010, OReilly2006} propose further that the basal ganglia can gate not just the transfer of information from motor cortex to downstream actuators, but also the transfer of information between cortical areas. To do so, the basal ganglia would dis-inhibit a thalamic relay~\citep{Sherman2005, Sherman2007} linking two cortical areas. Dopamine-related activity is thought to lead to temporal difference reinforcement learning of such gating policies in the basal ganglia~\citep{Frank2012}. Beyond the basal ganglia, there are also other, separate pathways involved in action selection, e.g., in the prefrontal cortex~\citep{daw2006actions}. Thus, multiple systems including basal ganglia and cortex could control the gating of long-range information transfer between cortical areas, with the thalamus perhaps largely constituting the switchboard itself.

How is such routing put to use in a learning context? One possibility is that the basal ganglia acts to orchestrate the training of the cortex. The basal ganglia  may exert tight control\footnote{It has been suggested that the basic role of the BG is to provide tonic inhibition to other circuits~\citep{Grillner2005}. Release of this inhibition can then activate a ``discrete'' action, such as a motor command. A core function of the BG is thus to choose, based on patterns detected in its input, which of a finite set of actions to initiate via such release of inhibition. In many models of the basal ganglia’s role in cognitive control, the targets of inhibition are thalamic relays~\citep{Sherman2005}, which are set in a default ``off'' state by tonic inhibition from the basal ganglia. Upon disinhibition of a relay, information is transferred from one cortical location to another -- a form of conditional ``gating'' of information transfer. For example, the BG might be able to selectively ``clamp'' particular groups of cortical neurons in a fixed state, while leaving others free to learn and adapt. It could thereby enforce complex training routines, perhaps similar to those used to force the emergence of disentangled representations in~\citep{Kulkarni2015}. The idea that the basal ganglia can train the cortex is not new, and already appears to have considerable experimental and anatomical support~\citep{ashby2007neurobiological, Ashby2010, Pasupathy2005}.} over the cortex, helping to determine when and how it is trained. Indeed, because the basal ganglia pre-dates the cortex evolutionarily, it is possible that the cortex evolved as a flexible, trainable resource that could be harnessed by existing basal ganglia circuitry. All of the main regions and circuits of the basal ganglia are conserved from our common ancestor with the lamprey more than five hundred million years ago. The major part of the basal ganglia even seems to be conserved from our common ancestor with insects~\citep{Strausfeld2013}. Thus, in addition to its real-time action selection and routing functions, the basal ganglia may sculpt how the cortex learns.

\subsection{Structured state representations to enable efficient algorithms}

Certain algorithmic problems benefit greatly from particular types of representation and transformation, such as a grid-like representation of space. In some cases, rather than just waiting for them to emerge via gradient descent optimization of appropriate cost functions, the brain may be pre-structured to facilitate their creation.

\subsubsection{Continuous predictive control}

We often have to plan and execute complicated sequences of actions on the fly, in response to a new situation. At the lowest level, that of motor control, our body and our immediate environment change all the time. As such, it is important for us to maintain knowledge about this environment in a continuous way. The deviations between our planned movements and those movements that we actually execute continuously provide information about the properties of the environment. Therefore it seems important to have a specialized system that takes all our motor errors and uses them to update a dynamical model of our body and our immediate environment that can predict the delayed sensory results of our motor actions~\citep{McKinstry2006}. 

It appears that the cerebellum is such a structure, and lesions to it abolish our way of dealing successfully with a changing body. Incidentally, the cerebellum has more connections than the rest of the brain taken together, apparently in a largely feedforward architecture, and the tiny cerebellar granule cells, which may form a randomized high-dimensional input representation~\citep{Marr1969, Jacobson2013}, outnumber all other neurons. The brain clearly needs a way of continuously correcting movements to minimize errors.

Newer research shows that the cerebellum is involved in a broad range of cognitive problems~\citep{Moberget2014} as well, potentially because they share computational problems with motor control. For example, when subjects estimate time intervals, which are naturally important for movement, it appears that the brain uses the cerebellum even if no movements are involved~\citep{Gooch2010}. Even individual cerebellar Purkinjie cells may learn to generate precise timings of their outputs~\citep{Johansson2014}. The brain also appears to use inverse models to rapidly predict motor activity that would give rise to a given sensory target~\citep{Giret2014, Hanuschkin2013}. Such mechanisms could be put to use far beyond motor control, in bootstrapping the training of a larger architecture by exploiting continuously changing error signals to update a real-time model of the system state.

\subsubsection{Hierarchical control}

Importantly, many of the control problems we appear to be solving are hierarchical. We have a spinal cord, which deals with the fast signals coming from our muscles and proprioception. Within neuroscience, it is generally assumed that this system deals with fast feedback loops and that this behavior is learned to optimize its own cost function. The nature of cost functions in motor control is still under debate. In particular, the timescale over which cost functions operate remains unclear: motor optimization may occur via real-time responses to a cost function that is computed and optimized online, or via policy choices that change over time more slowly in response to the cost function~\citep{Kording2007}. Nevertheless, the effect is that central processing in the brain has an effectively simplified physical system to control, e.g., one that is far more linear. So the spinal cord itself already suggests the existence of two levels of a hierarchy, each trained using different cost functions. 

However, within the computational motor control literature (see e.g., \citep{DeWolf2011a}), this idea can be pushed far further, e.g., with a hierarchy including spinal cord, M1, PMd, frontal, prefrontal areas. A low level may deal with muscles, the next level may deal with getting our limbs to places or moving objects, a next layer may deal with solving simple local problems (e.g., navigating across a room) while the highest levels may deal with us planning our path through life. This factorization of the problem comes with multiple aspects: First, each level can be solved with its own cost functions, and second, every layer has a characteristic timescale. Some levels, e.g., the spinal cord, must run at a high speed. Other levels, e.g., high-level planning, only need to be touched much more rarely. Converting the computationally hard optimal control problem into a hierarchical approximation promises to make it dramatically easier.

Does the brain solve control problems hierarchically? There is evidence that the brain uses such a strategy~\citep{Botvinick2014, Botvinick2009}, beside neural network demonstrations~\citep{Wayne2014}. The brain may use specialized structures at each hierarchical level to ensure that each  operates efficiently given the nature of its problem space and available training signals. At higher levels, these systems may use an abstract syntax for combining sequences of actions in pursuit of goals~\citep{Allen2010}. Subroutines in such processes could be derived by a process of chunking sequences of actions into single actions~\citep{Botvinick2014, graybiel1998basal}. Some brain areas like Broca's area, known for its involvement in language, also appear to be specifically involved in processing the hierarchical structure of behavior, as such, as opposed to its detailed temporal structure~\citep{koechlin2006broca}.

At the highest level of the decision making and control hierarchy, human reward systems reflect changing goals and subgoals, and we are only beginning to understand how goals are actually coded in the brain, how we switch between goals, and how the cost functions used in learning depend on goal state~\citep{OReilly2014a, Buschman2014, Pezzulo2014}. Goal hierarchies are beginning to be incorporated into deep learning~\citep{Kulkarni2016}. 

Given this hierarchical structure, the optimization algorithms can be fine-tuned. For the low levels, there is sheer unlimited training data. For the high levels, a simulation of the world may be simple, with a tractable number of high-level actions to choose from. Finally, each area needs to give reinforcement to other areas, e.g., high levels need to punish lower levels for making planning complicated. Thus this type of architecture can simplify the learning of control problems.

Progress is being made in both neuroscience and machine learning on finding potential mechanisms for this type of hierarchical planning and goal-seeking. This is beginning to reveal mechanisms for chunking goals and actions and for searching and pruning decision trees~\citep{Balaguer2016, Krishnamurthy2016,tamar2016value, OReilly2014, huys2015interplay}. The study of model-based hierarchical reinforcement learning and prospective optimization~\citep{Sejnowski2014}, which concerns the planning and evaluation of nested sequences of actions, implicates a network coupling the dorsolateral prefontral and orbitofrontal cortex, and the ventral and dorsolateral striatum~\citep{Botvinick2009}. Hierarchical RL relies on a hierarchical representation of state and action spaces, and it has been suggested that error-driven learning of an optimal such representation in the hippocampus\footnote{Like many brain areas, the hippocampus is richly innervated by a variety of reward-related and other neuromodulatory systems~\citep{Hasselmo1997, verney1985morphological, colino1987differential}.} gives rise to place and grid cell properties~\citep{Stachenfeld2014}, with goal representations themselves emerging in the amygdala, prefrontal cortex and other areas~\citep{OReilly2014}. 

The question of how control problems can be successfully divided into component problems remains one of the central questions in neuroscience and machine learning, and the cost functions involved in learning to create such decompositions are still unknown. These considerations may begin to make plausible, however, how the brain could not only achieve its remarkable feats of motor learning -- such as generating complex ``innate'' motor programs, like walking in the newborn gazelle almost immediately after birth -- but also the kind of planning that allows a human to prepare a meal or travel from London to Chicago.

\subsubsection{Spatial planning}

Spatial planning requires solving shortest-path problems subject to constraints. If we want to get from one location to another, there are an arbitrarily large number of simple paths that could be taken. Most naive implementations of such shortest paths problems are grossly inefficient. It appears that, in animals, the hippocampus aids -- at least in part through place cell and grid cell systems -- in efficient learning about new environments and in targeted navigation in such environments~\citep{Brown1323}. In some simple models, targeted navigation in the hippocampus is achieved via the dynamics of ``bump attractors'' or propagating waves in a place cell network with Hebbian plasticity and adaptation~\citep{Buzsaki2013, Hopfield2009, Ponulak2013}, which allows the network to effectively chart out a path in the space of place cell representations. 

Higher-level cognitive tasks such as prospective planning appear to share computational sub-problems with path-finding~\citep{Hassabis2009}\footnote{Other spatial problems such as mental rotation may require learning architectures specialized for geometric coordinate transformations~\citep{Hinton2011, Jaderberg2015} or binding mechanisms that support structural, compositional, parametric descriptions of a scene~\citep{Hayworth2011}.}. Interaction between hippocampus and prefrontal cortex could perhaps support a more abstract notion of ``navigation'' in a space of goals and sub-goals. Having specialized structures for path-finding simplifies these problems.

\subsubsection{Variable binding}

Language and reasoning appear to present a problem for neural networks~\citep{Minsky1991, Hadley2009, Marcus2001}: we seem to be able to apply common grammatical rules to sentences regardless of the content of those sentences, and regardless of whether we have ever seen even remotely similar sentences in the training data. While this is achieved automatically in a computer with fixed registers, location addressable memories, and hard-coded operations, how it could be achieved in a biological brain, or emerge from an optimization algorithm, has been under debate for decades. 

As the putative key capability underlying such operations, variable binding has been defined as ``the transitory or permanent tying together of two bits of information:  a variable (such as an X or Y in algebra, or a placeholder like subject or verb in a sentence) and an arbitrary instantiation of that variable (say, a single number, symbol, vector, or word)''~\citep{Marcus2014, Marcus2014a}. A number of potential biologically plausible binding mechanisms~\citep{Hayworth2012, Kriete2013, Eliasmith2012, Goertzel2014} are reviewed in~\citep{Marcus2014, Marcus2014a}. Some, such as vector symbolic architectures\footnote{There is some direct fMRI evidence for anatomically separate registers representing the contents of different sentence roles in the human brain~\citep{Frankland2015}, which is suggestive of a possible anatomical binding mechanism, but also consistent with other mechanisms like vector symbolic architectures. More generally, the substrates of symbolic processing in the brain may bear an intimate connection with the representation of objects in working memory in the prefrontal cortex, and specifically with the question of how the PFC represents multiple objects in working memory simultaneously. This question is undergoing extensive study in primates~\citep{Warden2007, Warden2010, Siegel2009, Rigotti2013}.}, which were proposed in cognitive science~\citep{Eliasmith2013, Plate1995, Stewart2009}, are also being considered in the context of efficiently-trainable artificial neural networks~\citep{Danihelka2016} -- in effect, these systems learn how to use variable binding. 

Variable binding could potentially emerge from simpler memory systems. For example, the Scrub-Jay can remember the place and time of last visit for hundreds of different locations, e.g., to determine whether high-quality food is currently buried at any given location~\citep{clayton1998episodic}. It is conceivable that such spatially-grounded memory systems enabled a more general binding mechanism to emerge during evolution, perhaps through integration with routing systems or other content-addressable or working memory systems. 

\subsubsection{Hierarchical syntax}

Fixed, static hierarchies (e.g., the hierarchical organization of cortical areas~\citep{felleman1991distributed}) only take us so far: to deal with long chains of arbitrary nested references, we need \emph{dynamic} hierarchies that can implement recursion on the fly. Human language syntax has a hierarchical structure, which Berwick et al described as ``composition of smaller forms like words and phrases into larger ones''~\citep{Berwick2012, Miyagawa2013}. Specific fronto-temporal networks may be involved in representing and generating such hierarchies~\citep{Dehaene2015}\footnote{There is controversy around claims that recursive syntax is also present in songbirds~\citep{van2009simple}.}. 

Little is known about the underlying circuit mechanisms for such dynamic hierarchies, but it is clear that specific affordances for representing such hierarchies in an efficient way would be beneficial. This may be closely connected with the issue of variable binding, and it is possible that operations similar to pointers could be useful in this context, in both the brain and artificial neural networks~\citep{Kriete2013, Kurach2015}. Augmenting neural networks with a differentiable analog of a push-down stack is another such affordance being pursued in machine learning~\citep{joulin2015inferring}.

\subsubsection{Mental programs and imagination}

Humans excel at stitching together sub-actions to form larger actions~\citep{Sejnowski2014, acuna2014multifaceted, verwey1996buffer}. Structured, serial, hierarchical probabilistic programs have recently been shown to model aspects of human conceptual representation and compositional learning~\citep{Lake2015}. In particular, sequential programs were found to enable one-shot learning of new geometric/visual concepts~\citep{Lake2015}, a key capability that deep learning networks for object recognition seem to fundamentally lack. Generative programs have also been proposed in the context of scene understanding~\citep{Battaglia2013}. The ability to deal with problems in terms of sub-problems is central both in human thought and in many successful algorithms.

One possibility is that the hippocampus supports the construction and learning of sequential programs. The hippocampus appears to explore, in simulation, possible future trajectories to a goal, even those involving previously unvisited locations~\citep{Olafsdottir2015}. Hippocampal-prefrontal interaction has been suggested to allow rapid, subconscious evaluation of potential action sequences during decision-making, with the hippocampus in effect simulating the expected outcomes of potential actions that are generated and evaluated in the prefrontal~\citep{Wang2015, Mushiake2006}. The role of the hippocampus in imagination, concept generation~\citep{Kumaran2009}, scene construction~\citep{Hassabis2007}, mental exploration and goal-directed path planning~\citep{Olafsdottir2015, Brown1323, Hopfield2009} suggests that it could help to create generative models to underpin more complex inference such as program induction~\citep{Lake2015} or common-sense world simulation~\citep{Battaglia2013}\footnote{One common idea is that the hippocampus plays a key role in certain processes where learning must occur quickly, whereas the cortex learns more slowly~\citep{Leibo2015, Herd2013}. For example, a sequential, programmatic process, mediated jointly by the basal ganglia, hippocampus and prefrontal cortex might allow one-shot learning of a new concept, as in the sequential computations underlying Bayesian Program Learning~\citep{Lake2015}.}.

Another related possibility is that the cortex itself intrinsically supports the construction and learning of sequential programs~\citep{Bach2015a}. Recurrent neural networks have been used for image generation through a sequential, attention-based process~\citep{Gregor2015}, although their correspondence with the brain is unclear\footnote{The above mechanisms are spontaneous and subconscious. In conscious thought, too, the brain can clearly visit the multiple layers of a program one after the other. We make high-level plans that we fill with lower-level plans. Humans also have memory for their own thought processes. We have some ability to put ``on hold'' our current state of mind, start a new train of thought, and then come back to our original thought. We also are able to ask, introspectively, whether we have had a given thought before. The neural basis of these processes is unclear, although one may speculate that the hippocampus is involved.}.

\subsection{Other specialized structures}

Importantly, there are many other specialized structures known in neuroscience, which arguably receive less attention than they deserve, even for those interested in higher cognition. In the above, in addition to the hippocampus, basal ganglia and cortex, we emphasized the key roles of the thalamus in routing, of the cerebellum as a rapidly trainable control and modeling system, of the amygdala and other areas as a potential source of utility functions, of the retina or early visual areas as a means to generate detectors for motion and other features to bootstrap more complex visual learning, and of the frontal eye fields and other areas as a possible source of attention control. We ignored other structures entirely, whose functions are only beginning to be uncovered, such as the claustrum~\citep{crick2005function}, which has been speculated to be important for rapidly binding together information from many modalities. Our overall understanding of the functional decomposition of brain circuitry still seems very preliminary.

\subsection{Relationships with other cognitive frameworks involving specialized systems}

A recent analysis~\citep{Lake2016} suggested directions by which to modify and enhance existing neural-net-based machine learning towards more powerful and human-like cognitive capabilities, particularly by introducing new structures and systems which go beyond data-driven optimization. This analysis emphasized that systems should construct generative models of the world that incorporate compositionality (discrete construction from re-usable parts), inductive biases reflecting causality, intuitive physics and intuitive psychology, and the capacity for probabilistic inference over discrete structured models (e.g., structured as graphs, trees, or programs)~\citep{Tervo2016} to harness abstractions and enable transfer learning.

We view these ideas as consistent with and complementary to the framework of cost functions, optimization and specialized systems discussed here. One might seek to understand how optimization and specialized systems could be used to implement some of the mechanisms proposed in~\citep{Lake2016} inside neural networks. \citet{Lake2016} emphasize how incorporating additional structure into trainable neural networks can potentially give rise to systems that use compositional, causal and intuitive inductive biases and that ``learn to learn'' using structured models and shared data structures. For example, sub-dividing networks into units that can be modularly and dynamically combined, where representations can be copied and routed, may present a path towards improved compositionality and transfer learning~\citep{andreas2015deep}. The control flow for recombining pre-existing modules and representations could be learned via reinforcement learning~\citep{andreas2016learning}. How to implement the broad set of mechanisms discussed in~\citep{Lake2016} is a key computational problem, and it remains open at which levels (e.g., cost functions and training procedures vs. specialized computational structures vs. underlying neural primitives) architectural innovations will need to be introduced to capture these phenomena.

Primitives that are more complex than those used in conventional neural networks -- for instance, primitives that act as state machines with complex message passing~\citep{Bach2015a} or networks that intrinsically implement Bayesian inference~\citep{George2009} -- could potentially be useful, and it is plausible that some of these may be found in the brain. Recent findings on the power of generic optimization also do not rule out the idea that the brain may explicitly generate and use particular types of structured representations to constrain its inferences; indeed, the specialized brain systems discussed here might provide a means to enforce such constraints. It might be possible to further map the concepts of~\citet{Lake2016} onto neuroscience via an infrastructure of interacting cost functions and specialized brain systems under rich genetic control, coupled to a powerful and generic neurally implemented capacity for optimization. For example, it was recently shown that complex probabilistic population coding and inference can arise automatically from backpropagation-based training of simple neural networks~\citep{Orhan2016}, without needing to be built in by hand. The nature of the underlying primitives in the brain, on top of which learning can operate, is a key question for neuroscience.

\section{Machine learning inspired neuroscience}

Hypotheses are primarily useful if they lead to concrete, experimentally testable predictions. As such, we now want to go through the hypotheses and see to which level they can be directly tested, as well as refined, through neuroscience. 

\subsection{\textit{Hypothesis 1--} Existence of cost functions}

There are multiple general strategies for addressing whether and how the brain optimizes cost functions. A first strategy is based on observing the endpoint of learning. If the brain uses a cost function, and we can guess its identity, then the final state of the brain should be close to optimal for the cost function. If we know the statistics of natural environments, and know the cost function, we can compare receptive fields that are optimized in a simulation with the measured ones. This strategy is only beginning to be used at the moment because it has been difficult to measure the receptive fields or other representational properties across a large population of neurons, but this situation is beginning to improve technologically with the emergence of large-scale recording methods.

A second strategy could directly quantify how well a cost function describes learning. If the dynamics of learning minimize a cost function then the underlying vector field should have a strong gradient descent type component and a weak rotational component. If we could somehow continuously monitor the synaptic strengths, while externally manipulating them, then we could, in principle, measure the vector field in the space of synaptic weights, and calculate its divergence as well as its rotation. For at least the subset of synapses that are being trained via some approximation to gradient descent, the divergence component should be strong relative to the rotational component. This strategy has not been developed yet due to experimental difficulties with monitoring large numbers of synaptic weights\footnote{Fluorescent techniques like~\citep{hayashi2015labelling} might be helpful.}.

A third strategy is based on perturbations: cost function based learning should undo the effects of perturbations which disrupt optimality, i.e., the system should return to local minima after a perturbation, and indeed perhaps to the same local minimum after a sufficiently small perturbation. If we change synaptic connections, e.g., in the context of a brain machine interface, we should be able to produce a reorganization that can be predicted based on a guess of the relevant cost function. This strategy is starting to be feasible in motor areas. 

Lastly, if we knew structurally which cell types and connections mediated the delivery of error signals vs. input data or other types of connections, then we could stimulate specific connections so as to impose a user-defined cost function. In effect, we would use the brain's own networks as a trainable deep learning substrate, and then study how the network responds to training. Brain machine interfaces can be used to set up specific local learning problems, in which the brain is asked to create certain user-specified representations, and the dynamics of this process can be monitored~\citep{Sadtler2014}. In order to do this properly, we must first understand more about the system is wired to deliver cost signals. Much of the structure that would be found in connectomic circuit maps, for example, would not just be relevant for short-timescale computing, but also for creating the infrastructure that supports cost functions and their optimization. 

Many of the learning mechanisms that we have discussed in this paper make specific predictions about connectivity or dynamics. For example, the ``feedback alignment'' approach to biological backpropagation suggests that cortical feedback connections should, at some level of neuronal grouping, be largely sign-concordant with the corresponding feedforward connections, although not necessarily of concordant weight~\citep{Liao2015}, and feedback alignment also makes predictions for synaptic normalization mechanisms~\citep{Liao2015}. The Kickback model for biologically plausible backpropagation has a specific role for NMDA receptors~\citep{Balduzzi2014a}. Some models that incorporate dendritic coincidence detection for learning temporal sequences predict that a given axon should make only a small number of synapses on a given dendritic segment~\citep{Hawkins2015}. Models that involve STDP learning will make predictions about the dynamics of changing firing rates~\citep{Hinton2007, Hinton2016talk, Bengio2015, Bengio2015a, Bengio2015b}, as well as about the particular network structures, such as those based on autoencoders or recirculation, in which STDP can give rise to a form of backpropagation. 

It is critical to establish the unit of optimization. We want to know the scale of the modules that are trainable by some approximation of gradient descent optimization. How large are the networks which share a given error signal or cost function? On what scales can appropriate training signals be delivered?  It could be that the whole brain is optimized end-to-end, in principle. In this case we would expect to find connections that carry training signals from each layer to the preceding ones. On successively smaller scales, optimization could be within a brain area, a microcircuit\footnote{The use of structured microcircuits rather than individual neurons as the units of learning can ease the burden on the learning rules possessed by individual neurons, as exemplified by a study implementing Helmholtz machine learning in a network of spiking neurons using conventional plasticity rules~\citep{Sountsov2015, Roudi2015}. As a simpler example, the classical problem of how neurons with only one output axon could communicate both activation and error derivatives for backpropagation ceases to be a problem if the unit of optimization is not a single neuron. Similar considerations hold for the issue of weight symmetry, or approximate sign-concordance in the case of feedback alignment~\citep{Liao2015}.}, or an individual neuron \citep{Kording2000, Kording2001, Mel1992, Hawkins2015}. Importantly, optimization may co-exist across these scales. There may be some slow optimization end-to-end, with stronger optimization within a local area and very efficient algorithms within each cell. Careful experiments should be able to identify the scale of optimization, e.g., by quantifying the extent of learning induced by a local perturbation.

The tightness of the structure-function relationship is the hallmark of molecular and to some extent cellular biology, but in large connectionist learning systems, this relationship can become difficult to extract: the same initial network can be driven to compute many different functions by subjecting it to different training\footnote{Within this framework, networks that adhere to the basic statistics of neural connectivity, electrophysiology and morphology, such as the initial cortical column models from the Blue Brain Project~\citep{Markram2015}, would recapitulate some properties of the cortex, but -- just like untrained neural networks -- would not spontaneously generate complex functional computation without being subjected to a multi-stage training process, naturalistic sensory data, signals arising from other brain areas and action-driven reinforcement signals.}\footnote{Not only in applied machine learning, but also in today's most advanced neuro-cognitive models such as SPAUN~\citep{Eliasmith2013, Eliasmith2012}, the detailed local circuit connectivity is obtained through an optimization process of some kind to achieve a particular functionality. In the case of modern machine learning, training is often done via end-to-end backpropagation through an architecture that is only structured at the level of higher-level ``blocks'' of units, whereas in SPAUN each block is optimized~\citep{Eliasmith2004} separately according to a procedure that allows the blocks to subsequently be stitched together in a coherent way. Technically, the Neural Engineering Framework~\citep{Eliasmith2004} used in SPAUN uses singular value decomposition, rather than gradient descent, to compute the connections weights as optimal linear decoders. This is possible because of a nonlinear mapping into a high-dimensional space, in which approximating any desired function can be done via a hyperplane regression~\citep{tapson2013learning}.}. It can be hard to understand the way a neural network solves its problems.

How could one tell the difference, then, between a gradient-descent trained network vs. untrained or random networks vs. a network that has been trained against a different kind of task? One possibility would be to train artificial neural networks against various candidate cost functions, study the resulting neural tuning properties~\citep{todorov2002cosine}, and compare them with those found in the circuit of interest~\citep{Zipser1988}. This has already been done to aid the interpretation of the neural dynamics underlying decision making in the PFC~\citep{Sussillo2014}, working memory in the posterior parietal cortex~\citep{rajan2016recurrent} and object representation in the visual system~\citep{Yamins2016, yamins2016eight}. Some have gone on to suggest a direct correspondence between cortical circuits and optimized, appropriately regularized~\citep{Sussillo2015}, recurrent neural networks~\citep{Liao2016}. In any case, effective analytical methods to reverse engineer complex machine learning systems~\citep{Jonas055624}, and methods to reverse engineer biological brains, may have some commonalities.

Does this emphasis on function optimization and trainable substrates mean that we should give up on reverse engineering the brain based on detailed measurements and models of its specific connectivity and dynamics? On the contrary: we should use large-scale brain maps to try to better understand a) how the brain implements optimization, b) where the training signals come from and what cost functions they embody, and c) what structures exist, at different levels of organization, to constrain this optimization to efficiently find solutions to specific kinds of problems. The answers may be influenced by diverse local properties of neurons and networks, such as homeostatic rules of neural structure, gene expression and function~\citep{Marder2006}, the diversity of synapse types, cell-type-specific connectivity~\citep{Jiang2015}, patterns of inter-laminar projection, distributions of inhibitory neuron types, dendritic targeting and local dendritic physiology and plasticity~\citep{Markram2015, Bloss2016, sandler2016novel} or local glial networks~\citep{Perea2009}. They may also be influenced by the integrated nature of higher-level brain systems, including mechanisms for developmental bootstrapping~\citep{Ullman2012}, information routing~\citep{Gurney2001, Stocco2010}, attention~\citep{Buschman2010} and hierarchical decision making~\citep{Lee2015}. Mapping these systems in detail is of paramount importance to understanding how the brain works, down to the nanoscale dendritic organization of ion channels and up to the real-time global coordination of cortex, striatum and hippocampus, all of which are computationally relevant in the framework we have explicated here. We thus expect that large-scale, multi-resolution brain maps would be useful in testing these framework-level ideas, in inspiring their refinements, and in using them to guide more detailed analysis.

\subsection{\textit{Hypothesis 2--} Biological fine-structure of cost functions}

Clearly, we can map differences in structure, dynamics and representation across brain areas. When we find such differences, the question remains as to whether we can interpret these as resulting from differences in the internally-generated cost functions, as opposed to differences in the input data, or from differences that reflect other constraints unrelated to cost functions. If we can directly measure aspects of the cost function in different areas, then we can also compare them across areas. For example, methods from inverse reinforcement learning\footnote{There is a rich tradition of trying to estimate the cost function used by human beings~\citep{Ng2000}. The idea is that we observe (by stipulation) behavior that is optimal for the human's cost function. We can then search for the cost function that makes the observed behavior most probable and simultaneously makes the behaviors that could have been observed, but were not, least probable. Extensions of such approaches could perhaps be used to ask which cost functions the brain is optimizing.} might allow backing out the cost function from observed plasticity~\citep{Ng2000}. 

Moreover, as we begin to understand the ``neural correlates'' of particular cost functions -- perhaps encoded in particular synaptic or neuromodulatory learning rules, genetically-guided local wiring patterns, or patterns of interaction between brain areas -- we can also begin to understand when differences in observed neural circuit architecture reflect differences in cost functions. 

We expect that, for each distinct learning rule or cost function, there may be specific molecularly identifiable types of cells and/or synapses. Moreover, for each specialized system there may be specific molecularly identifiable developmental programs that tune it or otherwise set its parameters. This would make sense if evolution has needed to tune the parameters of one cost function without impacting others.

How many different types of internal training signals does the brain generate? When thinking about error signals, we are not just talking about dopamine and serotonin, or other classical reward-related pathways. The error signals that may be used to train specific sub-networks in the brain, via some approximation of gradient descent or otherwise, are not necessarily equivalent to reward signals. It is important to distinguish between cost functions that may be used to drive optimization of specific sub-circuits in the brain, and what are referred to as ``value functions'' or ``utility functions'', i.e., functions that predict the agent’s aggregate future reward. In both cases, similar reinforcement learning mechanisms may be used, but the interpretation of the cost functions is different. We have not emphasized global utility functions for the animal here, since they are extensively studied elsewhere (e.g., \citep{OReilly2014, Bach2015}), and since we argue that, though important, they are only a part of the picture, i.e., that the brain is not solely an end-to-end reinforcement trained system.

Progress in brain mapping could soon allow us to classify the types of reward signals in the brain, follow the detailed anatomy and connectivity of reward pathways throughout the brain, and map in detail how reward pathways are integrated into striatal, cortical, hippocampal and cerebellar microcircuits. This program is beginning to be carried out in the fly brain, in which twenty specific types of dopamine neuron project to distinct anatomical compartments of the mushroom body to train distinct odor classifiers operating on a set of high-dimensional odor representations~\citep{Aso2014a, Aso2014, Caron2013, Cohn2015}. It is known that, even within the same system, such as the fly olfactory pathway, some neuronal wiring is highly specific and molecularly programmed~\citep{Hong2014, Hattori2007}, while other wiring is effectively random~\citep{Caron2013}, and yet other wiring is learned~\citep{Aso2014a}. The interplay between such design principles could give rise to many forms of ``division of labor'' between genetics and learning. Likewise, it is believed that birdsong learning is driven by reinforcement learning using a specialized cost function that relies on comparison with a memorized version of a tutor's song~\citep{Fiete2007}, and also that it involves specialized structures for controlling song variability during learning~\citep{Aronov2011}. These detailed pathways underlying the construction of cost functions for vocal learning are beginning to be mapped~\citep{Mandelblat-Cerf2014}. Starting with simple systems, it should become possible to map the reward pathways and how they evolved and diversified, which would be a step on the way to understanding how the system learns.

\subsection{\textit{Hypothesis 3--} Embedding within a pre-structured architecture}

If different brain structures are performing distinct types of computations with a shared goal, then optimization of a joint cost function will take place with different dynamics in each area. If we focus on a higher level task, e.g., maximizing the probability of correctly detecting something, then we should find that basic feature detection circuits should learn when the features were insufficient for detection, that attentional routing structures should learn when a different allocation of attention would have improved detection and that memory structures should learn when items that matter for detection were not remembered. If we assume that multiple structures are participating in a joint computation, which optimizes an overall cost function (but see \textbf{Hypothesis 2}), then an understanding of the computational function of each area leads to a prediction of the measurable plasticity rules.

\section{Neuroscience inspired machine learning}

Machine learning may be equally transformed by neuroscience. Within the brain, a myriad of subsystems and layers work together to produce an agent that exhibits general intelligence. The brain is able to show intelligent behavior across a broad range of problems using only relatively small amounts of data. As such, progress at understanding the brain promises to improve machine learning. In this section, we review our three hypotheses about the brain and discuss how their elaboration might contribute to more powerful machine learning systems.

\subsection{\textit{Hypothesis 1--} Existence of cost functions}

A good practitioner of machine learning should have a broad range of optimization methods at their disposal as different problems ask for different approaches. The brain, we have argued, is an implicit machine learning mechanism which has been evolved over millions of years. Consequently, we should expect the brain to be able to optimize cost functions efficiently, across many domains and kinds of data. Indeed, across different animal phyla, we even see \emph{convergent} evolution of certain brain structures~\citep{shimizu2013multiple, gunturkun2016cognition}, e.g., the bird brain has no cortex yet has developed homologous structures which -- as the linguistic feats of the African Grey Parrot demonstrate -- can give rise to quite complex intelligence. It seems reasonable to hope to learn how to do truly general-purpose optimization by looking at the brain.

Indeed, there are multiple kinds of optimization that we may expect to discover by looking at the brain. At the hardware level, the brain clearly manages to optimize functions efficiently despite having slow hardware subject to molecular fluctuations, suggesting directions for improving the hardware of machine learning to be more energy efficient. At the level of learning rules, the brain solves an optimization problem in a highly nonlinear, non-differentiable, temporally stochastic, spiking system with massive numbers of feedback connections, a problem that we arguably still do not know how to efficiently solve for neural networks. At the architectural level, the brain can optimize certain kinds of functions based on very few stimulus presentations, operates over diverse timescales, and clearly uses advanced forms of active learning to infer causal structure in the world.

While we have discussed a range of theories~\citep{Hinton2007,Hinton2016talk, Bengio2015, Balduzzi2014a, Roelfsema2010, OReilly1996, OReilly2014, Kording2001,Lillicrap2014} for how the brain can carry out optimization, these theories are still preliminary. Thus, the first step is to understand whether the brain indeed performs multi-layer credit assignment in a manner that approximates full gradient descent, and if so, how it does this. Either way, we can expect that answer to impact machine learning. If the brain does not do some form of backpropagation, this suggests that machine learning may benefit from understanding the tricks that the brain uses to avoid having to do so. If, on the other hand, the brain does do backpropagation, then the underlying mechanisms clearly can support a very wide range of efficient optimization processes across many domains, including learning from rich temporal data-streams and via unsupervised mechanisms, and the architectures behind this will likely be of long-term value to machine learning\footnote{Successes of deep learning are already being used, speculatively, to rationalize features of the brain. It has been suggested that large networks, with many more neurons available than are strictly needed for the target computation, make learning easier~\citep{Goodfellow2014a}. In concordance with this, visual cortex appears to be a 100-fold over-complete representation of the retinal output~\citep{Lewicki2000}. Likewise, it has been suggested that biological neurons stabilized~\citep{Turrigiano2012} to operate far below their saturating firing rates mirror the successful use of rectified linear units in facilitating the training of artificial neural networks~\citep{Roudi2015}. Hinton and others have also suggested a biological motivation~\citep{Roudi2015} for ``dropout'' regularization~\citep{Srivastava2014}, in which a fraction of hidden units is stochastically set to zero during each round of training: such a procedure may correspond to the noisiness of neural spike trains, although other theories interpret spikes as sampling in probabilistic inference~\citep{Buesing2011}, or in many other ways. Randomness of spiking has some support in neuroscience~\citep{Softky1993}, although recent experiments suggest that spike trains in certain areas may be less noisy than previously thought~\citep{Hires2015}. The key role of proper initialization in enabling effective gradient descent is an important recent finding~\citep{Saxe2013, Sutskever2013} which may also be reflected by biological mechanisms of neural homeostasis or self-organization that would enforce appropriate initial conditions for learning. But making these speculative claims of biological relevance more rigorous will require researchers to first evaluate \emph{whether} biological neural circuits are performing multi-layer optimization of cost functions in the first place.}. Moreover, the search for biologically plausible forms of backpropagation has already led to interesting insights, such as the possibility of using random feedback weights (feedback alignment) in backpropagation~\citep{Lillicrap2014}, or the unexpected power of internal FORCE learning in chaotic, spontaneously active recurrent networks~\citep{Sussillo2009}. This and other findings discussed here suggest that there are still fundamental things we don't understand about backpropagation -- which could potentially lead not only to more biologically plausible ways to train recurrent neural networks, but also to fundamentally simpler and more powerful ones.

\subsection{ \textit{Hypothesis 2--} Biological fine-structure of cost functions}

A good practitioner of machine learning has access to a broad range of learning techniques and thus implicitly is able to use many different cost functions. Some problems ask for clustering, others for extracting sparse variables, and yet others for prediction quality to be maximized. The brain also needs to be able to deal with many different kinds of datasets. As such, it makes sense for the brain to use a broad range of cost functions appropriate for the diverse set of tasks it has to solve to thrive in this world.

Many of the most notable successes of deep learning, from language modeling~\citep{sutskever2011generating}, to vision~\citep{krizhevsky2012imagenet}, to motor control~\citep{levine2015end}, have been driven by end-to-end optimization of single task objectives. We have highlighted cases where machine learning has opened the door to multiplicities of cost functions that shape network modules into specialized roles. We expect that machine learning will increasingly adopt these practices in the future. 

In computer vision, we have begun to see researchers re-appropriate neural networks trained for one task (e.g., ImageNet classification) and then deploy them on new tasks other than the ones they were trained for or for which more limited training data is available~\citep{yosinski2014transferable, oquab2014learning, noroozi2016unsupervised}. We imagine this procedure will be generalized, whereby, in series and in parallel, diverse training problems, each with an associated cost function, are used to shape visual representations. For example, visual data streams can be segmented into elements like foreground vs. background, objects that can move of their own accord vs. those that cannot, all using diverse unsupervised criteria~\citep{Ullman2012, Poggio2015}. Networks so trained can then be shared, augmented, and retrained on new tasks. They can be introduced as front-ends for systems that perform more complex objectives or even serve to produce cost functions for training other circuits~\citep{watter2015embed}. As a simple example, a network that can discriminate between images of different kinds of architectural structures (pyramid, staircase, etc.) could act as a critic for a building-construction network. 

Scientifically, determining the order in which cost functions are engaged in the biological brain will inform machine learning about how to construct systems with intricate and hierarchical behaviors via divide-and-conquer approaches to learning problems, active learning, and more. 

\subsection{ \textit{Hypothesis 3--} Embedding within a pre-structured architecture}

A good practitioner of machine learning should have a broad range of algorithms at their disposal. Some problems are efficiently solved through dynamic programming, others through hashing, and yet others through multi-layer backpropagation. The brain needs to be able to solve a broad range of learning problems without the luxury of being reprogrammed. As such, it makes sense for the brain to have specialized structures that allow it to rapidly learn to approximate a broad range of algorithms.

The first neural networks were simple single-layer systems, either linear or with limited non-linearities~\citep{rashevsky1939mathematical}. The explosion of neural network research in the 1980s~\citep{mcclelland1986parallel} saw the advent of multilayer networks, followed by networks with layer-wise specializations as in convolutional networks~\citep{fukushima1980neocognitron, lecun1995convolutional}. In the last two decades, architectures with specializations for holding variables stable in memory like the LSTM~\citep{Hochreiter1997}, the control of content-addressable memory~\citep{Weston2014, Graves2014}, and game playing by reinforcement learning~\citep{Mnih2015} have been developed. These networks, though formerly exotic, are now becoming mainstream algorithms in the toolbox of any deep learning practitioner. There is no sign that progress in developing new varieties of structured architectures is halting, and the heterogeneity and modularity of the brain's circuitry suggests that diverse, specialized architectures are needed to solve the diverse challenges that confront a behaving animal. 

The brain combines a jumble of specialized structures in a way that works. Solving this problem \emph{de novo} in machine learning promises to be very difficult, making it attractive to be inspired by observations about how the brain does it. An understanding of the breadth of specialized structures, as well as the architecture that combines them, should be quite useful.

\section{Did evolution separate cost functions from optimization algorithms?}

Deep learning methods have taken the field of machine learning by storm. Driving the success is the separation of the problem of learning into two pieces: \textbf{(1)} An algorithm, backpropagation, that allows efficient distributed optimization, and \textbf{(2)} Approaches to turn any given problem into an optimization problem, by designing a cost function and training procedure which will result in the desired computation. If we want to apply deep learning to a new domain, e.g., playing Jeopardy, we do not need to change the optimization algorithm -- we just need to cleverly set up the right cost function. A lot of work in deep learning, perhaps the majority, is now focused on setting up the right cost functions. 

We hypothesize that the brain also acquired such a separation between optimization mechanisms and cost functions. If neural circuits, such as in cortex, implement a general-purpose optimization algorithm, then any improvement to that algorithm will improve function across the cortex. At the same time, different cortical areas solve different problems, so tinkering with each area's cost function is likely to improve its performance. As such, functionally and evolutionarily separating the problems of optimization and cost function generation could allow evolution to produce better computations, faster. For example, common unsupervised mechanisms could be combined with area-specific reinforcement-based or supervised mechanisms and error signals, much as recent advances in machine learning have found natural ways to combine supervised and unsupervised objectives in a single system~\citep{Rasmus2015a}.

This suggests interesting questions\footnote{It would be interesting to study these questions in specific brain systems. The primary visual cortex, for example, is still only understood very incompletely~\citep{olshausen2004other}. It serves as a key input modality to both the ventral and dorsal visual pathways, one of which seems to specialize in object identity and the other in motion and manipulation. Higher-level areas like STP draw on both streams to perform tasks like complex action recognition. In some models (e.g., \citep{jhuang2007biologically}), both ventral and dorsal streams are structured hierarchically, but the ventral stream primarily makes use of the spatial filtering properties of V1, whereas the dorsal stream primarily makes use of its spatio-\emph{temporal} filtering properties, e.g., temporal frequency filtering by the space-time receptive fields of V1 neurons. Given this, we can ask interesting questions about V1. Within a framework of multilayer optimization, do both dorsal and ventral pathways impose cost functions that help to shape V1's response properties? Or is V1 largely pre-structured by genetics and local self-organization, with different optimization principles in the ventral and dorsal streams only having effects at higher levels of the hierarchy? Or, more likely, is there some interplay between pre-structuring of the V1 circuitry and optimization according to multiple cost functions? Relatedly, what establishes the differing roles of the downstream ventral vs. dorsal cortical areas, and can their differences be attributed to differing cost functions? This relates to ongoing questions about the basic nature of cortical circuitry. For example,~\citet{dicarlo2012does} suggests that visual cortical regions containing on the order of $10000$ neurons are locally optimized to perform disentangling of the manifolds corresponding to their local views of the transformations of an object, allowing these manifolds to be linearly separated by readout areas. Yet,~\citet{dicarlo2012does} also emphasizes the possibility that certain computations such as normalization are pre-initialized in the circuitry prior to learning-based optimization.}: When did the division between cost functions and optimization algorithms occur? How is this separation implemented? How did innovations in cost functions and optimization algorithms evolve? And how do our own cost functions and learning algorithms differ from those of other animals? 

There are many possibilities for how such a separation might be achieved in the brain. Perhaps the six-layered cortex represents a common optimization algorithm, which in different cortical areas is supplied with different cost functions. This claim is different from the claim that all cortical areas use a single unsupervised learning algorithm and achieve functional specificity by tuning the inputs to that algorithm. In that case, both the optimization mechanism and the implicit unsupervised cost function would be the same across areas (e.g., minimization of prediction error), with only the training data differing between areas, whereas in our suggestion, the optimization mechanism would be the same across areas but the cost function, \emph{as well as} the training data, would differ. Thus the cost function itself would be like an ancillary input to a cortical area, in addition to its input and output data. Some cortical microcircuits could then, perhaps, compute the cost functions that are to be delivered to other cortical microcircuits. 

Another possibility is that, within the same circuitry, certain aspects of the wiring and learning rules specify an optimization mechanism and are relatively fixed across areas, while others specify the cost function and are more variable. This latter possibility would be similar to the notion of cortical microcircuits as molecularly and structurally configurable elements, akin to the cells in a field-programmable gate array (FPGA)~\citep{Marcus2014, Marcus2014a}, rather than a homogenous substrate. The biological nature of such a separation, if any exists, remains an open question. For example, individual parts of a neuron may separately deal with optimization and with the specification of the cost function, or different parts of a microcircuit may specialize in this way, or there may be specialized types cells, some of which deal with signal processing and others with cost functions.

\section{Conclusions}

Due to the complexity and variability of the brain, pure ``bottom up'' analysis of neural data faces potential challenges of interpretation~\citep{Robinson1992, Jonas055624}. Theoretical frameworks can potentially be used to constrain the space of hypotheses being evaluated, allowing researchers to first address higher-level principles and structures in the system, and then ``zoom in'' to address the details. Proposed ``top down'' frameworks for understanding neural computation include entropy maximization, efficient encoding, faithful approximation of Bayesian inference, minimization of prediction error, attractor dynamics, modularity, the ability to subserve symbolic operations, and many others~\citep{Bialek2002, Bialek2006, Friston2010, Knill2004, Marcus2001, Pinker1999}. Interestingly, many of the ``top down'' frameworks boil down to assuming that the brain simply optimizes a single, given cost function for a single computational architecture. We generalize these proposals assuming both a heterogeneous combination of cost functions unfolding over development, and a diversity of specialized sub-systems.

Much of neuroscience has focused on the search for ``the neural code'', i.e., it has asked which stimuli are good at driving activity in individual neurons, regions, or brain areas. But, if the brain is capable of generic optimization of cost functions, then we need to be aware that rather simple cost functions can give rise to complicated stimulus responses. This potentially leads to a different set of questions. Are differing cost functions indeed a useful way to think about the differing functions of brain areas? How does the optimization of cost functions in the brain actually occur, and how is this different from the implementations of gradient descent in artificial neural networks? What additional constraints are present in the circuitry that remain fixed while optimization occurs? How does optimization interact with a structured architecture, and is this architecture similar to what we have sketched? Which computations are wired into the architecture, which emerge through optimization, and which arise from a mixture of those two extremes? To what extent are cost functions explicitly computed in the brain, versus implicit in its local learning rules? Did the brain evolve to separate the mechanisms involved in cost function generation from those involved in the optimization of cost functions, and if so how? What kinds of meta-level learning might the brain apply, to learn when and how to invoke different cost functions or specialized systems, among the diverse options available, to solve a given task? What crucial mechanisms are left out of this framework? A more in-depth dialog between neuroscience and machine learning could help elucidate some of these questions.

Much of machine learning has focused on finding ever faster ways of doing end-to-end gradient descent in neural networks. Neuroscience may inform machine learning at multiple levels. The optimization algorithms in the brain have undergone a couple of hundred million years of evolution. Moreover, the brain may have found ways of using heterogeneous cost functions that interact over development so as to simplify learning problems by guiding and shaping the outcomes of unsupervised learning. Lastly, the specialized structures evolved in the brain may inform us about ways of making learning efficient in a world that requires a broad range of computational problems to be solved over multiple timescales. Looking at the insights from neuroscience may help machine learning move towards general intelligence in a structured heterogeneous world with access to only small amounts of supervised data.

In some ways our proposal is opposite to many popular theories of neural computation. There is not one mechanism of optimization but (potentially) many, not one cost function but a host of them, not one kind of a representation but a representation of whatever is useful, and not one homogeneous structure but a large number of them. All these elements are held together by the optimization of internally generated cost functions, which allows these systems to make good use of one another. Rejecting simple unifying theories is in line with a broad range of previous approaches in AI. For example, Minsky and Papert's work on the Society of Mind~\citep{Minsky1988} -- and more broadly on ideas of genetically staged and internally bootstrapped development in connectionist systems~\citep{Minsky1977} -- emphasizes the need for a system of internal monitors and critics, specialized communication and storage mechanisms, and a hierarchical organization of simple control systems. 

At the time these early works were written, it was not yet clear that gradient-based optimization could give rise to powerful feature representations and behavioral policies. One can view our proposal as a renewed argument against simple end-to-end training and in favor of a heterogeneous approach. In other words, this framework could be viewed as proposing a kind of ``society'' of cost functions and trainable networks, permitting internal bootstrapping processes reminiscent of the Society of Mind~\citep{Minsky1988}. In this view, intelligence is enabled by many computationally specialized structures, each trained with its own developmentally regulated cost function, where both the structures and the cost functions are themselves optimized by evolution like the hyperparameters in neural networks. 

\acks{We thank Ken Hayworth for key discussions that led to this paper. We thank Ed Boyden, Chris Eliasmith, Gary Marcus, Shimon Ullman, Tomaso Poggio, Josh Tenenbaum, Dario Amodei, Tom Dean, Joscha Bach, Mohammad Gheshlaghi Azar, Joshua Glaser, Ali Hummos, David Markowitz, David Rolnick, Sam Rodriques, Nick Barry, Darcy Wayne, Lyra and Neo Marblestone, and all of the participants of a Kavli Salon on Cortical Computation (Feb/Oct 2015) for helpful discussions. We thank Miles Brundage for an excellent Twitter feed of deep learning papers.}


\begin{thebibliography}{}
   \providecommand{\natexlab}[1]{#1}
\providecommand{\url}[1]{\texttt{#1}}
\expandafter\ifx\csname urlstyle\endcsname\relax
  \providecommand{\doi}[1]{doi: #1}\else
  \providecommand{\doi}{doi: \begingroup \urlstyle{rm}\Url}\fi

\bibitem[Abbott and Blum(1996)]{Abbott1996}
LF~Abbott and KI~Blum.
\newblock {Functional significance of long-term potentiation for sequence
  learning and prediction}.
\newblock \emph{Cerebral Cortex}, 1996.
\newblock URL \url{http://cercor.oxfordjournals.org/content/6/3/406.short}.

\bibitem[Abbott et~al.(2016)Abbott, DePasquale, and Memmesheimer]{Abbott}
LF~Abbott, B~DePasquale, and RM~Memmesheimer.
\newblock {Building Functional Networks of Spiking Model Neurons}.
\newblock \emph{neurotheory.columbia.edu}, 2016.
\newblock URL
  \url{http://www.neurotheory.columbia.edu/Larry/SpikingNetworkReview.pdf}.

\bibitem[Ackley et~al.(1985)Ackley, Hinton, and Sejnowski]{Ackley1985}
DH~Ackley, GE~Hinton, and TJ~Sejnowski.
\newblock {A learning algorithm for Boltzmann machines}.
\newblock \emph{Cognitive science}, 1985.
\newblock URL
  \url{http://www.sciencedirect.com/science/article/pii/S0364021385800124}.

\bibitem[Acuna et~al.(2014)Acuna, Wymbs, Reynolds, Picard, Turner, Strick,
  Grafton, and Kording]{acuna2014multifaceted}
Daniel~E Acuna, Nicholas~F Wymbs, Chelsea~A Reynolds, Nathalie Picard, Robert~S
  Turner, Peter~L Strick, Scott~T Grafton, and Konrad~P Kording.
\newblock Multifaceted aspects of chunking enable robust algorithms.
\newblock \emph{Journal of neurophysiology}, 112\penalty0 (8):\penalty0
  1849--1856, 2014.

\bibitem[Alain et~al.(2015)Alain, Lamb, Sankar, Courville, and
  Bengio]{alain2015variance}
Guillaume Alain, Alex Lamb, Chinnadhurai Sankar, Aaron Courville, and Yoshua
  Bengio.
\newblock Variance reduction in sgd by distributed importance sampling.
\newblock \emph{arXiv preprint arXiv:1511.06481}, 2015.

\bibitem[Allen et~al.(2010)Allen, Ibara, and Seymour]{Allen2010}
K~Allen, S~Ibara, and A~Seymour.
\newblock {Abstract structural representations of goal-directed behavior}.
\newblock \emph{Psychological {\ldots}}, 2010.
\newblock URL \url{http://pss.sagepub.com/content/21/10/1518.short}.

\bibitem[Anderson and Van~Essen(1987)]{anderson1987shifter}
Charles~H Anderson and David~C Van~Essen.
\newblock Shifter circuits: a computational strategy for dynamic aspects of
  visual processing.
\newblock \emph{Proceedings of the National Academy of Sciences}, 84\penalty0
  (17):\penalty0 6297--6301, 1987.

\bibitem[Anderson(2007)]{Anderson2007}
John~R Anderson.
\newblock \emph{{How Can the Human Mind Occur in the Physical Universe?}}
\newblock Oxford University Press, 2007.
\newblock ISBN 0198043538.
\newblock URL \url{http://books.google.com/books?id=eYZXEtfplyAC{\&}pgis=1}.

\bibitem[Andreas et~al.(2015)Andreas, Rohrbach, Darrell, and
  Klein]{andreas2015deep}
Jacob Andreas, Marcus Rohrbach, Trevor Darrell, and Dan Klein.
\newblock Deep compositional question answering with neural module networks.
\newblock \emph{arXiv preprint arXiv:1511.02799}, 2015.

\bibitem[Andreas et~al.(2016)Andreas, Rohrbach, Darrell, and
  Klein]{andreas2016learning}
Jacob Andreas, Marcus Rohrbach, Trevor Darrell, and Dan Klein.
\newblock Learning to compose neural networks for question answering.
\newblock \emph{arXiv preprint arXiv:1601.01705}, 2016.

\bibitem[Anselmi et~al.(2015)Anselmi, Leibo, Rosasco, Mutch, Tacchetti, and
  Poggio]{Anselmi2015}
Fabio Anselmi, Joel~Z. Leibo, Lorenzo Rosasco, Jim Mutch, Andrea Tacchetti, and
  Tomaso Poggio.
\newblock {Unsupervised learning of invariant representations}.
\newblock \emph{Theoretical Computer Science}, jun 2015.
\newblock ISSN 03043975.
\newblock \doi{10.1016/j.tcs.2015.06.048}.
\newblock URL
  \url{http://www.sciencedirect.com/science/article/pii/S0304397515005587}.

\bibitem[Arancio et~al.(1996)Arancio, Kiebler, Lee, Lev-Ram, Tsien, Kandel, and
  Hawkins]{Arancio1996}
O~Arancio, M~Kiebler, C~J Lee, V~Lev-Ram, R~Y Tsien, E~R Kandel, and R~D
  Hawkins.
\newblock {Nitric oxide acts directly in the presynaptic neuron to produce
  long-term potentiation in cultured hippocampal neurons.}
\newblock \emph{Cell}, 87\penalty0 (6):\penalty0 1025--35, dec 1996.
\newblock ISSN 0092-8674.
\newblock URL \url{http://www.ncbi.nlm.nih.gov/pubmed/8978607}.

\bibitem[Aronov et~al.(2011)Aronov, Veit, Goldberg, and Fee]{Aronov2011}
Dmitriy Aronov, Lena Veit, Jesse~H Goldberg, and Michale~S Fee.
\newblock {Two distinct modes of forebrain circuit dynamics underlie temporal
  patterning in the vocalizations of young songbirds.}
\newblock \emph{The Journal of neuroscience : the official journal of the
  Society for Neuroscience}, 31\penalty0 (45):\penalty0 16353--68, nov 2011.
\newblock ISSN 1529-2401.
\newblock \doi{10.1523/JNEUROSCI.3009-11.2011}.

\bibitem[Arora et~al.(2015)Arora, Liang, and Ma]{Arora2015}
Sanjeev Arora, Yingyu Liang, and Tengyu Ma.
\newblock {Why are deep nets reversible: A simple theory, with implications for
  training}.
\newblock nov 2015.
\newblock URL \url{http://arxiv.org/abs/1511.05653}.

\bibitem[Ashby et~al.(2007)Ashby, Ennis, and
  Spiering]{ashby2007neurobiological}
F~Gregory Ashby, John~M Ennis, and Brian~J Spiering.
\newblock A neurobiological theory of automaticity in perceptual
  categorization.
\newblock \emph{Psychological review}, 114\penalty0 (3):\penalty0 632, 2007.

\bibitem[Ashby et~al.(2010)Ashby, Turner, and Horvitz]{Ashby2010}
F~Gregory Ashby, Benjamin~O Turner, and Jon~C Horvitz.
\newblock {Cortical and basal ganglia contributions to habit learning and
  automaticity.}
\newblock \emph{Trends in cognitive sciences}, 14\penalty0 (5):\penalty0
  208--15, may 2010.
\newblock ISSN 1879-307X.
\newblock \doi{10.1016/j.tics.2010.02.001}.
\newblock URL \url{http://www.cell.com/article/S1364661310000306/fulltext}.

\bibitem[Aso et~al.(2014{\natexlab{a}})Aso, Hattori, Yu, Johnston, Iyer, Ngo,
  Dionne, Abbott, Axel, Tanimoto, and Rubin]{Aso2014a}
Yoshinori Aso, Daisuke Hattori, Yang Yu, Rebecca~M Johnston, Nirmala~A Iyer,
  Teri-T~B Ngo, Heather Dionne, L~F Abbott, Richard Axel, Hiromu Tanimoto, and
  Gerald~M Rubin.
\newblock {The neuronal architecture of the mushroom body provides a logic for
  associative learning.}
\newblock \emph{eLife}, 3:\penalty0 e04577, jan 2014{\natexlab{a}}.
\newblock ISSN 2050-084X.
\newblock \doi{10.7554/eLife.04577}.
\newblock URL \url{http://elifesciences.org/content/3/e04577.abstract}.

\bibitem[Aso et~al.(2014{\natexlab{b}})Aso, Sitaraman, Ichinose, Kaun, Vogt,
  Belliart-Gu{\'{e}}rin, Pla{\c{c}}ais, Robie, Yamagata, Schnaitmann, Rowell,
  Johnston, Ngo, Chen, Korff, Nitabach, Heberlein, Preat, Branson, Tanimoto,
  and Rubin]{Aso2014}
Yoshinori Aso, Divya Sitaraman, Toshiharu Ichinose, Karla~R Kaun, Katrin Vogt,
  Ghislain Belliart-Gu{\'{e}}rin, Pierre-Yves Pla{\c{c}}ais, Alice~A Robie,
  Nobuhiro Yamagata, Christopher Schnaitmann, William~J Rowell, Rebecca~M
  Johnston, Teri-T~B Ngo, Nan Chen, Wyatt Korff, Michael~N Nitabach, Ulrike
  Heberlein, Thomas Preat, Kristin~M Branson, Hiromu Tanimoto, and Gerald~M
  Rubin.
\newblock {Mushroom body output neurons encode valence and guide memory-based
  action selection in Drosophila.}
\newblock \emph{eLife}, 3:\penalty0 e04580, jan 2014{\natexlab{b}}.
\newblock ISSN 2050-084X.
\newblock \doi{10.7554/eLife.04580}.
\newblock URL \url{http://elifesciences.org/content/3/e04580.abstract}.

\bibitem[Ba and Caruana(2014)]{Ba2014}
J~Ba and R~Caruana.
\newblock {Do deep nets really need to be deep?}
\newblock \emph{Advances in neural information processing {\ldots}}, 2014.
\newblock URL
  \url{http://papers.nips.cc/paper/5484-do-deep-nets-really-need-to-be-deep}.

\bibitem[Bach(2015)]{Bach2015}
Joscha Bach.
\newblock {Modeling Motivation in MicroPsi 2}.
\newblock \emph{Artificial General Intelligence}, 2015.
\newblock URL
  \url{http://link.springer.com/chapter/10.1007/978-3-319-21365-1{\_}1}.

\bibitem[Bach and Herger(2015)]{Bach2015a}
Joscha Bach and Priska Herger.
\newblock {Request confirmation networks for neuro-symbolic script execution}.
\newblock In Tarek Besold, Artur d’Avila Garcez, Gary Marcus, and Risto
  Miikkulainen, editors, \emph{Workshop on Cognitive Computation: Integrating
  Neural and Symbolic Approaches at NIPS}, 2015.
\newblock URL \url{http://www.neural-symbolic.org/CoCo2015/}.

\bibitem[Balaguer et~al.(2016)Balaguer, Spiers, Hassabis, and
  Summerfield]{Balaguer2016}
Jan Balaguer, Hugo Spiers, Demis Hassabis, and Christopher Summerfield.
\newblock {Neural Mechanisms of Hierarchical Planning in a Virtual Subway
  Network}.
\newblock \emph{Neuron}, 90\penalty0 (4):\penalty0 893--903, may 2016.
\newblock ISSN 08966273.
\newblock \doi{10.1016/j.neuron.2016.03.037}.
\newblock URL \url{http://www.cell.com/article/S0896627316300575/fulltext}.

\bibitem[Baldauf and Desimone(2014)]{Baldauf2014}
Daniel Baldauf and Robert Desimone.
\newblock {Neural mechanisms of object-based attention.}
\newblock \emph{Science (New York, N.Y.)}, 344\penalty0 (6182):\penalty0
  424--7, apr 2014.
\newblock ISSN 1095-9203.
\newblock \doi{10.1126/science.1247003}.
\newblock URL \url{http://www.ncbi.nlm.nih.gov/pubmed/24763592}.

\bibitem[Baldi and Sadowski(2015)]{Baldi2015}
Pierre Baldi and Peter Sadowski.
\newblock {The Ebb and Flow of Deep Learning: a Theory of Local Learning}.
\newblock jun 2015.
\newblock URL \url{http://arxiv.org/abs/1506.06472}.

\bibitem[Balduzzi(2014)]{Balduzzi2014}
D~Balduzzi.
\newblock {Cortical prediction markets}.
\newblock \emph{Proceedings of the 2014 international conference on {\ldots}},
  2014.
\newblock URL \url{http://dl.acm.org/citation.cfm?id=2617449}.

\bibitem[Balduzzi et~al.(2014)Balduzzi, Vanchinathan, and
  Buhmann]{Balduzzi2014a}
David Balduzzi, Hastagiri Vanchinathan, and Joachim Buhmann.
\newblock {Kickback cuts Backprop's red-tape: Biologically plausible credit
  assignment in neural networks}.
\newblock page~7, nov 2014.
\newblock URL \url{http://arxiv.org/abs/1411.6191}.

\bibitem[Bargmann(2012)]{Bargmann2012}
Cornelia~I Bargmann.
\newblock {Beyond the connectome: how neuromodulators shape neural circuits.}
\newblock \emph{BioEssays : news and reviews in molecular, cellular and
  developmental biology}, 34\penalty0 (6):\penalty0 458--65, jun 2012.
\newblock ISSN 1521-1878.
\newblock \doi{10.1002/bies.201100185}.
\newblock URL \url{http://www.ncbi.nlm.nih.gov/pubmed/22396302}.

\bibitem[Bargmann and Marder(2013)]{Bargmann2013}
Cornelia~I Bargmann and Eve Marder.
\newblock {From the connectome to brain function}.
\newblock \emph{Nature Methods}, 10\penalty0 (6):\penalty0 483--490, may 2013.
\newblock ISSN 1548-7091.
\newblock \doi{10.1038/nmeth.2451}.
\newblock URL \url{http://dx.doi.org/10.1038/nmeth.2451}.

\bibitem[Battaglia et~al.(2013)Battaglia, Hamrick, and
  Tenenbaum]{Battaglia2013}
Peter~W Battaglia, Jessica~B Hamrick, and Joshua~B Tenenbaum.
\newblock {Simulation as an engine of physical scene understanding.}
\newblock \emph{Proceedings of the National Academy of Sciences of the United
  States of America}, 110\penalty0 (45):\penalty0 18327--32, nov 2013.
\newblock ISSN 1091-6490.
\newblock \doi{10.1073/pnas.1306572110}.
\newblock URL \url{http://www.pnas.org/content/110/45/18327.short}.

\bibitem[Becker and Hinton(1992)]{becker1992self}
Suzanna Becker and Geoffrey~E Hinton.
\newblock Self-organizing neural network that discovers surfaces in random-dot
  stereograms.
\newblock \emph{Nature}, 355\penalty0 (6356):\penalty0 161--163, 1992.

\bibitem[Bengio(2014)]{Bengio2014}
Yoshua Bengio.
\newblock {How Auto-Encoders Could Provide Credit Assignment in Deep Networks
  via Target Propagation}.
\newblock jul 2014.
\newblock URL \url{http://arxiv.org/abs/1407.7906}.

\bibitem[Bengio and Fischer(2015)]{Bengio2015a}
Yoshua Bengio and Asja Fischer.
\newblock {Early Inference in Energy-Based Models Approximates
  Back-Propagation}.
\newblock oct 2015.
\newblock URL \url{http://arxiv.org/abs/1510.02777}.

\bibitem[Bengio et~al.(2009)Bengio, Louradour, Collobert, and
  Weston]{bengio2009curriculum}
Yoshua Bengio, J{\'e}r{\^o}me Louradour, Ronan Collobert, and Jason Weston.
\newblock Curriculum learning.
\newblock In \emph{Proceedings of the 26th annual international conference on
  machine learning}, pages 41--48. ACM, 2009.

\bibitem[Bengio et~al.(2015{\natexlab{a}})Bengio, Lee, Bornschein, and
  Lin]{Bengio2015}
Yoshua Bengio, Dong-Hyun Lee, Jorg Bornschein, and Zhouhan Lin.
\newblock {Towards Biologically Plausible Deep Learning}.
\newblock feb 2015{\natexlab{a}}.
\newblock URL \url{http://arxiv.org/abs/1502.04156}.

\bibitem[Bengio et~al.(2015{\natexlab{b}})Bengio, Mesnard, Fischer, Zhang, and
  Wu]{Bengio2015b}
Yoshua Bengio, Thomas Mesnard, Asja Fischer, Saizheng Zhang, and Yuhuai Wu.
\newblock {STDP as presynaptic activity times rate of change of postsynaptic
  activity}.
\newblock sep 2015{\natexlab{b}}.
\newblock URL \url{http://arxiv.org/abs/1509.05936}.

\bibitem[Berwick et~al.(2012)Berwick, Beckers, Okanoya, and
  Bolhuis]{Berwick2012}
Robert~C Berwick, Gabri{\"{e}}l J~L Beckers, Kazuo Okanoya, and Johan~J
  Bolhuis.
\newblock {A Bird's Eye View of Human Language Evolution.}
\newblock \emph{Frontiers in evolutionary neuroscience}, 4:\penalty0 5, jan
  2012.
\newblock ISSN 1663-070X.
\newblock \doi{10.3389/fnevo.2012.00005}.
\newblock URL
  \url{http://journal.frontiersin.org/article/10.3389/fnevo.2012.00005/abstract}.

\bibitem[Bialek(2002)]{Bialek2002}
W~Bialek.
\newblock {Thinking about the brain}.
\newblock \emph{Physics of bio-molecules and cells. Physique des {\ldots}},
  2002.
\newblock URL
  \url{http://link.springer.com/chapter/10.1007/3-540-45701-1{\_}12}.

\bibitem[Bialek et~al.(2006)Bialek, {De Ruyter Van Steveninck}, and
  Tishby]{Bialek2006}
William Bialek, Rob {De Ruyter Van Steveninck}, and Naftali Tishby.
\newblock {Efficient representation as a design principle for neural coding and
  computation}.
\newblock In \emph{2006 IEEE International Symposium on Information Theory},
  pages 659--663. IEEE, jul 2006.
\newblock ISBN 1-4244-0505-X.
\newblock \doi{10.1109/ISIT.2006.261867}.
\newblock URL
  \url{http://ieeexplore.ieee.org/articleDetails.jsp?arnumber=4036045}.

\bibitem[Biilthoff et~al.(1989)Biilthoff, Little, and
  Poggio]{biilthoff1989parallel}
H~Biilthoff, J~Little, and T~Poggio.
\newblock A parallel algorithm for real-time computation of optical flow.
\newblock \emph{Nature}, 337\penalty0 (6207):\penalty0 549--553, 1989.

\bibitem[Bloss et~al.(2016)Bloss, Cembrowski, Karsh, Colonell, Fetter, and
  Spruston]{Bloss2016}
Erik B. Bloss, Mark S. Cembrowski, Bill Karsh, Jennifer Colonell, Richard D.
  Fetter, and Nelson Spruston.
\newblock {Structured Dendritic Inhibition Supports Branch-Selective
  Integration in CA1 Pyramidal Cells}.
\newblock \emph{Neuron}, 89\penalty0 (5):\penalty0 1016--30, feb 2016.
\newblock ISSN 08966273.
\newblock \doi{10.1016/j.neuron.2016.01.029}.
\newblock URL \url{http://www.cell.com/article/S0896627316000544/fulltext}.

\bibitem[Bostrom(1996)]{Bostrom1996}
Nick Bostrom.
\newblock {Cortical integration: Possible solutions to the binding and linking
  problems in perception, reasoning and long term memory}.
\newblock 1996.
\newblock URL \url{http://philpapers.org/rec/BOSCIP}.

\bibitem[Botvinick and Weinstein(2014)]{Botvinick2014}
Matthew Botvinick and Ari Weinstein.
\newblock {Model-based hierarchical reinforcement learning and human action
  control.}
\newblock \emph{Philosophical transactions of the Royal Society of London.
  Series B, Biological sciences}, 369\penalty0 (1655):\penalty0 20130480--, nov
  2014.
\newblock ISSN 1471-2970.
\newblock \doi{10.1098/rstb.2013.0480}.
\newblock URL
  \url{http://rstb.royalsocietypublishing.org/content/369/1655/20130480}.

\bibitem[Botvinick et~al.(2009)Botvinick, Niv, and Barto]{Botvinick2009}
MM~Botvinick, Y~Niv, and AC~Barto.
\newblock {Hierarchically organized behavior and its neural foundations: A
  reinforcement learning perspective}.
\newblock \emph{Cognition}, 2009.
\newblock URL
  \url{http://www.sciencedirect.com/science/article/pii/S0010027708002059}.

\bibitem[Bouchard et~al.(2015)Bouchard, Trouillon, Perez, and
  Gaidon]{bouchard2015accelerating}
Guillaume Bouchard, Th{\'e}o Trouillon, Julien Perez, and Adrien Gaidon.
\newblock Accelerating stochastic gradient descent via online learning to
  sample.
\newblock \emph{arXiv preprint arXiv:1506.09016}, 2015.

\bibitem[Bourdoukan and Den{\`{e}}ve(2015)]{Bourdoukan2015}
Ralph Bourdoukan and Sophie Den{\`{e}}ve.
\newblock {Enforcing balance allows local supervised learning in spiking
  recurrent networks}.
\newblock In \emph{Advances in Neural Information Processing Systems}, pages
  982--990, 2015.

\bibitem[Braitenberg and Schutz(1991)]{braitenberg1991anatomy}
V~Braitenberg and A~Schutz.
\newblock Anatomy of the cortex: studies of brain function, 1991.

\bibitem[Bremner et~al.(2015)Bremner, Slater, and Johnson]{Bremner2015}
J.~Gavin Bremner, Alan~M. Slater, and Scott~P. Johnson.
\newblock {Perception of Object Persistence: The Origins of Object Permanence
  in Infancy}.
\newblock \emph{Child Development Perspectives}, 9\penalty0 (1):\penalty0
  7--13, mar 2015.
\newblock ISSN 17508592.
\newblock \doi{10.1111/cdep.12098}.
\newblock URL \url{http://doi.wiley.com/10.1111/cdep.12098}.

\bibitem[Brosch et~al.(2015)Brosch, Neumann, and Roelfsema]{Brosch2015}
Tobias Brosch, Heiko Neumann, and Pieter~R Roelfsema.
\newblock {Reinforcement Learning of Linking and Tracing Contours in Recurrent
  Neural Networks.}
\newblock \emph{PLoS computational biology}, 11\penalty0 (10):\penalty0
  e1004489, oct 2015.
\newblock ISSN 1553-7358.
\newblock \doi{10.1371/journal.pcbi.1004489}.

\bibitem[Brown et~al.(2016)Brown, Carr, LaRocque, Favila, Gordon, Bowles,
  Bailenson, and Wagner]{Brown1323}
Thackery~I. Brown, Valerie~A. Carr, Karen~F. LaRocque, Serra~E. Favila, Alan~M.
  Gordon, Ben Bowles, Jeremy~N. Bailenson, and Anthony~D. Wagner.
\newblock Prospective representation of navigational goals in the human
  hippocampus.
\newblock \emph{Science}, 352\penalty0 (6291):\penalty0 1323--1326, 2016.
\newblock ISSN 0036-8075.
\newblock \doi{10.1126/science.aaf0784}.
\newblock URL \url{http://science.sciencemag.org/content/352/6291/1323}.

\bibitem[Buesing et~al.(2011)Buesing, Bill, Nessler, and Maass]{Buesing2011}
Lars Buesing, Johannes Bill, Bernhard Nessler, and Wolfgang Maass.
\newblock {Neural dynamics as sampling: a model for stochastic computation in
  recurrent networks of spiking neurons.}
\newblock \emph{PLoS computational biology}, 7\penalty0 (11):\penalty0
  e1002211, nov 2011.
\newblock ISSN 1553-7358.
\newblock \doi{10.1371/journal.pcbi.1002211}.

\bibitem[Buonomano and Merzenich(1995)]{Buonomano1995}
D~V Buonomano and M~M Merzenich.
\newblock {Temporal information transformed into a spatial code by a neural
  network with realistic properties.}
\newblock \emph{Science (New York, N.Y.)}, 267\penalty0 (5200):\penalty0
  1028--30, feb 1995.
\newblock ISSN 0036-8075.
\newblock URL \url{http://www.ncbi.nlm.nih.gov/pubmed/7863330}.

\bibitem[Buschman and Miller(2010)]{Buschman2010}
Timothy~J Buschman and Earl~K Miller.
\newblock {Shifting the spotlight of attention: evidence for discrete
  computations in cognition.}
\newblock \emph{Frontiers in human neuroscience}, 4:\penalty0 194, jan 2010.
\newblock ISSN 1662-5161.
\newblock \doi{10.3389/fnhum.2010.00194}.

\bibitem[Buschman and Miller(2014)]{Buschman2014}
Timothy~J Buschman and Earl~K Miller.
\newblock {Goal-direction and top-down control.}
\newblock \emph{Philosophical transactions of the Royal Society of London.
  Series B, Biological sciences}, 369\penalty0 (1655), nov 2014.
\newblock ISSN 1471-2970.
\newblock \doi{10.1098/rstb.2013.0471}.
\newblock URL \url{http://www.ncbi.nlm.nih.gov/pubmed/25267814}.

\bibitem[Bush(2007)]{bush2007echo}
Keith~A Bush.
\newblock An echo state model of non-markovian reinforcement learning.
\newblock 2007.

\bibitem[Buzs{\'{a}}ki and Moser(2013)]{Buzsaki2013}
Gy{\"{o}}rgy Buzs{\'{a}}ki and Edvard~I Moser.
\newblock {Memory, navigation and theta rhythm in the hippocampal-entorhinal
  system.}
\newblock \emph{Nature neuroscience}, 16\penalty0 (2):\penalty0 130--8, feb
  2013.
\newblock ISSN 1546-1726.
\newblock \doi{10.1038/nn.3304}.
\newblock URL \url{http://dx.doi.org/10.1038/nn.3304}.

\bibitem[Cai et~al.(2016)Cai, Aharoni, Shuman, Shobe, Biane, Song, Wei,
  Veshkini, La-Vu, Lou, et~al.]{cai2016shared}
Denise~J Cai, Daniel Aharoni, Tristan Shuman, Justin Shobe, Jeremy Biane,
  Weilin Song, Brandon Wei, Michael Veshkini, Mimi La-Vu, Jerry Lou, et~al.
\newblock A shared neural ensemble links distinct contextual memories encoded
  close in time.
\newblock \emph{Nature}, 534\penalty0 (7605):\penalty0 115--118, 2016.

\bibitem[Callaway(2004)]{Callaway2004}
EM~Callaway.
\newblock {Feedforward, feedback and inhibitory connections in primate visual
  cortex}.
\newblock \emph{Neural Networks}, 2004.
\newblock URL
  \url{http://www.sciencedirect.com/science/article/pii/S0893608004000887}.

\bibitem[Caron et~al.(2013)Caron, Ruta, Abbott, and Axel]{Caron2013}
Sophie J~C Caron, Vanessa Ruta, L~F Abbott, and Richard Axel.
\newblock {Random convergence of olfactory inputs in the Drosophila mushroom
  body.}
\newblock \emph{Nature}, 497\penalty0 (7447):\penalty0 113--7, may 2013.
\newblock ISSN 1476-4687.
\newblock \doi{10.1038/nature12063}.
\newblock URL \url{http://dx.doi.org/10.1038/nature12063}.

\bibitem[Chen et~al.(2014)Chen, Schwing, Yuille, and Urtasun]{Chen2014}
Liang-Chieh Chen, Alexander~G. Schwing, Alan~L. Yuille, and Raquel Urtasun.
\newblock {Learning Deep Structured Models}.
\newblock page~11, jul 2014.
\newblock URL \url{http://arxiv.org/abs/1407.2538}.

\bibitem[Chikkerur et~al.(2010)Chikkerur, Serre, Tan, and
  Poggio]{Chikkerur2010}
Sharat Chikkerur, Thomas Serre, Cheston Tan, and Tomaso Poggio.
\newblock {What and where: a Bayesian inference theory of attention.}
\newblock \emph{Vision research}, 50\penalty0 (22):\penalty0 2233--47, oct
  2010.
\newblock ISSN 1878-5646.
\newblock \doi{10.1016/j.visres.2010.05.013}.
\newblock URL \url{http://www.ncbi.nlm.nih.gov/pubmed/20493206}.

\bibitem[{Choo and Eliasmith}(2010)]{ChooandEliasmith2010}
{Choo and Eliasmith}.
\newblock {A Spiking Neuron Model of Serial-Order Recall}.
\newblock \emph{32nd Annual Conference of the Cognitive Science Society}, 2010.
\newblock URL
  \url{http://citeseerx.ist.psu.edu/viewdoc/summary?doi=10.1.1.353.1190}.

\bibitem[Chubykin et~al.(2013)Chubykin, Roach, Bear, and Shuler]{Chubykin2013}
Alexander~A Chubykin, Emma~B Roach, Mark~F Bear, and Marshall G~Hussain Shuler.
\newblock {A cholinergic mechanism for reward timing within primary visual
  cortex.}
\newblock \emph{Neuron}, 77\penalty0 (4):\penalty0 723--35, feb 2013.
\newblock ISSN 1097-4199.
\newblock \doi{10.1016/j.neuron.2012.12.039}.
\newblock URL \url{http://www.cell.com/article/S0896627313000470/fulltext}.

\bibitem[Chung et~al.(2014)Chung, Gulcehre, Cho, and Bengio]{Chung2014}
Junyoung Chung, Caglar Gulcehre, KyungHyun Cho, and Yoshua Bengio.
\newblock {Empirical Evaluation of Gated Recurrent Neural Networks on Sequence
  Modeling}.
\newblock dec 2014.
\newblock URL \url{http://arxiv.org/abs/1412.3555}.

\bibitem[Clayton and Dickinson(1998)]{clayton1998episodic}
Nicola~S Clayton and Anthony Dickinson.
\newblock Episodic-like memory during cache recovery by scrub jays.
\newblock \emph{Nature}, 395\penalty0 (6699):\penalty0 272--274, 1998.

\bibitem[Clopath and Gerstner(2010)]{clopath2010voltage}
Claudia Clopath and Wulfram Gerstner.
\newblock Voltage and spike timing interact in stdp--a unified model.
\newblock \emph{Spike-timing dependent plasticity}, page 294, 2010.

\bibitem[Cohn et~al.(2015)Cohn, Morantte, and Ruta]{Cohn2015}
Raphael Cohn, Ianessa Morantte, and Vanessa Ruta.
\newblock {Coordinated and Compartmentalized Neuromodulation Shapes Sensory
  Processing in Drosophila}.
\newblock \emph{Cell}, 163\penalty0 (7):\penalty0 1742--1755, dec 2015.
\newblock ISSN 00928674.
\newblock \doi{10.1016/j.cell.2015.11.019}.
\newblock URL \url{http://www.cell.com/article/S0092867415014993/fulltext}.

\bibitem[Colino and Halliwell(1987)]{colino1987differential}
A~Colino and JV~Halliwell.
\newblock Differential modulation of three separate k-conductances in
  hippocampal ca1 neurons by serotonin.
\newblock \emph{Nature}, 328\penalty0 (6125):\penalty0 73--77, 1987.

\bibitem[Cox and Dean(2014)]{Cox2014}
David~Daniel Cox and Thomas Dean.
\newblock {Neural networks and neuroscience-inspired computer vision.}
\newblock \emph{Current biology : CB}, 24\penalty0 (18):\penalty0 R921--9, sep
  2014.
\newblock ISSN 1879-0445.
\newblock \doi{10.1016/j.cub.2014.08.026}.
\newblock URL
  \url{http://www.sciencedirect.com/science/article/pii/S0960982214010392}.

\bibitem[Crick(1989)]{Crick1989}
F~Crick.
\newblock {The recent excitement about neural networks.}
\newblock \emph{Nature}, 337\penalty0 (6203):\penalty0 129--32, jan 1989.
\newblock ISSN 0028-0836.
\newblock \doi{10.1038/337129a0}.
\newblock URL \url{http://dx.doi.org/10.1038/337129a0}.

\bibitem[Crick and Koch(2005)]{crick2005function}
Francis~C Crick and Christof Koch.
\newblock What is the function of the claustrum?
\newblock \emph{Philosophical Transactions of the Royal Society of London B:
  Biological Sciences}, 360\penalty0 (1458):\penalty0 1271--1279, 2005.

\bibitem[Crouzet and Thorpe(2011)]{Crouzet2011}
S{\'{e}}bastien~M Crouzet and Simon~J Thorpe.
\newblock {Low-level cues and ultra-fast face detection.}
\newblock \emph{Frontiers in psychology}, 2:\penalty0 342, jan 2011.
\newblock ISSN 1664-1078.
\newblock \doi{10.3389/fpsyg.2011.00342}.

\bibitem[Cui et~al.(2015)Cui, Surpur, Ahmad, and Hawkins]{Cui2015}
Yuwei Cui, Chetan Surpur, Subutai Ahmad, and Jeff Hawkins.
\newblock {Continuous online sequence learning with an unsupervised neural
  network model}.
\newblock dec 2015.
\newblock URL \url{http://arxiv.org/abs/1512.05463}.

\bibitem[Danihelka et~al.(2016)Danihelka, Wayne, Uria, Kalchbrenner, and
  Graves]{Danihelka2016}
Ivo Danihelka, Greg Wayne, Benigno Uria, Nal Kalchbrenner, and Alex Graves.
\newblock {Associative Long Short-Term Memory}.
\newblock feb 2016.
\newblock URL \url{http://arxiv.org/abs/1602.03032}.

\bibitem[Daw et~al.(2006)Daw, Niv, and Dayan]{daw2006actions}
Nathaniel~D Daw, Yael Niv, and Peter Dayan.
\newblock Actions, policies, values and the basal ganglia.
\newblock \emph{Recent breakthroughs in basal ganglia research}, pages 91--106,
  2006.

\bibitem[Dayan(2012)]{Dayan2012}
Peter Dayan.
\newblock {Twenty-five lessons from computational neuromodulation.}
\newblock \emph{Neuron}, 76\penalty0 (1):\penalty0 240--56, oct 2012.
\newblock ISSN 1097-4199.
\newblock \doi{10.1016/j.neuron.2012.09.027}.
\newblock URL
  \url{http://www.sciencedirect.com/science/article/pii/S0896627312008628}.

\bibitem[Dean(2005)]{Dean2005}
T~Dean.
\newblock {A computational model of the cerebral cortex}.
\newblock \emph{Proceedings of the National Conference on Artificial {\ldots}},
  2005.
\newblock URL \url{http://www.aaai.org/Papers/AAAI/2005/AAAI05-148.pdf}.

\bibitem[Dehaene et~al.(2015)Dehaene, Meyniel, Wacongne, Wang, and
  Pallier]{Dehaene2015}
Stanislas Dehaene, Florent Meyniel, Catherine Wacongne, Liping Wang, and
  Christophe Pallier.
\newblock {The Neural Representation of Sequences: From Transition
  Probabilities to Algebraic Patterns and Linguistic Trees}.
\newblock \emph{Neuron}, 88\penalty0 (1):\penalty0 2--19, oct 2015.
\newblock ISSN 08966273.
\newblock \doi{10.1016/j.neuron.2015.09.019}.
\newblock URL
  \url{http://www.sciencedirect.com/science/article/pii/S089662731500776X}.

\bibitem[Delalleau and Bengio(2011)]{Delalleau2011}
Olivier Delalleau and Yoshua Bengio.
\newblock {Shallow vs. Deep Sum-Product Networks}.
\newblock In \emph{Advances in Neural Information Processing Systems}, pages
  666--674, 2011.
\newblock URL
  \url{http://papers.nips.cc/paper/4350-shallow-vs-deep-sum-product-networks}.

\bibitem[DePasquale et~al.(2016)DePasquale, Churchland, and
  Abbott]{DePasquale2016}
B~DePasquale, MM~Churchland, and LF~Abbott.
\newblock {Using firing-rate dynamics to train recurrent networks of spiking
  model neurons}.
\newblock \emph{arXiv preprint arXiv: {\ldots}}, 2016.
\newblock URL \url{http://arxiv.org/abs/1601.07620}.

\bibitem[DeWolf and Eliasmith(2011)]{DeWolf2011a}
T~DeWolf and C~Eliasmith.
\newblock {The neural optimal control hierarchy for motor control.}
\newblock \emph{Journal of neural engineering}, 8\penalty0 (6):\penalty0
  065009, dec 2011.
\newblock ISSN 1741-2552.
\newblock \doi{10.1088/1741-2560/8/6/065009}.
\newblock URL \url{http://www.ncbi.nlm.nih.gov/pubmed/22056418}.

\bibitem[DiCarlo et~al.(2012)DiCarlo, Zoccolan, and Rust]{dicarlo2012does}
James~J DiCarlo, Davide Zoccolan, and Nicole~C Rust.
\newblock How does the brain solve visual object recognition?
\newblock \emph{Neuron}, 73\penalty0 (3):\penalty0 415--434, 2012.

\bibitem[Douglas and Martin(2004)]{Douglas2004}
Rodney~J Douglas and Kevan A~C Martin.
\newblock {Neuronal circuits of the neocortex.}
\newblock \emph{Annual review of neuroscience}, 27:\penalty0 419--51, jan 2004.
\newblock ISSN 0147-006X.
\newblock \doi{10.1146/annurev.neuro.27.070203.144152}.
\newblock URL \url{http://www.ncbi.nlm.nih.gov/pubmed/15217339}.

\bibitem[Doya(1999)]{doya1999computations}
Kenji Doya.
\newblock What are the computations of the cerebellum, the basal ganglia and
  the cerebral cortex?
\newblock \emph{Neural networks}, 12\penalty0 (7):\penalty0 961--974, 1999.

\bibitem[Dudman et~al.(2007)Dudman, Tsay, and Siegelbaum]{dudman2007role}
Joshua~T Dudman, David Tsay, and Steven~A Siegelbaum.
\newblock A role for synaptic inputs at distal dendrites: instructive signals
  for hippocampal long-term plasticity.
\newblock \emph{Neuron}, 56\penalty0 (5):\penalty0 866--879, 2007.

\bibitem[Eliasmith(2013)]{Eliasmith2013}
Chris Eliasmith.
\newblock \emph{{How to Build a Brain: A Neural Architecture for Biological
  Cognition}}.
\newblock Oxford University Press, 2013.
\newblock ISBN 0199794545.
\newblock URL \url{http://books.google.com/books?id=BK0YRJPmuzgC{\&}pgis=1}.

\bibitem[Eliasmith and Anderson(2004)]{Eliasmith2004}
Chris Eliasmith and Charles~H. Anderson.
\newblock \emph{{Neural Engineering: Computation, Representation, and Dynamics
  in Neurobiological Systems}}.
\newblock MIT Press, 2004.
\newblock ISBN 0262550601.
\newblock URL \url{http://books.google.com/books?id=J6jz9s4kbfIC{\&}pgis=1}.

\bibitem[Eliasmith and Martens(2011)]{eliasmith2011normalization}
Chris Eliasmith and James Martens.
\newblock Normalization for probabilistic inference with neurons.
\newblock \emph{Biological cybernetics}, 104\penalty0 (4-5):\penalty0 251--262,
  2011.

\bibitem[Eliasmith et~al.(2012)Eliasmith, Stewart, Choo, Bekolay, DeWolf, Tang,
  Tang, and Rasmussen]{Eliasmith2012}
Chris Eliasmith, Terrence~C Stewart, Xuan Choo, Trevor Bekolay, Travis DeWolf,
  Yichuan Tang, Charlie Tang, and Daniel Rasmussen.
\newblock {A large-scale model of the functioning brain.}
\newblock \emph{Science (New York, N.Y.)}, 338\penalty0 (6111):\penalty0
  1202--5, nov 2012.
\newblock ISSN 1095-9203.
\newblock \doi{10.1126/science.1225266}.
\newblock URL \url{http://www.sciencemag.org/content/338/6111/1202}.

\bibitem[Emlen(1967)]{emlen1967migratory}
Stephen~T Emlen.
\newblock Migratory orientation in the indigo bunting, passerina cyanea: part
  i: evidence for use of celestial cues.
\newblock \emph{The Auk}, 84\penalty0 (3):\penalty0 309--342, 1967.

\bibitem[Erhan and Manzagol(2009)]{Erhan2009}
D~Erhan and PA~Manzagol.
\newblock {The difficulty of training deep architectures and the effect of
  unsupervised pre-training}.
\newblock \emph{International {\ldots}}, 2009.
\newblock URL
  \url{http://machinelearning.wustl.edu/mlpapers/paper{\_}files/AISTATS09{\_}ErhanMBBV.pdf}.

\bibitem[Eslami et~al.(2016)Eslami, Heess, and Weber]{Eslami2016}
SM~Eslami, N~Heess, and T~Weber.
\newblock {Attend, Infer, Repeat: Fast Scene Understanding with Generative
  Models}.
\newblock \emph{arXiv preprint arXiv: {\ldots}}, 2016.
\newblock URL \url{http://arxiv.org/abs/1603.08575}.

\bibitem[Fausey et~al.(2016)Fausey, Jayaraman, and Smith]{fausey2016faces}
Caitlin~M Fausey, Swapnaa Jayaraman, and Linda~B Smith.
\newblock From faces to hands: Changing visual input in the first two years.
\newblock \emph{Cognition}, 152:\penalty0 101--107, 2016.

\bibitem[Felleman and Van~Essen(1991)]{felleman1991distributed}
Daniel~J Felleman and David~C Van~Essen.
\newblock Distributed hierarchical processing in the primate cerebral cortex.
\newblock \emph{Cerebral cortex}, 1\penalty0 (1):\penalty0 1--47, 1991.

\bibitem[Ferster and Miller(2003)]{Ferster2003}
David Ferster and Kenneth~D. Miller.
\newblock {Neural Mechanisms of Orientation Selectivity in the Visual Cortex}.
\newblock nov 2003.
\newblock URL
  \url{http://www.annualreviews.org/doi/abs/10.1146/annurev.neuro.23.1.441}.

\bibitem[Fetz(1969)]{fetz1969operant}
Eberhard~E Fetz.
\newblock Operant conditioning of cortical unit activity.
\newblock \emph{Science}, 163\penalty0 (3870):\penalty0 955--958, 1969.

\bibitem[Fetz(2007)]{fetz2007volitional}
Eberhard~E Fetz.
\newblock Volitional control of neural activity: implications for
  brain--computer interfaces.
\newblock \emph{The Journal of physiology}, 579\penalty0 (3):\penalty0
  571--579, 2007.

\bibitem[Fiete and Seung(2006)]{Fiete2006}
Ila~R Fiete and H~Sebastian Seung.
\newblock {Gradient learning in spiking neural networks by dynamic perturbation
  of conductances.}
\newblock \emph{Physical review letters}, 97\penalty0 (4):\penalty0 048104, jul
  2006.
\newblock ISSN 0031-9007.
\newblock \doi{10.1103/PhysRevLett.97.048104}.
\newblock URL
  \url{http://journals.aps.org/prl/abstract/10.1103/PhysRevLett.97.048104}.

\bibitem[Fiete et~al.(2007)Fiete, Fee, and Seung]{Fiete2007}
Ila~R Fiete, Michale~S Fee, and H~Sebastian Seung.
\newblock {Model of birdsong learning based on gradient estimation by dynamic
  perturbation of neural conductances.}
\newblock \emph{Journal of neurophysiology}, 98\penalty0 (4):\penalty0
  2038--57, oct 2007.
\newblock ISSN 0022-3077.
\newblock \doi{10.1152/jn.01311.2006}.
\newblock URL \url{http://www.ncbi.nlm.nih.gov/pubmed/17652414}.

\bibitem[Fiete et~al.(2010)Fiete, Senn, Wang, and Hahnloser]{Fiete2010}
Ila~R Fiete, Walter Senn, Claude Z~H Wang, and Richard H~R Hahnloser.
\newblock {Spike-time-dependent plasticity and heterosynaptic competition
  organize networks to produce long scale-free sequences of neural activity.}
\newblock \emph{Neuron}, 65\penalty0 (4):\penalty0 563--76, feb 2010.
\newblock ISSN 1097-4199.
\newblock \doi{10.1016/j.neuron.2010.02.003}.
\newblock URL \url{http://www.ncbi.nlm.nih.gov/pubmed/20188660}.

\bibitem[Finnerty and Shadlen(2015)]{Finnerty2015}
GT~Finnerty and MN~Shadlen.
\newblock {Time in Cortical Circuits}.
\newblock \emph{The Journal of {\ldots}}, 2015.
\newblock URL \url{https://www.jneurosci.org/content/35/41/13912.full}.

\bibitem[Fodor and Crowther(2002)]{fodor2002understanding}
Janet~Dean Fodor and Carrie Crowther.
\newblock Understanding stimulus poverty arguments.
\newblock \emph{The Linguistic Review}, 18\penalty0 (1-2):\penalty0 105--145,
  2002.

\bibitem[F{\"{o}}ldi{\'{a}}k(2008)]{Foldiak2008}
Peter F{\"{o}}ldi{\'{a}}k.
\newblock {Learning Invariance from Transformation Sequences}, mar 2008.
\newblock URL
  \url{http://www.mitpressjournals.org/doi/abs/10.1162/neco.1991.3.2.194{\#}.VofznO8rKHo}.

\bibitem[Foster et~al.(2000)Foster, Morris, Dayan, and
  Neuroscience]{Foster2000}
D.~J. Foster, R.~G.~M. Morris, Peter Dayan, and Centre~For Neuroscience.
\newblock {Models of Hippocampally Dependent Navigation, Using The Temporal
  Difference Learning Rule}.
\newblock sep 2000.

\bibitem[Frank and Badre(2012)]{Frank2012}
Michael~J Frank and David Badre.
\newblock {Mechanisms of hierarchical reinforcement learning in corticostriatal
  circuits 1: computational analysis.}
\newblock \emph{Cerebral cortex (New York, N.Y. : 1991)}, 22\penalty0
  (3):\penalty0 509--26, mar 2012.
\newblock ISSN 1460-2199.
\newblock \doi{10.1093/cercor/bhr114}.

\bibitem[Frankland and Greene(2015)]{Frankland2015}
Steven~M Frankland and Joshua~D Greene.
\newblock {An architecture for encoding sentence meaning in left mid-superior
  temporal cortex.}
\newblock \emph{Proceedings of the National Academy of Sciences of the United
  States of America}, 112\penalty0 (37):\penalty0 11732--11737, aug 2015.
\newblock ISSN 1091-6490.
\newblock \doi{10.1073/pnas.1421236112}.
\newblock URL \url{http://www.pnas.org/content/112/37/11732.full}.

\bibitem[Franzius et~al.(2007)Franzius, Sprekeler, and Wiskott]{Franzius2007}
Mathias Franzius, Henning Sprekeler, and Laurenz Wiskott.
\newblock {Slowness and sparseness lead to place, head-direction, and
  spatial-view cells.}
\newblock \emph{PLoS computational biology}, 3\penalty0 (8):\penalty0 e166, aug
  2007.
\newblock ISSN 1553-7358.
\newblock \doi{10.1371/journal.pcbi.0030166}.
\newblock URL
  \url{http://journals.plos.org/ploscompbiol/article?id=10.1371/journal.pcbi.0030166}.

\bibitem[Friston(2010)]{Friston2010}
K~Friston.
\newblock {The free-energy principle: a unified brain theory?}
\newblock \emph{Nature Reviews Neuroscience}, 2010.
\newblock URL \url{http://www.nature.com/nrn/journal/v11/n2/abs/nrn2787.html}.

\bibitem[Friston and Stephan(2007)]{Friston2007}
KJ~Friston and KE~Stephan.
\newblock {Free-energy and the brain}.
\newblock \emph{Synthese}, 2007.
\newblock URL \url{http://link.springer.com/article/10.1007/s11229-007-9237-y}.

\bibitem[Fukushima(1980)]{fukushima1980neocognitron}
Kunihiko Fukushima.
\newblock Neocognitron: A self-organizing neural network model for a mechanism
  of pattern recognition unaffected by shift in position.
\newblock \emph{Biological cybernetics}, 36\penalty0 (4):\penalty0 193--202,
  1980.

\bibitem[Gao et~al.(2014)Gao, Harari, Tenenbaum, and Ullman]{Gao2014}
Tao Gao, Daniel Harari, Joshua Tenenbaum, and Shimon Ullman.
\newblock {When Computer Vision Gazes at Cognition}.
\newblock dec 2014.
\newblock URL \url{http://arxiv.org/abs/1412.2672}.

\bibitem[George and Hawkins(2009)]{George2009}
Dileep George and Jeff Hawkins.
\newblock {Towards a mathematical theory of cortical micro-circuits.}
\newblock \emph{PLoS computational biology}, 5\penalty0 (10):\penalty0
  e1000532, oct 2009.
\newblock ISSN 1553-7358.
\newblock \doi{10.1371/journal.pcbi.1000532}.

\bibitem[Gershman and Beck(2016)]{gershman2016complex}
Samuel~J Gershman and Jeffrey~M Beck.
\newblock Complex probabilistic inference: From cognition to neural
  computation.
\newblock 2016.

\bibitem[Gershman et~al.(2012)Gershman, Moore, Todd, Norman, and
  Sederberg]{gershman2012successor}
Samuel~J Gershman, Christopher~D Moore, Michael~T Todd, Kenneth~A Norman, and
  Per~B Sederberg.
\newblock The successor representation and temporal context.
\newblock \emph{Neural Computation}, 24\penalty0 (6):\penalty0 1553--1568,
  2012.

\bibitem[Gershman et~al.(2014)Gershman, Moustafa, and Ludvig]{gershman2014time}
Samuel~J Gershman, Ahmed~A Moustafa, and Elliot~A Ludvig.
\newblock Time representation in reinforcement learning models of the basal
  ganglia.
\newblock 2014.

\bibitem[Giret et~al.(2014)Giret, Kornfeld, Ganguli, and Hahnloser]{Giret2014}
Nicolas Giret, Joergen Kornfeld, Surya Ganguli, and Richard H~R Hahnloser.
\newblock {Evidence for a causal inverse model in an avian cortico-basal
  ganglia circuit.}
\newblock \emph{Proceedings of the National Academy of Sciences of the United
  States of America}, 111\penalty0 (16):\penalty0 6063--8, apr 2014.
\newblock ISSN 1091-6490.
\newblock \doi{10.1073/pnas.1317087111}.
\newblock URL \url{http://www.pnas.org/content/111/16/6063.full}.

\bibitem[Goertzel(2014)]{Goertzel2014}
B~Goertzel.
\newblock {How might the brain represent complex symbolic knowledge?}
\newblock \emph{Neural Networks (IJCNN), 2014 International Joint {\ldots}},
  2014.
\newblock URL
  \url{http://ieeexplore.ieee.org/xpls/abs{\_}all.jsp?arnumber=6889662}.

\bibitem[Goldman et~al.(2003)Goldman, Levine, Major, Tank, and
  Seung]{goldman2003robust}
Mark~S Goldman, Joseph~H Levine, Guy Major, David~W Tank, and HS~Seung.
\newblock Robust persistent neural activity in a model integrator with multiple
  hysteretic dendrites per neuron.
\newblock \emph{Cerebral cortex}, 13\penalty0 (11):\penalty0 1185--1195, 2003.

\bibitem[{Gonzalez Andino} and {Grave de Peralta
  Menendez}(2012)]{GonzalezAndino2012}
S~L {Gonzalez Andino} and R~{Grave de Peralta Menendez}.
\newblock {Coding of saliency by ensemble bursting in the amygdala of
  primates.}
\newblock \emph{Frontiers in behavioral neuroscience}, 6:\penalty0 38, jan
  2012.
\newblock ISSN 1662-5153.
\newblock \doi{10.3389/fnbeh.2012.00038}.
\newblock URL
  \url{http://journal.frontiersin.org/article/10.3389/fnbeh.2012.00038/abstract}.

\bibitem[Gooch et~al.(2010)Gooch, Wiener, Wencil, and Coslett]{Gooch2010}
Cynthia~M Gooch, Martin Wiener, Elaine~B Wencil, and H~Branch Coslett.
\newblock {Interval timing disruptions in subjects with cerebellar lesions.}
\newblock \emph{Neuropsychologia}, 48\penalty0 (4):\penalty0 1022--31, mar
  2010.
\newblock ISSN 1873-3514.
\newblock \doi{10.1016/j.neuropsychologia.2009.11.028}.

\bibitem[Goodfellow et~al.(2014{\natexlab{a}})Goodfellow, Pouget-Abadie, Mirza,
  Xu, Warde-Farley, Ozair, Courville, and Bengio]{Goodfellow2014}
Ian~J. Goodfellow, Jean Pouget-Abadie, Mehdi Mirza, Bing Xu, David
  Warde-Farley, Sherjil Ozair, Aaron Courville, and Yoshua Bengio.
\newblock {Generative Adversarial Networks}.
\newblock jun 2014{\natexlab{a}}.
\newblock URL \url{http://arxiv.org/abs/1406.2661}.

\bibitem[Goodfellow et~al.(2014{\natexlab{b}})Goodfellow, Vinyals, and
  Saxe]{Goodfellow2014a}
Ian~J. Goodfellow, Oriol Vinyals, and Andrew~M. Saxe.
\newblock {Qualitatively characterizing neural network optimization problems}.
\newblock dec 2014{\natexlab{b}}.
\newblock URL \url{http://arxiv.org/abs/1412.6544}.

\bibitem[Gopnik et~al.(2000)Gopnik, Meltzoff, and Kuhl]{gopnik2000scientist}
Alison Gopnik, Andrew~N Meltzoff, and Patricia~K Kuhl.
\newblock \emph{The scientist in the crib: What early learning tells us about
  the mind}.
\newblock Harper Paperbacks, 2000.

\bibitem[Graves et~al.(2014)Graves, Wayne, and Danihelka]{Graves2014}
Alex Graves, Greg Wayne, and Ivo Danihelka.
\newblock {Neural Turing Machines}.
\newblock \emph{ArXiv}, oct 2014.
\newblock URL \url{http://arxiv.org/abs/1410.5401}.

\bibitem[Graybiel(1998)]{graybiel1998basal}
Ann~M Graybiel.
\newblock The basal ganglia and chunking of action repertoires.
\newblock \emph{Neurobiology of learning and memory}, 70\penalty0 (1):\penalty0
  119--136, 1998.

\bibitem[Gregor et~al.(2015)Gregor, Danihelka, Graves, Rezende, and
  Wierstra]{Gregor2015}
Karol Gregor, Ivo Danihelka, Alex Graves, Danilo~Jimenez Rezende, and Daan
  Wierstra.
\newblock {DRAW: A Recurrent Neural Network For Image Generation}.
\newblock feb 2015.
\newblock URL \url{http://arxiv.org/abs/1502.04623}.

\bibitem[Grillner et~al.(2005)Grillner, Hellgren, M{\'{e}}nard, Saitoh, and
  Wikstr{\"{o}}m]{Grillner2005}
Sten Grillner, Jeanette Hellgren, Ariane M{\'{e}}nard, Kazuya Saitoh, and
  Martin~A Wikstr{\"{o}}m.
\newblock {Mechanisms for selection of basic motor programs--roles for the
  striatum and pallidum.}
\newblock \emph{Trends in neurosciences}, 28\penalty0 (7):\penalty0 364--70,
  jul 2005.
\newblock ISSN 0166-2236.
\newblock \doi{10.1016/j.tins.2005.05.004}.
\newblock URL \url{http://www.ncbi.nlm.nih.gov/pubmed/15935487}.

\bibitem[Grossberg(2013)]{Grossberg2013}
Stephen Grossberg.
\newblock {Adaptive Resonance Theory: how a brain learns to consciously attend,
  learn, and recognize a changing world.}
\newblock \emph{Neural networks : the official journal of the International
  Neural Network Society}, 37:\penalty0 1--47, jan 2013.
\newblock ISSN 1879-2782.
\newblock \doi{10.1016/j.neunet.2012.09.017}.
\newblock URL
  \url{http://www.sciencedirect.com/science/article/pii/S0893608012002584}.

\bibitem[Guez et~al.(2012)Guez, Silver, and Dayan]{guez2012efficient}
Arthur Guez, David Silver, and Peter Dayan.
\newblock Efficient bayes-adaptive reinforcement learning using sample-based
  search.
\newblock In \emph{Advances in Neural Information Processing Systems}, pages
  1025--1033, 2012.

\bibitem[G{\"{u}}l{\c{c}}ehre and Bengio(2016)]{Gulcehre2016}
{\c{C}}ağlar G{\"{u}}l{\c{c}}ehre and Yoshua Bengio.
\newblock {Knowledge Matters: Importance of Prior Information for
  Optimization}.
\newblock \emph{Journal of Machine Learning Research}, 17\penalty0
  (8):\penalty0 1--32, 2016.
\newblock URL \url{http://jmlr.org/papers/v17/gulchere16a.html}.

\bibitem[G{\"u}nt{\"u}rk{\"u}n and Bugnyar(2016)]{gunturkun2016cognition}
Onur G{\"u}nt{\"u}rk{\"u}n and Thomas Bugnyar.
\newblock Cognition without cortex.
\newblock \emph{Trends in cognitive sciences}, 20\penalty0 (4):\penalty0
  291--303, 2016.

\bibitem[Gurney et~al.(2001)Gurney, Prescott, and Redgrave]{Gurney2001}
K~Gurney, T~J Prescott, and P~Redgrave.
\newblock {A computational model of action selection in the basal ganglia. I. A
  new functional anatomy.}
\newblock \emph{Biological cybernetics}, 84\penalty0 (6):\penalty0 401--10, jun
  2001.
\newblock ISSN 0340-1200.
\newblock URL \url{http://www.ncbi.nlm.nih.gov/pubmed/11417052}.

\bibitem[Hadley(2009)]{Hadley2009}
Robert~F Hadley.
\newblock {The problem of rapid variable creation.}
\newblock \emph{Neural computation}, 21\penalty0 (2):\penalty0 510--32, mar
  2009.
\newblock ISSN 0899-7667.
\newblock \doi{10.1162/neco.2008.07-07-572}.
\newblock URL \url{http://www.ncbi.nlm.nih.gov/pubmed/19431268}.

\bibitem[Hamlin et~al.(2007)Hamlin, Wynn, and Bloom]{Hamlin2007}
J~Kiley Hamlin, Karen Wynn, and Paul Bloom.
\newblock {Social evaluation by preverbal infants.}
\newblock \emph{Nature}, 450\penalty0 (7169):\penalty0 557--9, nov 2007.
\newblock ISSN 1476-4687.
\newblock \doi{10.1038/nature06288}.
\newblock URL \url{http://dx.doi.org/10.1038/nature06288}.

\bibitem[Hangya et~al.(2015)Hangya, Ranade, Lorenc, and Kepecs]{Hangya2015}
Bal{\'{a}}zs Hangya, Sachin P. Ranade, Maja Lorenc, and Adam Kepecs.
\newblock {Central Cholinergic Neurons Are Rapidly Recruited by Reinforcement
  Feedback}.
\newblock \emph{Cell}, 162\penalty0 (5):\penalty0 1155--1168, aug 2015.
\newblock ISSN 00928674.
\newblock \doi{10.1016/j.cell.2015.07.057}.
\newblock URL
  \url{http://www.sciencedirect.com/science/article/pii/S0092867415009733}.

\bibitem[Hanuschkin et~al.(2013)Hanuschkin, Ganguli, and
  Hahnloser]{Hanuschkin2013}
A~Hanuschkin, S~Ganguli, and R~H~R Hahnloser.
\newblock {A Hebbian learning rule gives rise to mirror neurons and links them
  to control theoretic inverse models.}
\newblock \emph{Frontiers in neural circuits}, 7:\penalty0 106, jan 2013.
\newblock ISSN 1662-5110.
\newblock \doi{10.3389/fncir.2013.00106}.

\bibitem[Harris and Wolpert(1998)]{harris1998signal}
Christopher~M Harris and Daniel~M Wolpert.
\newblock Signal-dependent noise determines motor planning.
\newblock \emph{Nature}, 394\penalty0 (6695):\penalty0 780--784, 1998.

\bibitem[Harris(2008)]{Harris2008}
KD~Harris.
\newblock {Stability of the fittest: organizing learning through retroaxonal
  signals}.
\newblock \emph{Trends in neurosciences}, 2008.
\newblock URL
  \url{http://www.sciencedirect.com/science/article/pii/S0166223608000180}.

\bibitem[Hassabis and Maguire(2009)]{Hassabis2009}
D~Hassabis and EA~Maguire.
\newblock {The construction system of the brain}.
\newblock \emph{{\ldots} of the Royal {\ldots}}, 2009.
\newblock URL
  \url{http://rstb.royalsocietypublishing.org/content/364/1521/1263.short}.

\bibitem[Hassabis and Maguire(2007)]{Hassabis2007}
Demis Hassabis and Eleanor~A Maguire.
\newblock {Deconstructing episodic memory with construction.}
\newblock \emph{Trends in cognitive sciences}, 11\penalty0 (7):\penalty0
  299--306, jul 2007.
\newblock ISSN 1364-6613.
\newblock \doi{10.1016/j.tics.2007.05.001}.
\newblock URL
  \url{http://www.sciencedirect.com/science/article/pii/S1364661307001258}.

\bibitem[Hasselmo(2006)]{Hasselmo2006}
Michael~E Hasselmo.
\newblock {The role of acetylcholine in learning and memory.}
\newblock \emph{Current opinion in neurobiology}, 16\penalty0 (6):\penalty0
  710--5, dec 2006.
\newblock ISSN 0959-4388.
\newblock \doi{10.1016/j.conb.2006.09.002}.

\bibitem[Hasselmo and Wyble(1997)]{Hasselmo1997}
Michael~E Hasselmo and Bradley~P Wyble.
\newblock {Free recall and recognition in a network model of the hippocampus:
  simulating effects of scopolamine on human memory function}.
\newblock \emph{Behavioural Brain Research}, 89\penalty0 (1-2):\penalty0 1--34,
  dec 1997.
\newblock ISSN 01664328.
\newblock \doi{10.1016/S0166-4328(97)00048-X}.
\newblock URL
  \url{http://www.sciencedirect.com/science/article/pii/S016643289700048X}.

\bibitem[Hattori et~al.(2007)Hattori, Demir, Kim, Viragh, Zipursky, and
  Dickson]{Hattori2007}
Daisuke Hattori, Ebru Demir, Ho~Won Kim, Erika Viragh, S~Lawrence Zipursky, and
  Barry~J Dickson.
\newblock {Dscam diversity is essential for neuronal wiring and
  self-recognition.}
\newblock \emph{Nature}, 449\penalty0 (7159):\penalty0 223--7, sep 2007.
\newblock ISSN 1476-4687.
\newblock \doi{10.1038/nature06099}.

\bibitem[Hawkins and Ahmad(2015)]{Hawkins2015}
Jeff Hawkins and Subutai Ahmad.
\newblock {Why Neurons Have Thousands of Synapses, A Theory of Sequence Memory
  in Neocortex}.
\newblock oct 2015.
\newblock URL \url{http://arxiv.org/abs/1511.00083}.

\bibitem[Hawkins and Blakeslee(2007)]{Hawkins2007}
Jeff Hawkins and Sandra Blakeslee.
\newblock \emph{{On Intelligence}}.
\newblock Henry Holt and Company, 2007.
\newblock ISBN 1429900458.
\newblock URL \url{http://books.google.com/books?id=Qg2dmntfxmQC{\&}pgis=1}.

\bibitem[Hayashi-Takagi et~al.(2015)Hayashi-Takagi, Yagishita, Nakamura,
  Shirai, Wu, Loshbaugh, Kuhlman, Hahn, and Kasai]{hayashi2015labelling}
Akiko Hayashi-Takagi, Sho Yagishita, Mayumi Nakamura, Fukutoshi Shirai, Yi~I
  Wu, Amanda~L Loshbaugh, Brian Kuhlman, Klaus~M Hahn, and Haruo Kasai.
\newblock Labelling and optical erasure of synaptic memory traces in the motor
  cortex.
\newblock \emph{Nature}, 2015.

\bibitem[Haykin(1994)]{Haykin1994}
Simon~S. Haykin.
\newblock \emph{{Neural networks: a comprehensive foundation}}.
\newblock Macmillan, 1994.
\newblock ISBN 0023527617.
\newblock URL
  \url{https://books.google.com/books/about/Neural{\_}networks.html?id=OZJKAQAAIAAJ{\&}pgis=1}.

\bibitem[Hayworth(2012)]{Hayworth2012}
Kenneth~J Hayworth.
\newblock {Dynamically partitionable autoassociative networks as a solution to
  the neural binding problem.}
\newblock \emph{Frontiers in computational neuroscience}, 6:\penalty0 73, jan
  2012.
\newblock ISSN 1662-5188.
\newblock \doi{10.3389/fncom.2012.00073}.

\bibitem[Hayworth et~al.(2011)Hayworth, Lescroart, and Biederman]{Hayworth2011}
Kenneth~J Hayworth, Mark~D Lescroart, and Irving Biederman.
\newblock {Neural encoding of relative position.}
\newblock \emph{Journal of experimental psychology. Human perception and
  performance}, 37\penalty0 (4):\penalty0 1032--50, aug 2011.
\newblock ISSN 1939-1277.
\newblock \doi{10.1037/a0022338}.
\newblock URL \url{http://www.ncbi.nlm.nih.gov/pubmed/21517211}.

\bibitem[Herd et~al.(2013)Herd, Krueger, and Kriete]{Herd2013}
SA~Herd, KA~Krueger, and TE~Kriete.
\newblock {Strategic cognitive sequencing: a computational cognitive
  neuroscience approach}.
\newblock \emph{Computational {\ldots}}, 2013.
\newblock URL \url{http://dl.acm.org/citation.cfm?id=2537972}.

\bibitem[Hinton(2007)]{Hinton2007}
G~Hinton.
\newblock {How to do backpropagation in a brain}.
\newblock \emph{Invited talk at the NIPS'2007 Deep Learning Workshop}, 2007.
\newblock URL
  \url{http://www.cs.utoronto.ca/{~}hinton/backpropincortex2014.pdf}.

\bibitem[Hinton(2016)]{Hinton2016talk}
G~Hinton.
\newblock Can the brain do back-propagation?
\newblock \emph{Invited talk at Stanford University Colloquium on Computer
  Systems}, 2016.
\newblock URL \url{https://www.youtube.com/watch?v=VIRCybGgHts}.

\bibitem[Hinton(1989)]{Hinton1989}
GE~Hinton.
\newblock {Connectionist learning procedures}.
\newblock \emph{Artificial intelligence}, 1989.
\newblock URL
  \url{http://www.sciencedirect.com/science/article/pii/0004370289900490}.

\bibitem[Hinton and McClelland(1988)]{Hinton1988}
GE~Hinton and JL~McClelland.
\newblock {Learning representations by recirculation}.
\newblock \emph{Neural information processing {\ldots}}, 1988.

\bibitem[Hinton et~al.(1995)Hinton, Dayan, Frey, and Neal]{Hinton1995}
GE~Hinton, P~Dayan, BJ~Frey, and RM~Neal.
\newblock {The" wake-sleep" algorithm for unsupervised neural networks}.
\newblock \emph{Science}, 1995.
\newblock URL \url{http://www.sciencemag.org/content/268/5214/1158.short}.

\bibitem[Hinton et~al.(2011)Hinton, Krizhevsky, and Wang]{Hinton2011}
GE~Hinton, A~Krizhevsky, and SD~Wang.
\newblock {Transforming auto-encoders}.
\newblock \emph{Artificial Neural Networks and {\ldots}}, 2011.
\newblock URL
  \url{http://link.springer.com/chapter/10.1007/978-3-642-21735-7{\_}6}.

\bibitem[Hinton et~al.(2006)Hinton, Osindero, and Teh]{Hinton2006}
Geoffrey~E Hinton, Simon Osindero, and Yee-Whye Teh.
\newblock {A fast learning algorithm for deep belief nets.}
\newblock \emph{Neural computation}, 18\penalty0 (7):\penalty0 1527--54, jul
  2006.
\newblock ISSN 0899-7667.
\newblock \doi{10.1162/neco.2006.18.7.1527}.
\newblock URL \url{http://www.ncbi.nlm.nih.gov/pubmed/16764513}.

\bibitem[Hires et~al.(2015)Hires, Gutnisky, Yu, O'Connor, and
  Svoboda]{Hires2015}
Samuel~Andrew Hires, Diego~A Gutnisky, Jianing Yu, Daniel~H O'Connor, and Karel
  Svoboda.
\newblock {Low-noise encoding of active touch by layer 4 in the somatosensory
  cortex.}
\newblock \emph{eLife}, 4, jan 2015.
\newblock ISSN 2050-084X.
\newblock \doi{10.7554/eLife.06619}.

\bibitem[Histed et~al.(2013)Histed, Ni, and Maunsell]{histed2013insights}
Mark~H Histed, Amy~M Ni, and John~HR Maunsell.
\newblock Insights into cortical mechanisms of behavior from microstimulation
  experiments.
\newblock \emph{Progress in neurobiology}, 103:\penalty0 115--130, 2013.

\bibitem[Hochreiter and Schmidhuber(1997)]{Hochreiter1997}
S~Hochreiter and J~Schmidhuber.
\newblock {Long short-term memory.}
\newblock \emph{Neural computation}, 9\penalty0 (8):\penalty0 1735--80, nov
  1997.
\newblock ISSN 0899-7667.
\newblock URL \url{http://www.ncbi.nlm.nih.gov/pubmed/9377276}.

\bibitem[Hoerzer et~al.(2014)Hoerzer, Legenstein, and Maass]{Hoerzer2014}
Gregor~M Hoerzer, Robert Legenstein, and Wolfgang Maass.
\newblock {Emergence of complex computational structures from chaotic neural
  networks through reward-modulated Hebbian learning.}
\newblock \emph{Cerebral cortex (New York, N.Y. : 1991)}, 24\penalty0
  (3):\penalty0 677--90, mar 2014.
\newblock ISSN 1460-2199.
\newblock \doi{10.1093/cercor/bhs348}.
\newblock URL \url{http://www.ncbi.nlm.nih.gov/pubmed/23146969}.

\bibitem[Hong and Luo(2014)]{Hong2014}
Weizhe Hong and Liqun Luo.
\newblock {Genetic control of wiring specificity in the fly olfactory system.}
\newblock \emph{Genetics}, 196\penalty0 (1):\penalty0 17--29, jan 2014.
\newblock ISSN 1943-2631.
\newblock \doi{10.1534/genetics.113.154336}.

\bibitem[Hopfield(1982)]{Hopfield1982}
J~J Hopfield.
\newblock {Neural networks and physical systems with emergent collective
  computational abilities.}
\newblock \emph{Proceedings of the National Academy of Sciences of the United
  States of America}, 79\penalty0 (8):\penalty0 2554--8, apr 1982.
\newblock ISSN 0027-8424.

\bibitem[Hopfield(1984)]{Hopfield1984}
J~J Hopfield.
\newblock {Neurons with graded response have collective computational
  properties like those of two-state neurons.}
\newblock \emph{Proceedings of the National Academy of Sciences of the United
  States of America}, 81\penalty0 (10):\penalty0 3088--92, may 1984.
\newblock ISSN 0027-8424.
\newblock URL \url{http://www.pnas.org/content/81/10/3088.abstract}.

\bibitem[Hopfield(2009)]{Hopfield2009}
J.~J. Hopfield.
\newblock {Neurodynamics of mental exploration}.
\newblock \emph{Proceedings of the National Academy of Sciences}, 107\penalty0
  (4):\penalty0 1648--1653, dec 2009.
\newblock ISSN 0027-8424.
\newblock \doi{10.1073/pnas.0913991107}.
\newblock URL \url{http://www.pnas.org/content/107/4/1648.abstract}.

\bibitem[Huang and Rao(2011)]{Huang2011}
Y~Huang and RPN Rao.
\newblock {Predictive coding}.
\newblock \emph{Wiley Interdisciplinary Reviews: Cognitive {\ldots}}, 2011.
\newblock URL \url{http://onlinelibrary.wiley.com/doi/10.1002/wcs.142/pdf}.

\bibitem[Huys et~al.(2015)Huys, Lally, Faulkner, Eshel, Seifritz, Gershman,
  Dayan, and Roiser]{huys2015interplay}
Quentin~JM Huys, N{\'\i}all Lally, Paul Faulkner, Neir Eshel, Erich Seifritz,
  Samuel~J Gershman, Peter Dayan, and Jonathan~P Roiser.
\newblock Interplay of approximate planning strategies.
\newblock \emph{Proceedings of the National Academy of Sciences}, 112\penalty0
  (10):\penalty0 3098--3103, 2015.

\bibitem[Isik et~al.(2012)Isik, Leibo, and Poggio]{Isik2012}
Leyla Isik, Joel~Z Leibo, and Tomaso Poggio.
\newblock {Learning and disrupting invariance in visual recognition with a
  temporal association rule.}
\newblock \emph{Frontiers in computational neuroscience}, 6:\penalty0 37, jan
  2012.
\newblock ISSN 1662-5188.
\newblock \doi{10.3389/fncom.2012.00037}.
\newblock URL
  \url{http://journal.frontiersin.org/article/10.3389/fncom.2012.00037/abstract}.

\bibitem[Izhikevich(2007)]{Izhikevich2007}
Eugene~M Izhikevich.
\newblock {Solving the distal reward problem through linkage of STDP and
  dopamine signaling.}
\newblock \emph{Cerebral cortex (New York, N.Y. : 1991)}, 17\penalty0
  (10):\penalty0 2443--52, oct 2007.
\newblock ISSN 1047-3211.
\newblock \doi{10.1093/cercor/bhl152}.
\newblock URL \url{http://www.ncbi.nlm.nih.gov/pubmed/17220510}.

\bibitem[Jacobson and Friedrich(2013)]{Jacobson2013}
Gilad~A Jacobson and Rainer~W Friedrich.
\newblock {Neural circuits: random design of a higher-order olfactory
  projection.}
\newblock \emph{Current biology : CB}, 23\penalty0 (10):\penalty0 R448--51, may
  2013.
\newblock ISSN 1879-0445.
\newblock \doi{10.1016/j.cub.2013.04.016}.
\newblock URL
  \url{http://www.sciencedirect.com/science/article/pii/S0960982213004247}.

\bibitem[Jaderberg et~al.(2015{\natexlab{a}})Jaderberg, Simonyan, and
  Zisserman]{Jaderberg2015}
M~Jaderberg, K~Simonyan, and A~Zisserman.
\newblock {Spatial transformer networks}.
\newblock \emph{Advances in Neural {\ldots}}, 2015{\natexlab{a}}.
\newblock URL
  \url{http://papers.nips.cc/paper/5854-spatial-transformer-networks}.

\bibitem[Jaderberg et~al.(2015{\natexlab{b}})Jaderberg, Simonyan, Zisserman,
  and Kavukcuoglu]{Jaderberg2015a}
Max Jaderberg, Karen Simonyan, Andrew Zisserman, and Koray Kavukcuoglu.
\newblock {Spatial Transformer Networks}.
\newblock jun 2015{\natexlab{b}}.
\newblock URL \url{http://arxiv.org/abs/1506.02025}.

\bibitem[Jaeger and Haas(2004)]{Jaeger2004}
Herbert Jaeger and Harald Haas.
\newblock {Harnessing nonlinearity: predicting chaotic systems and saving
  energy in wireless communication.}
\newblock \emph{Science (New York, N.Y.)}, 304\penalty0 (5667):\penalty0
  78--80, apr 2004.
\newblock ISSN 1095-9203.
\newblock \doi{10.1126/science.1091277}.
\newblock URL \url{http://www.sciencemag.org/content/304/5667/78.abstract}.

\bibitem[Jaramillo and Pearlmutter(2004)]{jaramillo2004normative}
Santiago Jaramillo and Barak~A Pearlmutter.
\newblock A normative model of attention: Receptive field modulation.
\newblock \emph{Neurocomputing}, 58:\penalty0 613--618, 2004.

\bibitem[Jhuang et~al.(2007)Jhuang, Serre, Wolf, and
  Poggio]{jhuang2007biologically}
Hueihan Jhuang, Thomas Serre, Lior Wolf, and Tomaso Poggio.
\newblock A biologically inspired system for action recognition.
\newblock In \emph{Computer Vision, 2007. ICCV 2007. IEEE 11th International
  Conference on}, pages 1--8. Ieee, 2007.

\bibitem[Ji and Wilson(2007)]{Ji2007}
Daoyun Ji and Matthew~A Wilson.
\newblock {Coordinated memory replay in the visual cortex and hippocampus
  during sleep.}
\newblock \emph{Nature neuroscience}, 10\penalty0 (1):\penalty0 100--7, jan
  2007.
\newblock ISSN 1097-6256.
\newblock \doi{10.1038/nn1825}.
\newblock URL \url{http://dx.doi.org/10.1038/nn1825}.

\bibitem[Jiang et~al.(2015)Jiang, Shen, Cadwell, Berens, Sinz, Ecker, Patel,
  and Tolias]{Jiang2015}
X.~Jiang, S.~Shen, C.~R. Cadwell, P.~Berens, F.~Sinz, A.~S. Ecker, S.~Patel,
  and A.~S. Tolias.
\newblock {Principles of connectivity among morphologically defined cell types
  in adult neocortex}.
\newblock \emph{Science}, 350\penalty0 (6264):\penalty0 aac9462--aac9462, nov
  2015.
\newblock ISSN 0036-8075.
\newblock \doi{10.1126/science.aac9462}.
\newblock URL
  \url{http://science.sciencemag.org/content/350/6264/aac9462.abstract}.

\bibitem[Johansson et~al.(2014)Johansson, Jirenhed, Rasmussen, Zucca, and
  Hesslow]{Johansson2014}
Fredrik Johansson, Dan-Anders Jirenhed, Anders Rasmussen, Riccardo Zucca, and
  Germund Hesslow.
\newblock {Memory trace and timing mechanism localized to cerebellar Purkinje
  cells.}
\newblock \emph{Proceedings of the National Academy of Sciences of the United
  States of America}, 111\penalty0 (41):\penalty0 14930--4, oct 2014.
\newblock ISSN 1091-6490.
\newblock \doi{10.1073/pnas.1415371111}.
\newblock URL \url{http://www.pnas.org/content/111/41/14930.abstract}.

\bibitem[Jonas and Kording(2016)]{Jonas055624}
Eric Jonas and Konrad Kording.
\newblock Could a neuroscientist understand a microprocessor?
\newblock \emph{bioRxiv}, 2016.
\newblock \doi{10.1101/055624}.
\newblock URL \url{http://biorxiv.org/content/early/2016/05/26/055624}.

\bibitem[Joulin and Mikolov(2015)]{joulin2015inferring}
Armand Joulin and Tomas Mikolov.
\newblock Inferring algorithmic patterns with stack-augmented recurrent nets.
\newblock In \emph{Advances in Neural Information Processing Systems}, pages
  190--198, 2015.

\bibitem[Kalisman et~al.(2005)Kalisman, Silberberg, and Markram]{Kalisman2005}
Nir Kalisman, Gilad Silberberg, and Henry Markram.
\newblock {The neocortical microcircuit as a tabula rasa.}
\newblock \emph{Proceedings of the National Academy of Sciences of the United
  States of America}, 102\penalty0 (3):\penalty0 880--5, jan 2005.
\newblock ISSN 0027-8424.
\newblock \doi{10.1073/pnas.0407088102}.

\bibitem[Kanwisher et~al.(1997)Kanwisher, McDermott, and
  Chun]{kanwisher1997fusiform}
Nancy Kanwisher, Josh McDermott, and Marvin~M Chun.
\newblock The fusiform face area: a module in human extrastriate cortex
  specialized for face perception.
\newblock \emph{The Journal of Neuroscience}, 17\penalty0 (11):\penalty0
  4302--4311, 1997.

\bibitem[Kappel et~al.(2014)Kappel, Nessler, and Maass]{Kappel2014}
David Kappel, Bernhard Nessler, and Wolfgang Maass.
\newblock {STDP installs in Winner-Take-All circuits an online approximation to
  hidden Markov model learning.}
\newblock \emph{PLoS computational biology}, 10\penalty0 (3):\penalty0
  e1003511, mar 2014.
\newblock ISSN 1553-7358.
\newblock \doi{10.1371/journal.pcbi.1003511}.

\bibitem[Kempter et~al.(2001)Kempter, Gerstner, and van Hemmen]{Kempter2001}
R~Kempter, W~Gerstner, and J~L van Hemmen.
\newblock {Intrinsic stabilization of output rates by spike-based Hebbian
  learning.}
\newblock \emph{Neural computation}, 13\penalty0 (12):\penalty0 2709--41, dec
  2001.
\newblock ISSN 0899-7667.
\newblock \doi{10.1162/089976601317098501}.
\newblock URL \url{http://www.ncbi.nlm.nih.gov/pubmed/11705408}.

\bibitem[Kingma and Welling(2013)]{Kingma2013}
Diederik~P Kingma and Max Welling.
\newblock {Auto-Encoding Variational Bayes}.
\newblock dec 2013.
\newblock URL \url{http://arxiv.org/abs/1312.6114}.

\bibitem[Knill and Pouget(2004)]{Knill2004}
DC~Knill and A~Pouget.
\newblock {The Bayesian brain: the role of uncertainty in neural coding and
  computation}.
\newblock \emph{TRENDS in Neurosciences}, 2004.
\newblock URL
  \url{http://www.sciencedirect.com/science/article/pii/S0166223604003352}.

\bibitem[Koechlin and Jubault(2006)]{koechlin2006broca}
Etienne Koechlin and Thomas Jubault.
\newblock Broca's area and the hierarchical organization of human behavior.
\newblock \emph{Neuron}, 50\penalty0 (6):\penalty0 963--974, 2006.

\bibitem[Komer and Eliasmith(2016)]{Komer2016}
Brent Komer and Chris Eliasmith.
\newblock {A unified theoretical approach for biological cognition and
  learning}.
\newblock \emph{Current Opinion in Behavioral Sciences}, 11:\penalty0 14--20,
  mar 2016.
\newblock ISSN 23521546.
\newblock \doi{10.1016/j.cobeha.2016.03.006}.
\newblock URL
  \url{http://www.sciencedirect.com/science/article/pii/S2352154616300651}.

\bibitem[K{\"{o}}rding(2007)]{Kording2007}
K~K{\"{o}}rding.
\newblock {Decision theory: what" should" the nervous system do?}
\newblock \emph{Science}, 2007.
\newblock URL \url{http://science.sciencemag.org/content/318/5850/606.short}.

\bibitem[K{\"{o}}rding et~al.(2004)K{\"{o}}rding, Kayser, Einh{\"{a}}user, and
  K{\"{o}}nig]{Kording2004}
Konrad~P K{\"{o}}rding, Christoph Kayser, Wolfgang Einh{\"{a}}user, and Peter
  K{\"{o}}nig.
\newblock {How are complex cell properties adapted to the statistics of natural
  stimuli?}
\newblock \emph{Journal of neurophysiology}, 91\penalty0 (1):\penalty0 206--12,
  jan 2004.
\newblock ISSN 0022-3077.
\newblock \doi{10.1152/jn.00149.2003}.
\newblock URL \url{http://jn.physiology.org/content/91/1/206.short}.

\bibitem[K{\"{o}}rding and K{\"{o}}nig(2000)]{Kording2000}
K.P K{\"{o}}rding and P~K{\"{o}}nig.
\newblock {A learning rule for dynamic recruitment and decorrelation}.
\newblock \emph{Neural Networks}, 13\penalty0 (1):\penalty0 1--9, jan 2000.
\newblock ISSN 08936080.
\newblock \doi{10.1016/S0893-6080(99)00088-X}.
\newblock URL
  \url{http://www.sciencedirect.com/science/article/pii/S089360809900088X}.

\bibitem[K{\"{o}}rding and K{\"{o}}nig(2001)]{Kording2001}
KP~K{\"{o}}rding and P~K{\"{o}}nig.
\newblock {Supervised and unsupervised learning with two sites of synaptic
  integration}.
\newblock \emph{Journal of Computational Neuroscience}, 2001.
\newblock URL \url{http://link.springer.com/article/10.1023/A:1013776130161}.

\bibitem[Kraus et~al.(2013)Kraus, Robinson, White, Eichenbaum, and
  Hasselmo]{kraus2013hippocampal}
Benjamin~J Kraus, Robert~J Robinson, John~A White, Howard Eichenbaum, and
  Michael~E Hasselmo.
\newblock Hippocampal “time cells”: time versus path integration.
\newblock \emph{Neuron}, 78\penalty0 (6):\penalty0 1090--1101, 2013.

\bibitem[Kriete et~al.(2013)Kriete, Noelle, Cohen, and O'Reilly]{Kriete2013}
Trenton Kriete, David~C Noelle, Jonathan~D Cohen, and Randall~C O'Reilly.
\newblock {Indirection and symbol-like processing in the prefrontal cortex and
  basal ganglia.}
\newblock \emph{Proceedings of the National Academy of Sciences of the United
  States of America}, 110\penalty0 (41):\penalty0 16390--5, oct 2013.
\newblock ISSN 1091-6490.
\newblock \doi{10.1073/pnas.1303547110}.
\newblock URL \url{http://www.pnas.org/content/110/41/16390.short}.

\bibitem[Krishnamurthy et~al.(2016)Krishnamurthy, Lakshminarayanan, Kumar, and
  Ravindran]{Krishnamurthy2016}
Ramnandan Krishnamurthy, Aravind~S. Lakshminarayanan, Peeyush Kumar, and
  Balaraman Ravindran.
\newblock {Hierarchical Reinforcement Learning using Spatio-Temporal
  Abstractions and Deep Neural Networks}.
\newblock page~13, may 2016.
\newblock URL \url{http://arxiv.org/abs/1605.05359}.

\bibitem[Krizhevsky et~al.(2012)Krizhevsky, Sutskever, and
  Hinton]{krizhevsky2012imagenet}
Alex Krizhevsky, Ilya Sutskever, and Geoffrey~E Hinton.
\newblock Imagenet classification with deep convolutional neural networks.
\newblock In \emph{Advances in neural information processing systems}, pages
  1097--1105, 2012.

\bibitem[Kulkarni et~al.(2015)Kulkarni, Whitney, Kohli, and
  Tenenbaum]{Kulkarni2015}
Tejas~D. Kulkarni, Will Whitney, Pushmeet Kohli, and Joshua~B. Tenenbaum.
\newblock {Deep Convolutional Inverse Graphics Network}.
\newblock mar 2015.
\newblock URL \url{http://arxiv.org/abs/1503.03167}.

\bibitem[Kulkarni et~al.(2016)Kulkarni, Narasimhan, Saeedi, and
  Tenenbaum]{Kulkarni2016}
Tejas~D. Kulkarni, Karthik~R. Narasimhan, Ardavan Saeedi, and Joshua~B.
  Tenenbaum.
\newblock {Hierarchical Deep Reinforcement Learning: Integrating Temporal
  Abstraction and Intrinsic Motivation}.
\newblock page~13, apr 2016.
\newblock URL \url{http://arxiv.org/abs/1604.06057}.

\bibitem[Kumaran et~al.(2009)Kumaran, Summerfield, Hassabis, and
  Maguire]{Kumaran2009}
Dharshan Kumaran, Jennifer~J Summerfield, Demis Hassabis, and Eleanor~A
  Maguire.
\newblock {Tracking the emergence of conceptual knowledge during human decision
  making.}
\newblock \emph{Neuron}, 63\penalty0 (6):\penalty0 889--901, sep 2009.
\newblock ISSN 1097-4199.
\newblock \doi{10.1016/j.neuron.2009.07.030}.
\newblock URL \url{http://www.cell.com/article/S0896627309006187/fulltext}.

\bibitem[Kurach et~al.(2015)Kurach, Andrychowicz, and Sutskever]{Kurach2015}
Karol Kurach, Marcin Andrychowicz, and Ilya Sutskever.
\newblock {Neural Random-Access Machines}.
\newblock page~13, nov 2015.
\newblock URL \url{http://arxiv.org/abs/1511.06392}.

\bibitem[Lake et~al.(2015)Lake, Salakhutdinov, and Tenenbaum]{Lake2015}
Brenden~M. Lake, Ruslan Salakhutdinov, and Joshua~B. Tenenbaum.
\newblock {Human-level concept learning through probabilistic program
  induction}.
\newblock \emph{Science}, 350\penalty0 (6266):\penalty0 1332--1338, dec 2015.
\newblock \doi{10.1126/science.aab3050}.
\newblock URL \url{http://www.sciencemag.org/content/350/6266/1332.full}.

\bibitem[Lake et~al.(2016)Lake, Ullman, Tenenbaum, and Gershman]{Lake2016}
Brenden~M Lake, Tomer~D Ullman, Joshua~B Tenenbaum, and Samuel~J Gershman.
\newblock Building machines that learn and think like people.
\newblock \emph{arXiv preprint arXiv:1604.00289}, 2016.

\bibitem[Larkum(2013)]{Larkum2013}
Matthew Larkum.
\newblock {A cellular mechanism for cortical associations: an organizing
  principle for the cerebral cortex.}
\newblock \emph{Trends in neurosciences}, 36\penalty0 (3):\penalty0 141--51,
  mar 2013.
\newblock ISSN 1878-108X.
\newblock \doi{10.1016/j.tins.2012.11.006}.
\newblock URL \url{http://www.ncbi.nlm.nih.gov/pubmed/23273272}.

\bibitem[Le et~al.(2011)Le, Ranzato, Monga, Devin, Chen, Corrado, Dean, and
  Ng]{Le2011}
Quoc~V. Le, Marc'Aurelio Ranzato, Rajat Monga, Matthieu Devin, Kai Chen,
  Greg~S. Corrado, Jeff Dean, and Andrew~Y. Ng.
\newblock {Building high-level features using large scale unsupervised
  learning}.
\newblock dec 2011.
\newblock URL \url{http://arxiv.org/abs/1112.6209}.

\bibitem[LeCun and Bengio(1995)]{lecun1995convolutional}
Yann LeCun and Yoshua Bengio.
\newblock Convolutional networks for images, speech, and time series.
\newblock \emph{The handbook of brain theory and neural networks},
  3361\penalty0 (10):\penalty0 1995, 1995.

\bibitem[LeCun et~al.(2015)LeCun, Bengio, and Hinton]{LeCun2015}
Yann LeCun, Yoshua Bengio, and Geoffrey Hinton.
\newblock {Deep learning}.
\newblock \emph{Nature}, 521\penalty0 (7553):\penalty0 436--444, may 2015.
\newblock ISSN 0028-0836.
\newblock \doi{10.1038/nature14539}.
\newblock URL \url{http://dx.doi.org/10.1038/nature14539}.

\bibitem[Lee et~al.(2015)Lee, Tai, Zador, and Wilbrecht]{Lee2015}
A~M Lee, L-H Tai, A~Zador, and L~Wilbrecht.
\newblock {Between the primate and 'reptilian' brain: Rodent models demonstrate
  the role of corticostriatal circuits in decision making.}
\newblock \emph{Neuroscience}, 296:\penalty0 66--74, jun 2015.
\newblock ISSN 1873-7544.
\newblock \doi{10.1016/j.neuroscience.2014.12.042}.
\newblock URL
  \url{http://www.sciencedirect.com/science/article/pii/S0306452214010914}.

\bibitem[Lee and Mumford(2003)]{Lee2003}
Tai~Sing Lee and David Mumford.
\newblock {Hierarchical Bayesian inference in the visual cortex.}
\newblock \emph{Journal of the Optical Society of America. A, Optics, image
  science, and vision}, 20\penalty0 (7):\penalty0 1434--48, jul 2003.
\newblock ISSN 1084-7529.
\newblock URL \url{http://www.ncbi.nlm.nih.gov/pubmed/12868647}.

\bibitem[Lee and Yuille(2011)]{Lee2011}
TS~Lee and AL~Yuille.
\newblock {Efficient coding of visual scenes by grouping and segmentation:
  theoretical predictions and biological evidence}.
\newblock \emph{Department of Statistics, UCLA}, 2011.
\newblock URL \url{http://escholarship.org/uc/item/1mc5v1b6.pdf}.

\bibitem[Legenstein and Maass(2011)]{Legenstein2011a}
Robert Legenstein and Wolfgang Maass.
\newblock {Branch-specific plasticity enables self-organization of nonlinear
  computation in single neurons.}
\newblock \emph{The Journal of neuroscience : the official journal of the
  Society for Neuroscience}, 31\penalty0 (30):\penalty0 10787--802, jul 2011.
\newblock ISSN 1529-2401.
\newblock \doi{10.1523/JNEUROSCI.5684-10.2011}.
\newblock URL
  \url{http://www.jneurosci.org.libproxy.mit.edu/content/31/30/10787.short}.

\bibitem[Leibo et~al.(2015{\natexlab{a}})Leibo, Cornebise, G{\'{o}}mez, and
  Hassabis]{Leibo2015}
Joel~Z. Leibo, Julien Cornebise, Sergio G{\'{o}}mez, and Demis Hassabis.
\newblock {Approximate Hubel-Wiesel Modules and the Data Structures of Neural
  Computation}.
\newblock page~13, dec 2015{\natexlab{a}}.
\newblock URL \url{http://arxiv.org/abs/1512.08457}.

\bibitem[Leibo et~al.(2015{\natexlab{b}})Leibo, Liao, Anselmi, and
  Poggio]{leibo2015invariance}
Joel~Z Leibo, Qianli Liao, Fabio Anselmi, and Tomaso Poggio.
\newblock The invariance hypothesis implies domain-specific regions in visual
  cortex.
\newblock \emph{PLoS Comput Biol}, 11\penalty0 (10):\penalty0 e1004390,
  2015{\natexlab{b}}.

\bibitem[Lettvin et~al.(1959)Lettvin, Maturana, McCulloch, and
  Pitts]{Lettvin1959}
J.~Lettvin, H.~Maturana, W.~McCulloch, and W.~Pitts.
\newblock {What the Frog's Eye Tells the Frog's Brain}.
\newblock \emph{Proceedings of the IRE}, 47\penalty0 (11):\penalty0 1940--1951,
  nov 1959.
\newblock ISSN 0096-8390.
\newblock \doi{10.1109/JRPROC.1959.287207}.
\newblock URL
  \url{http://ieeexplore.ieee.org/articleDetails.jsp?arnumber=4065609}.

\bibitem[Levine et~al.(2015)Levine, Finn, Darrell, and Abbeel]{levine2015end}
Sergey Levine, Chelsea Finn, Trevor Darrell, and Pieter Abbeel.
\newblock End-to-end training of deep visuomotor policies.
\newblock \emph{arXiv preprint arXiv:1504.00702}, 2015.

\bibitem[Lewicki and Sejnowski(2000)]{Lewicki2000}
Michael~S. Lewicki and Terrence~J. Sejnowski.
\newblock {Learning Overcomplete Representations}.
\newblock \emph{Neural Computation}, 12\penalty0 (2):\penalty0 337--365, feb
  2000.
\newblock ISSN 0899-7667.
\newblock \doi{10.1162/089976600300015826}.

\bibitem[Lewis and Harris(2014)]{Lewis2014}
S.~N. Lewis and K.~D. Harris.
\newblock {The Neural Marketplace: I. General Formalismand Linear Theory}.
\newblock Technical report, dec 2014.
\newblock URL
  \url{http://biorxiv.org/content/early/2014/12/23/013185.abstract}.

\bibitem[Li and Dicarlo(2012)]{Li2012}
Nuo Li and James~J Dicarlo.
\newblock {Neuronal learning of invariant object representation in the ventral
  visual stream is not dependent on reward.}
\newblock \emph{The Journal of neuroscience : the official journal of the
  Society for Neuroscience}, 32\penalty0 (19):\penalty0 6611--20, may 2012.
\newblock ISSN 1529-2401.
\newblock \doi{10.1523/JNEUROSCI.3786-11.2012}.
\newblock URL \url{http://www.jneurosci.org/content/32/19/6611.full}.

\bibitem[Liao and Poggio(2016)]{Liao2016}
Qianli Liao and Tomaso Poggio.
\newblock {Bridging the Gaps Between Residual Learning, Recurrent Neural
  Networks and Visual Cortex}.
\newblock apr 2016.
\newblock URL \url{http://arxiv.org/abs/1604.03640}.

\bibitem[Liao et~al.(2015)Liao, Leibo, and Poggio]{Liao2015}
Qianli Liao, Joel~Z. Leibo, and Tomaso Poggio.
\newblock {How Important is Weight Symmetry in Backpropagation?}
\newblock oct 2015.
\newblock URL \url{http://arxiv.org/abs/1510.05067}.

\bibitem[Lillicrap et~al.(2014)Lillicrap, Cownden, Tweed, and
  Akerman]{Lillicrap2014}
Timothy~P. Lillicrap, Daniel Cownden, Douglas~B. Tweed, and Colin~J. Akerman.
\newblock {Random feedback weights support learning in deep neural networks}.
\newblock page~14, nov 2014.
\newblock URL \url{http://arxiv.org/abs/1411.0247}.

\bibitem[Liu and Buonomano(2009)]{Liu2009}
Jian~K Liu and Dean~V Buonomano.
\newblock {Embedding multiple trajectories in simulated recurrent neural
  networks in a self-organizing manner.}
\newblock \emph{The Journal of neuroscience : the official journal of the
  Society for Neuroscience}, 29\penalty0 (42):\penalty0 13172--81, oct 2009.
\newblock ISSN 1529-2401.
\newblock \doi{10.1523/JNEUROSCI.2358-09.2009}.
\newblock URL \url{http://www.jneurosci.org/content/29/42/13172.short}.

\bibitem[Livni et~al.(2013)Livni, Shalev-Shwartz, and Shamir]{Livni2013}
Roi Livni, Shai Shalev-Shwartz, and Ohad Shamir.
\newblock {An Algorithm for Training Polynomial Networks}.
\newblock apr 2013.
\newblock URL \url{http://arxiv.org/abs/1304.7045}.

\bibitem[Lotter et~al.(2015)Lotter, Kreiman, and Cox]{Lotter2015}
William Lotter, Gabriel Kreiman, and David Cox.
\newblock {Unsupervised Learning of Visual Structure using Predictive
  Generative Networks}.
\newblock nov 2015.
\newblock URL \url{http://arxiv.org/abs/1511.06380}.

\bibitem[Lotter et~al.(2016)Lotter, Kreiman, and Cox]{lotter2016deep}
William Lotter, Gabriel Kreiman, and David Cox.
\newblock Deep predictive coding networks for video prediction and unsupervised
  learning.
\newblock \emph{arXiv preprint arXiv:1605.08104}, 2016.

\bibitem[Ma et~al.(2006)Ma, Beck, Latham, and Pouget]{Ma2006}
Wei~Ji Ma, Jeffrey~M Beck, Peter~E Latham, and Alexandre Pouget.
\newblock {Bayesian inference with probabilistic population codes.}
\newblock \emph{Nature neuroscience}, 9\penalty0 (11):\penalty0 1432--8, nov
  2006.
\newblock ISSN 1097-6256.
\newblock \doi{10.1038/nn1790}.
\newblock URL \url{http://dx.doi.org/10.1038/nn1790}.

\bibitem[Maass et~al.(2002)Maass, Natschl{\"{a}}ger, and Markram]{Maass2002}
Wolfgang Maass, Thomas Natschl{\"{a}}ger, and Henry Markram.
\newblock {Real-time computing without stable states: a new framework for
  neural computation based on perturbations.}
\newblock \emph{Neural computation}, 14\penalty0 (11):\penalty0 2531--60, nov
  2002.
\newblock ISSN 0899-7667.
\newblock \doi{10.1162/089976602760407955}.

\bibitem[Maass et~al.(2007)Maass, Joshi, and Sontag]{Maass2007}
Wolfgang Maass, Prashant Joshi, and Eduardo~D Sontag.
\newblock {Computational aspects of feedback in neural circuits.}
\newblock \emph{PLoS computational biology}, 3\penalty0 (1):\penalty0 e165, jan
  2007.
\newblock ISSN 1553-7358.
\newblock \doi{10.1371/journal.pcbi.0020165}.
\newblock URL
  \url{http://journals.plos.org/ploscompbiol/article?id=10.1371/journal.pcbi.0020165}.

\bibitem[MacDonald et~al.(2011)MacDonald, Lepage, Eden, and
  Eichenbaum]{macdonald2011hippocampal}
Christopher~J MacDonald, Kyle~Q Lepage, Uri~T Eden, and Howard Eichenbaum.
\newblock Hippocampal “time cells” bridge the gap in memory for
  discontiguous events.
\newblock \emph{Neuron}, 71\penalty0 (4):\penalty0 737--749, 2011.

\bibitem[Maclaurin et~al.(2015)Maclaurin, Duvenaud, and Adams]{Maclaurin2015}
D~Maclaurin, D~Duvenaud, and RP~Adams.
\newblock {Gradient-based hyperparameter optimization through reversible
  learning}.
\newblock \emph{arXiv preprint arXiv:1502.03492}, 2015.
\newblock URL \url{http://arxiv.org/abs/1502.03492}.

\bibitem[Mandelblat-Cerf et~al.(2014)Mandelblat-Cerf, Las, Denisenko, and
  Fee]{Mandelblat-Cerf2014}
Yael Mandelblat-Cerf, Liora Las, Natalia Denisenko, and Michale~S Fee.
\newblock {A role for descending auditory cortical projections in songbird
  vocal learning.}
\newblock \emph{eLife}, 3:\penalty0 e02152, jan 2014.
\newblock ISSN 2050-084X.
\newblock \doi{10.7554/eLife.02152}.
\newblock URL \url{http://elifesciences.org/content/3/e02152.abstract}.

\bibitem[Marblestone and Boyden(2014)]{Marblestone2014a}
Adam~H Marblestone and Edward~S Boyden.
\newblock {Designing tools for assumption-proof brain mapping.}
\newblock \emph{Neuron}, 83\penalty0 (6):\penalty0 1239--41, sep 2014.
\newblock ISSN 1097-4199.
\newblock \doi{10.1016/j.neuron.2014.09.004}.
\newblock URL \url{http://www.cell.com/article/S0896627314007922/fulltext}.

\bibitem[Marcus(2001)]{Marcus2001}
Gary Marcus.
\newblock \emph{{The Algebraic Mind: Integrating Connectionism and Cognitive
  Science}}.
\newblock MIT Press, 2001.
\newblock ISBN 0262632683.
\newblock URL \url{http://books.google.com/books?id=7YpuRUlFLm8C{\&}pgis=1}.

\bibitem[Marcus(2004)]{marcus2004birth}
Gary Marcus.
\newblock \emph{The birth of the mind: How a tiny number of genes creates the
  complexities of human thought}.
\newblock Basic Books, 2004.

\bibitem[Marcus et~al.(2014{\natexlab{a}})Marcus, Marblestone, and
  Dean]{Marcus2014}
Gary Marcus, Adam Marblestone, and Thomas Dean.
\newblock {Frequently asked question for: the atoms of neural computation.}
\newblock oct 2014{\natexlab{a}}.
\newblock URL \url{http://arxiv.org/abs/1410.8826}.

\bibitem[Marcus et~al.(2014{\natexlab{b}})Marcus, Marblestone, and
  Dean]{Marcus2014a}
Gary Marcus, Adam Marblestone, and Thomas Dean.
\newblock {The atoms of neural computation}.
\newblock \emph{Science}, 346\penalty0 (6209):\penalty0 551--552,
  2014{\natexlab{b}}.
\newblock \doi{10.1126/science.1261661}.

\bibitem[Marder and Goaillard(2006)]{Marder2006}
Eve Marder and Jean-Marc Goaillard.
\newblock {Variability, compensation and homeostasis in neuron and network
  function.}
\newblock \emph{Nature reviews. Neuroscience}, 7\penalty0 (7):\penalty0
  563--74, jul 2006.
\newblock ISSN 1471-003X.
\newblock \doi{10.1038/nrn1949}.
\newblock URL \url{http://dx.doi.org/10.1038/nrn1949}.

\bibitem[Markowitz et~al.(2015)Markowitz, Curtis, and Pesaran]{Markowitz2015}
David~A Markowitz, Clayton~E Curtis, and Bijan Pesaran.
\newblock {Multiple component networks support working memory in prefrontal
  cortex.}
\newblock \emph{Proceedings of the National Academy of Sciences of the United
  States of America}, 112\penalty0 (35):\penalty0 11084--11089, aug 2015.
\newblock ISSN 1091-6490.
\newblock \doi{10.1073/pnas.1504172112}.
\newblock URL \url{http://www.pnas.org/content/112/35/11084.abstract}.

\bibitem[Markram et~al.(1997)Markram, L{\"{u}}bke, Frotscher, and
  Sakmann]{Markram1997}
H~Markram, J~L{\"{u}}bke, M~Frotscher, and B~Sakmann.
\newblock {Regulation of synaptic efficacy by coincidence of postsynaptic APs
  and EPSPs.}
\newblock \emph{Science (New York, N.Y.)}, 275\penalty0 (5297):\penalty0
  213--5, jan 1997.
\newblock ISSN 0036-8075.
\newblock URL \url{http://www.ncbi.nlm.nih.gov/pubmed/8985014}.

\bibitem[Markram et~al.(2015)Markram, Muller, Ramaswamy, Reimann, Abdellah,
  Sanchez, Ailamaki, Alonso-Nanclares, Antille, Arsever, Kahou, Berger,
  Bilgili, Buncic, Chalimourda, Chindemi, Courcol, Delalondre, Delattre,
  Druckmann, Dumusc, Dynes, Eilemann, Gal, Gevaert, Ghobril, Gidon, Graham,
  Gupta, Haenel, Hay, Heinis, Hernando, Hines, Kanari, Keller, Kenyon, Khazen,
  Kim, King, Kisvarday, Kumbhar, Lasserre, {Le B{\'{e}}}, Magalh{\~{a}}es,
  Merch{\'{a}}n-P{\'{e}}rez, Meystre, Morrice, Muller,
  Mu{\~{n}}oz-C{\'{e}}spedes, Muralidhar, Muthurasa, Nachbaur, Newton, Nolte,
  Ovcharenko, Palacios, Pastor, Perin, Ranjan, Riachi, Rodr{\'{\i}}guez,
  Riquelme, R{\"{o}}ssert, Sfyrakis, Shi, Shillcock, Silberberg, Silva,
  Tauheed, Telefont, Toledo-Rodriguez, Tr{\"{a}}nkler, {Van Geit},
  D{\'{\i}}az, Walker, Wang, Zaninetta, DeFelipe, Hill, Segev, and
  Sch{\"{u}}rmann]{Markram2015}
Henry Markram, Eilif Muller, Srikanth Ramaswamy, Michael W. Reimann, Marwan
  Abdellah, Carlos Aguado Sanchez, Anastasia Ailamaki, Lidia Alonso-Nanclares,
  Nicolas Antille, Selim Arsever, Guy Antoine Atenekeng Kahou, Thomas K.
  Berger, Ahmet Bilgili, Nenad Buncic, Athanassia Chalimourda, Giuseppe
  Chindemi, Jean-Denis Courcol, Fabien Delalondre, Vincent Delattre, Shaul
  Druckmann, Raphael Dumusc, James Dynes, Stefan Eilemann, Eyal Gal,
  Michael Emiel Gevaert, Jean-Pierre Ghobril, Albert Gidon, Joe W. Graham,
  Anirudh Gupta, Valentin Haenel, Etay Hay, Thomas Heinis, Juan B. Hernando,
  Michael Hines, Lida Kanari, Daniel Keller, John Kenyon, Georges Khazen, Yihwa
  Kim, James G. King, Zoltan Kisvarday, Pramod Kumbhar, S{\'{e}}bastien
  Lasserre, Jean-Vincent {Le B{\'{e}}}, Bruno R.C. Magalh{\~{a}}es, Angel
  Merch{\'{a}}n-P{\'{e}}rez, Julie Meystre, Benjamin Roy Morrice, Jeffrey
  Muller, Alberto Mu{\~{n}}oz-C{\'{e}}spedes, Shruti Muralidhar, Keerthan
  Muthurasa, Daniel Nachbaur, Taylor H. Newton, Max Nolte, Aleksandr
  Ovcharenko, Juan Palacios, Luis Pastor, Rodrigo Perin, Rajnish Ranjan, Imad
  Riachi, Jos{\'{e}}-Rodrigo Rodr{\'{\i}}guez, Juan Luis Riquelme, Christian
  R{\"{o}}ssert, Konstantinos Sfyrakis, Ying Shi, Julian C. Shillcock, Gilad
  Silberberg, Ricardo Silva, Farhan Tauheed, Martin Telefont, Maria
  Toledo-Rodriguez, Thomas Tr{\"{a}}nkler, Werner {Van Geit},
  Jafet Villafranca D{\'{\i}}az, Richard Walker, Yun Wang, Stefano M.
  Zaninetta, Javier DeFelipe, Sean L. Hill, Idan Segev, and Felix
  Sch{\"{u}}rmann.
\newblock {Reconstruction and Simulation of Neocortical Microcircuitry}.
\newblock \emph{Cell}, 163\penalty0 (2):\penalty0 456--492, oct 2015.
\newblock ISSN 00928674.
\newblock \doi{10.1016/j.cell.2015.09.029}.
\newblock URL \url{http://www.cell.com/article/S0092867415011915/fulltext}.

\bibitem[Marr(1969)]{Marr1969}
D~Marr.
\newblock {A theory of cerebellar cortex.}
\newblock \emph{The Journal of physiology}, 202\penalty0 (2):\penalty0 437--70,
  jun 1969.
\newblock ISSN 0022-3751.

\bibitem[Martens and Sutskever(2011)]{Martens2011}
J~Martens and I~Sutskever.
\newblock {Learning recurrent neural networks with hessian-free optimization}.
\newblock \emph{Proceedings of the 28th {\ldots}}, 2011.
\newblock URL
  \url{http://machinelearning.wustl.edu/mlpapers/paper{\_}files/ICML2011Martens{\_}532.pdf}.

\bibitem[McCandliss et~al.(2003)McCandliss, Cohen, and
  Dehaene]{mccandliss2003visual}
Bruce~D McCandliss, Laurent Cohen, and Stanislas Dehaene.
\newblock The visual word form area: expertise for reading in the fusiform
  gyrus.
\newblock \emph{Trends in cognitive sciences}, 7\penalty0 (7):\penalty0
  293--299, 2003.

\bibitem[McClelland et~al.(1986)McClelland, Rumelhart,
  et~al.]{mcclelland1986parallel}
James~L McClelland, David~E Rumelhart, et~al.
\newblock Parallel distributed processing, vol. 1.
\newblock \emph{Cambridge, Ma., MIT Press. Zipser D.(1986),Feature Discovery by
  Competitive Learning, in DE Rumel hart-JL McClelland (eds), voi}, 1:\penalty0
  151--163, 1986.

\bibitem[McCulloch and Pitts(1943)]{McCulloch1943}
Warren~S. McCulloch and Walter Pitts.
\newblock {A logical calculus of the ideas immanent in nervous activity}.
\newblock \emph{The Bulletin of Mathematical Biophysics}, 5\penalty0
  (4):\penalty0 115--133, dec 1943.
\newblock ISSN 0007-4985.
\newblock \doi{10.1007/BF02478259}.
\newblock URL \url{http://link.springer.com/10.1007/BF02478259}.

\bibitem[McKinstry et~al.(2006)McKinstry, Edelman, and Krichmar]{McKinstry2006}
Jeffrey~L McKinstry, Gerald~M Edelman, and Jeffrey~L Krichmar.
\newblock {A cerebellar model for predictive motor control tested in a
  brain-based device.}
\newblock \emph{Proceedings of the National Academy of Sciences of the United
  States of America}, 103\penalty0 (9):\penalty0 3387--92, feb 2006.
\newblock ISSN 0027-8424.
\newblock \doi{10.1073/pnas.0511281103}.
\newblock URL \url{http://www.pnas.org/content/103/9/3387.full{\#}ref-4}.

\bibitem[McKone et~al.(2009)McKone, Crookes, Kanwisher,
  et~al.]{mckone2009cognitive}
Elinor McKone, Kate Crookes, Nancy Kanwisher, et~al.
\newblock The cognitive and neural development of face recognition in humans.
\newblock \emph{The cognitive neurosciences}, 4:\penalty0 467--482, 2009.

\bibitem[Mel(1992)]{Mel1992}
BW~Mel.
\newblock {The clusteron: toward a simple abstraction for a complex neuron}.
\newblock \emph{Advances in neural information processing systems}, 1992.

\bibitem[Miller(1956)]{Miller1956}
George~A. Miller.
\newblock {The magical number seven, plus or minus two: some limits on our
  capacity for processing information.}
\newblock 1956.

\bibitem[Miller et~al.(1989)Miller, Keller, and Stryker]{Miller1989}
K.~Miller, J.~Keller, and M.~Stryker.
\newblock {Ocular dominance column development: analysis and simulation}.
\newblock \emph{Science}, 245\penalty0 (4918):\penalty0 605--615, aug 1989.
\newblock ISSN 0036-8075.
\newblock \doi{10.1126/science.2762813}.
\newblock URL \url{http://www.sciencemag.org/content/245/4918/605.short}.

\bibitem[Miller and MacKay(1994)]{Miller1994}
Kenneth~D. Miller and David J.~C. MacKay.
\newblock {The Role of Constraints in Hebbian Learning}.
\newblock \emph{Neural Computation}, 6\penalty0 (1):\penalty0 100--126, jan
  1994.
\newblock ISSN 0899-7667.
\newblock \doi{10.1162/neco.1994.6.1.100}.
\newblock URL
  \url{http://ieeexplore.ieee.org/articleDetails.jsp?arnumber=6797012}.

\bibitem[Minsky(1977)]{Minsky1977}
M~Minsky.
\newblock {Plain talk about neurodevelopmental epistemology}.
\newblock 1977.
\newblock URL \url{http://dspace.mit.edu/handle/1721.1/5763}.

\bibitem[Minsky(1988)]{Minsky1988}
Marvin Minsky.
\newblock \emph{{Society Of Mind}}.
\newblock Simon and Schuster, 1988.
\newblock ISBN 0671657135.
\newblock URL
  \url{http://books.google.com/books/about/Society{\_}Of{\_}Mind.html?id=bLDLllfRpdkC{\&}pgis=1}.

\bibitem[Minsky(1991)]{Minsky1991}
Marvin~L. Minsky.
\newblock {Logical Versus Analogical or Symbolic Versus Connectionist or Neat
  Versus Scruffy}, jun 1991.
\newblock ISSN 0738-4602.
\newblock URL
  \url{http://www.aaai.org/ojs/index.php/aimagazine/article/view/894}.

\bibitem[Minsky and Papert(1972)]{Minsky1972}
Marvin~Lee Minsky and Seymour Papert.
\newblock \emph{{Perceptrons: An Introduction to Computational Geometry}}.
\newblock Mit Press, 1972.
\newblock ISBN 0262630222.
\newblock URL
  \url{https://books.google.com/books/about/Perceptrons.html?id=Ow1OAQAAIAAJ{\&}pgis=1}.

\bibitem[Mishra et~al.(2016)Mishra, Kim, Guzman, and Jonas]{Mishra2016}
Rajiv~K. Mishra, Sooyun Kim, Segundo~J. Guzman, and Peter Jonas.
\newblock Symmetric spike timing-dependent plasticity at ca3-ca3 synapses
  optimizes storage and recall in autoassociative networks.
\newblock \emph{Nat Commun}, 7, May 2016.
\newblock URL \url{http://dx.doi.org/10.1038/ncomms11552}.
\newblock Article.

\bibitem[Mitchell(1980)]{mitchell1980need}
Tom~M Mitchell.
\newblock \emph{The need for biases in learning generalizations}.
\newblock Department of Computer Science, Laboratory for Computer Science
  Research, Rutgers Univ. New Jersey, 1980.

\bibitem[Miyagawa et~al.(2013)Miyagawa, Berwick, and Okanoya]{Miyagawa2013}
Shigeru Miyagawa, Robert~C Berwick, and Kazuo Okanoya.
\newblock {The emergence of hierarchical structure in human language.}
\newblock \emph{Frontiers in psychology}, 4:\penalty0 71, jan 2013.
\newblock ISSN 1664-1078.
\newblock \doi{10.3389/fpsyg.2013.00071}.

\bibitem[Mnih et~al.(2014)Mnih, Heess, Graves, et~al.]{mnih2014recurrent}
Volodymyr Mnih, Nicolas Heess, Alex Graves, et~al.
\newblock Recurrent models of visual attention.
\newblock In \emph{Advances in Neural Information Processing Systems}, pages
  2204--2212, 2014.

\bibitem[Mnih et~al.(2015)Mnih, Kavukcuoglu, Silver, Rusu, Veness, Bellemare,
  Graves, Riedmiller, Fidjeland, Ostrovski, Petersen, Beattie, Sadik,
  Antonoglou, King, Kumaran, Wierstra, Legg, and Hassabis]{Mnih2015}
Volodymyr Mnih, Koray Kavukcuoglu, David Silver, Andrei~A. Rusu, Joel Veness,
  Marc~G. Bellemare, Alex Graves, Martin Riedmiller, Andreas~K. Fidjeland,
  Georg Ostrovski, Stig Petersen, Charles Beattie, Amir Sadik, Ioannis
  Antonoglou, Helen King, Dharshan Kumaran, Daan Wierstra, Shane Legg, and
  Demis Hassabis.
\newblock {Human-level control through deep reinforcement learning}.
\newblock \emph{Nature}, 518\penalty0 (7540):\penalty0 529--533, feb 2015.
\newblock ISSN 0028-0836.
\newblock \doi{10.1038/nature14236}.
\newblock URL \url{http://dx.doi.org/10.1038/nature14236}.

\bibitem[Mobahi et~al.(2009)Mobahi, Collobert, and Weston]{Mobahi2009}
Hossein Mobahi, Ronan Collobert, and Jason Weston.
\newblock {Deep learning from temporal coherence in video}.
\newblock In \emph{Proceedings of the 26th Annual International Conference on
  Machine Learning - ICML '09}, pages 1--8, New York, New York, USA, jun 2009.
  ACM Press.
\newblock ISBN 9781605585161.
\newblock \doi{10.1145/1553374.1553469}.
\newblock URL \url{http://dl.acm.org/citation.cfm?id=1553374.1553469}.

\bibitem[Moberget et~al.(2014)Moberget, Gullesen, Andersson, Ivry, and
  Endestad]{Moberget2014}
Torgeir Moberget, Eva~Hilland Gullesen, Stein Andersson, Richard~B Ivry, and
  Tor Endestad.
\newblock {Generalized role for the cerebellum in encoding internal models:
  evidence from semantic processing.}
\newblock \emph{The Journal of neuroscience : the official journal of the
  Society for Neuroscience}, 34\penalty0 (8):\penalty0 2871--8, feb 2014.
\newblock ISSN 1529-2401.
\newblock \doi{10.1523/JNEUROSCI.2264-13.2014}.
\newblock URL \url{http://www.jneurosci.org/content/34/8/2871.short}.

\bibitem[Mohamed and Rezende(2015)]{Mohamed2015}
Shakir Mohamed and Danilo~Jimenez Rezende.
\newblock {Variational Information Maximisation for Intrinsically Motivated
  Reinforcement Learning}.
\newblock sep 2015.
\newblock URL \url{http://arxiv.org/abs/1509.08731}.

\bibitem[Mordatch et~al.(2012)Mordatch, Todorov, and
  Popovi{\'c}]{mordatch2012discovery}
Igor Mordatch, Emanuel Todorov, and Zoran Popovi{\'c}.
\newblock Discovery of complex behaviors through contact-invariant
  optimization.
\newblock \emph{ACM Transactions on Graphics (TOG)}, 31\penalty0 (4):\penalty0
  43, 2012.

\bibitem[Murray et~al.(2012)Murray, Wallace, Cappe, Rouiller, and
  Barone]{murray2012cortical}
Micah~M Murray, Mark~T Wallace, C{\'e}line Cappe, Eric~M Rouiller, and Pascal
  Barone.
\newblock Cortical and thalamic pathways for multisensory and sensorimotor
  interplay.
\newblock 2012.

\bibitem[Mushiake et~al.(2006)Mushiake, Saito, Sakamoto, Itoyama, and
  Tanji]{Mushiake2006}
Hajime Mushiake, Naohiro Saito, Kazuhiro Sakamoto, Yasuto Itoyama, and Jun
  Tanji.
\newblock {Activity in the lateral prefrontal cortex reflects multiple steps of
  future events in action plans.}
\newblock \emph{Neuron}, 50\penalty0 (4):\penalty0 631--41, may 2006.
\newblock ISSN 0896-6273.
\newblock \doi{10.1016/j.neuron.2006.03.045}.
\newblock URL \url{http://www.ncbi.nlm.nih.gov/pubmed/16701212}.

\bibitem[Neelakantan et~al.(2015)Neelakantan, Le, and
  Sutskever]{Neelakantan2015}
Arvind Neelakantan, Quoc~V. Le, and Ilya Sutskever.
\newblock {Neural Programmer: Inducing Latent Programs with Gradient Descent}.
\newblock nov 2015.
\newblock URL \url{http://arxiv.org/abs/1511.04834}.

\bibitem[Nessler et~al.(2013)Nessler, Pfeiffer, Buesing, and
  Maass]{Nessler2013}
Bernhard Nessler, Michael Pfeiffer, Lars Buesing, and Wolfgang Maass.
\newblock {Bayesian computation emerges in generic cortical microcircuits
  through spike-timing-dependent plasticity.}
\newblock \emph{PLoS computational biology}, 9\penalty0 (4):\penalty0 e1003037,
  apr 2013.
\newblock ISSN 1553-7358.
\newblock \doi{10.1371/journal.pcbi.1003037}.

\bibitem[Ng and Russell(2000)]{Ng2000}
AY~Ng and SJ~Russell.
\newblock {Algorithms for inverse reinforcement learning.}
\newblock \emph{Icml}, 2000.
\newblock URL \url{http://ai.stanford.edu/{~}ang/papers/icml00-irl.pdf}.

\bibitem[Noroozi and Favaro(2016)]{noroozi2016unsupervised}
Mehdi Noroozi and Paolo Favaro.
\newblock Unsupervised learning of visual representations by solving jigsaw
  puzzles.
\newblock \emph{arXiv preprint arXiv:1603.09246}, 2016.

\bibitem[{\'{O}}lafsd{\'{o}}ttir et~al.(2015){\'{O}}lafsd{\'{o}}ttir, Barry,
  Saleem, Hassabis, and Spiers]{Olafsdottir2015}
H~Freyja {\'{O}}lafsd{\'{o}}ttir, Caswell Barry, Aman~B Saleem, Demis Hassabis,
  and Hugo~J Spiers.
\newblock {Hippocampal place cells construct reward related sequences through
  unexplored space}.
\newblock \emph{eLife}, 4:\penalty0 e06063, jun 2015.
\newblock ISSN 2050-084X.
\newblock \doi{10.7554/eLife.06063}.
\newblock URL \url{http://elifesciences.org/content/4/e06063.abstract}.

\bibitem[Ollivier and Charpiat(2015)]{Ollivier2015}
Y~Ollivier and G~Charpiat.
\newblock {Training recurrent networks online without backtracking}.
\newblock \emph{arXiv preprint arXiv:1507.07680}, 2015.
\newblock URL \url{http://arxiv.org/abs/1507.07680}.

\bibitem[Olshausen et~al.(1993)Olshausen, Anderson, and {Van
  Essen}]{Olshausen1993}
B~A Olshausen, C~H Anderson, and D~C {Van Essen}.
\newblock {A neurobiological model of visual attention and invariant pattern
  recognition based on dynamic routing of information.}
\newblock \emph{The Journal of neuroscience : the official journal of the
  Society for Neuroscience}, 13\penalty0 (11):\penalty0 4700--19, nov 1993.
\newblock ISSN 0270-6474.
\newblock URL \url{http://www.ncbi.nlm.nih.gov/pubmed/8229193}.

\bibitem[Olshausen and Field(1996)]{Olshausen1996}
Bruno~A. Olshausen and David~J. Field.
\newblock {Emergence of simple-cell receptive field properties by learning a
  sparse code for natural images}.
\newblock \emph{Nature}, 381\penalty0 (6583):\penalty0 607--609, jun 1996.
\newblock ISSN 0028-0836.
\newblock \doi{10.1038/381607a0}.
\newblock URL \url{http://www.ncbi.nlm.nih.gov/pubmed/8637596}.

\bibitem[Olshausen and Field(1997)]{Olshausen1997}
Bruno~A. Olshausen and David~J. Field.
\newblock {Sparse coding with an overcomplete basis set: A strategy employed by
  V1?}
\newblock \emph{Vision Research}, 37\penalty0 (23):\penalty0 3311--3325, dec
  1997.
\newblock ISSN 00426989.
\newblock \doi{10.1016/S0042-6989(97)00169-7}.
\newblock URL
  \url{http://www.sciencedirect.com/science/article/pii/S0042698997001697}.

\bibitem[Olshausen and Field(2004)]{olshausen2004other}
Bruno~A Olshausen and David~J Field.
\newblock What is the other 85\% of v1 doing.
\newblock \emph{Problems in Systems Neuroscience}, 4\penalty0 (5):\penalty0
  182--211, 2004.

\bibitem[Oquab et~al.(2014)Oquab, Bottou, Laptev, and Sivic]{oquab2014learning}
Maxime Oquab, Leon Bottou, Ivan Laptev, and Josef Sivic.
\newblock Learning and transferring mid-level image representations using
  convolutional neural networks.
\newblock In \emph{Proceedings of the IEEE Conference on Computer Vision and
  Pattern Recognition}, pages 1717--1724, 2014.

\bibitem[O'Reilly(1996)]{OReilly1996}
Randall~C. O'Reilly.
\newblock {Biologically Plausible Error-Driven Learning Using Local Activation
  Differences: The Generalized Recirculation Algorithm}.
\newblock \emph{Neural Computation}, 8\penalty0 (5):\penalty0 895--938, jul
  1996.
\newblock ISSN 0899-7667.
\newblock \doi{10.1162/neco.1996.8.5.895}.
\newblock URL
  \url{http://ieeexplore.ieee.org/articleDetails.jsp?arnumber=6796552}.

\bibitem[O'Reilly(2006)]{OReilly2006a}
Randall~C O'Reilly.
\newblock {Biologically based computational models of high-level cognition.}
\newblock \emph{Science (New York, N.Y.)}, 314\penalty0 (5796):\penalty0 91--4,
  oct 2006.
\newblock ISSN 1095-9203.
\newblock \doi{10.1126/science.1127242}.
\newblock URL \url{http://www.sciencemag.org/content/314/5796/91}.

\bibitem[O'Reilly and Frank(2006)]{OReilly2006}
Randall~C O'Reilly and Michael~J Frank.
\newblock {Making working memory work: a computational model of learning in the
  prefrontal cortex and basal ganglia.}
\newblock \emph{Neural computation}, 18\penalty0 (2):\penalty0 283--328, mar
  2006.
\newblock ISSN 0899-7667.
\newblock \doi{10.1162/089976606775093909}.
\newblock URL \url{http://www.ncbi.nlm.nih.gov/pubmed/16378516}.

\bibitem[O'Reilly et~al.(2014{\natexlab{a}})O'Reilly, Hazy, Mollick, Mackie,
  and Herd]{OReilly2014}
Randall~C. O'Reilly, Thomas~E. Hazy, Jessica Mollick, Prescott Mackie, and Seth
  Herd.
\newblock {Goal-Driven Cognition in the Brain: A Computational Framework}.
\newblock apr 2014{\natexlab{a}}.
\newblock URL \url{http://arxiv.org/abs/1404.7591}.

\bibitem[O'Reilly et~al.(2014{\natexlab{b}})O'Reilly, Wyatte, and
  Rohrlich]{OReilly2014a}
Randall~C. O'Reilly, Dean Wyatte, and John Rohrlich.
\newblock {Learning Through Time in the Thalamocortical Loops}.
\newblock page~37, jul 2014{\natexlab{b}}.
\newblock URL \url{http://arxiv.org/abs/1407.3432}.

\bibitem[O'Reilly et~al.(2012)O'Reilly, Munakata, Frank, and Hazy]{OReilly2012}
RC~O'Reilly, Y~Munakata, MJ~Frank, and TE~Hazy.
\newblock \emph{{Computational cognitive neuroscience}}.
\newblock 2012.
\newblock URL
  \url{grey.colorado.edu/mediawiki/sites/CompCogNeuro/images/a/a2/Cecn{\_}oreilly{\_}intro.pdf}.

\bibitem[Orhan and Ma(2016)]{Orhan2016}
A.~Emin Orhan and Wei~Ji Ma.
\newblock {The Inevitability of Probability: Probabilistic Inference in Generic
  Neural Networks Trained with Non-Probabilistic Feedback}.
\newblock page~26, jan 2016.
\newblock URL \url{http://arxiv.org/abs/1601.03060}.

\bibitem[Palmer et~al.(2014)Palmer, Shai, Reeve, Anderson, Paulsen, and
  Larkum]{Palmer2014}
Lucy~M Palmer, Adam~S Shai, James~E Reeve, Harry~L Anderson, Ole Paulsen, and
  Matthew~E Larkum.
\newblock {NMDA spikes enhance action potential generation during sensory
  input.}
\newblock \emph{Nature neuroscience}, 17\penalty0 (3):\penalty0 383--90, mar
  2014.
\newblock ISSN 1546-1726.
\newblock \doi{10.1038/nn.3646}.
\newblock URL \url{http://dx.doi.org/10.1038/nn.3646}.

\bibitem[Parisien et~al.(2008)Parisien, Anderson, and Eliasmith]{Parisien2008}
C~Parisien, CH~Anderson, and C~Eliasmith.
\newblock {Solving the problem of negative synaptic weights in cortical
  models}.
\newblock \emph{Neural computation}, 2008.
\newblock URL
  \url{http://ieeexplore.ieee.org/xpls/abs{\_}all.jsp?arnumber=6796691}.

\bibitem[Pasupathy and Miller(2005)]{Pasupathy2005}
Anitha Pasupathy and Earl~K Miller.
\newblock {Different time courses of learning-related activity in the
  prefrontal cortex and striatum.}
\newblock \emph{Nature}, 433\penalty0 (7028):\penalty0 873--6, feb 2005.
\newblock ISSN 1476-4687.
\newblock \doi{10.1038/nature03287}.
\newblock URL \url{http://dx.doi.org/10.1038/nature03287}.

\bibitem[Patel et~al.(2015)Patel, Nguyen, and Baraniuk]{Patel2015}
AB~Patel, T~Nguyen, and RG~Baraniuk.
\newblock {A Probabilistic Theory of Deep Learning}.
\newblock \emph{arXiv preprint arXiv:1504.00641}, 2015.
\newblock URL \url{http://arxiv.org/abs/1504.00641}.

\bibitem[Perea et~al.(2009)Perea, Navarrete, and Araque]{Perea2009}
Gertrudis Perea, Marta Navarrete, and Alfonso Araque.
\newblock {Tripartite synapses: astrocytes process and control synaptic
  information.}
\newblock \emph{Trends in neurosciences}, 32\penalty0 (8):\penalty0 421--31,
  aug 2009.
\newblock ISSN 1878-108X.
\newblock \doi{10.1016/j.tins.2009.05.001}.
\newblock URL \url{http://www.ncbi.nlm.nih.gov/pubmed/19615761}.

\bibitem[Petrov et~al.(2010)Petrov, Jilk, and O'Reilly]{Petrov2010}
Alexander~A. Petrov, David~J. Jilk, and Randall~C. O'Reilly.
\newblock {The Leabra architecture: Specialization without modularity}.
\newblock \emph{Behavioral and Brain Sciences}, 33\penalty0 (04):\penalty0
  286--287, oct 2010.
\newblock ISSN 0140-525X.
\newblock \doi{10.1017/S0140525X10001160}.
\newblock URL
  \url{http://journals.cambridge.org/abstract{\_}S0140525X10001160}.

\bibitem[Pezzulo et~al.(2014)Pezzulo, Verschure, Balkenius, and
  Pennartz]{Pezzulo2014}
Giovanni Pezzulo, Paul F M~J Verschure, Christian Balkenius, and Cyriel M~A
  Pennartz.
\newblock {The principles of goal-directed decision-making: from neural
  mechanisms to computation and robotics.}
\newblock \emph{Philosophical transactions of the Royal Society of London.
  Series B, Biological sciences}, 369\penalty0 (1655):\penalty0 20130470--, nov
  2014.
\newblock ISSN 1471-2970.
\newblock \doi{10.1098/rstb.2013.0470}.
\newblock URL
  \url{http://rstb.royalsocietypublishing.org/content/369/1655/20130470.short}.

\bibitem[Pfister and Gerstner(2006)]{pfister2006triplets}
Jean-Pascal Pfister and Wulfram Gerstner.
\newblock Triplets of spikes in a model of spike timing-dependent plasticity.
\newblock \emph{The Journal of neuroscience}, 26\penalty0 (38):\penalty0
  9673--9682, 2006.

\bibitem[Pinker(1999)]{Pinker1999}
S~Pinker.
\newblock {How the mind works}.
\newblock \emph{Annals of the New York Academy of Sciences}, 1999.
\newblock URL
  \url{http://onlinelibrary.wiley.com/doi/10.1111/j.1749-6632.1999.tb08538.x/full}.

\bibitem[Plate(1995)]{Plate1995}
T~A Plate.
\newblock {Holographic reduced representations.}
\newblock \emph{IEEE transactions on neural networks / a publication of the
  IEEE Neural Networks Council}, 6\penalty0 (3):\penalty0 623--41, jan 1995.
\newblock ISSN 1045-9227.
\newblock \doi{10.1109/72.377968}.
\newblock URL
  \url{http://ieeexplore.ieee.org/articleDetails.jsp?arnumber=377968}.

\bibitem[Poggio(2015)]{Poggio2015}
Tomaso Poggio.
\newblock {What if...}
\newblock 2015.
\newblock URL \url{http://cbmm.mit.edu/views-reviews/article/what-if}.

\bibitem[Ponulak and Hopfield(2013)]{Ponulak2013}
Filip Ponulak and John~J Hopfield.
\newblock {Rapid, parallel path planning by propagating wavefronts of spiking
  neural activity.}
\newblock \emph{Frontiers in computational neuroscience}, 7:\penalty0 98, jan
  2013.
\newblock ISSN 1662-5188.
\newblock \doi{10.3389/fncom.2013.00098}.

\bibitem[Radford et~al.(2015)Radford, Metz, and Chintala]{Radford2015}
Alec Radford, Luke Metz, and Soumith Chintala.
\newblock {Unsupervised Representation Learning with Deep Convolutional
  Generative Adversarial Networks}.
\newblock nov 2015.
\newblock URL \url{http://arxiv.org/abs/1511.06434}.

\bibitem[Rajan et~al.(2016)Rajan, Harvey, and Tank]{rajan2016recurrent}
Kanaka Rajan, Christopher~D Harvey, and David~W Tank.
\newblock Recurrent network models of sequence generation and memory.
\newblock \emph{Neuron}, 90\penalty0 (1):\penalty0 128--142, 2016.

\bibitem[Rao(2004)]{rao2004bayesian}
Rajesh~PN Rao.
\newblock Bayesian computation in recurrent neural circuits.
\newblock \emph{Neural computation}, 16\penalty0 (1):\penalty0 1--38, 2004.

\bibitem[Rashevsky(1939)]{rashevsky1939mathematical}
Nicolas Rashevsky.
\newblock Mathematical biophysics: physico-mathematical foundations of biology.
\newblock \emph{Bull. Amer. Math. Soc. 45 (1939), 223-224 DOI: http://dx. doi.
  org/10.1090/S0002-9904-1939-06963-2 PII}, pages 0002--9904, 1939.

\bibitem[Rasmus and Berglund(2015)]{Rasmus2015a}
A~Rasmus and M~Berglund.
\newblock {Semi-Supervised Learning with Ladder Networks}.
\newblock \emph{Advances in Neural {\ldots}}, 2015.
\newblock URL
  \url{papers.nips.cc/paper/5947-semi-supervised-learning-with-ladder-networks}.

\bibitem[Reynolds and Desimone(1999)]{Reynolds1999}
John~H. Reynolds and Robert Desimone.
\newblock {The Role of Neural Mechanisms of Attention in Solving the Binding
  Problem}.
\newblock \emph{Neuron}, 24\penalty0 (1):\penalty0 19--29, sep 1999.
\newblock ISSN 08966273.
\newblock \doi{10.1016/S0896-6273(00)80819-3}.
\newblock URL
  \url{http://www.sciencedirect.com/science/article/pii/S0896627300808193}.

\bibitem[Rezende et~al.(2016)Rezende, Mohamed, Danihelka, Gregor, and
  Wierstra]{Rezende2016}
Danilo~Jimenez Rezende, Shakir Mohamed, Ivo Danihelka, Karol Gregor, and Daan
  Wierstra.
\newblock {One-Shot Generalization in Deep Generative Models}.
\newblock mar 2016.
\newblock URL \url{http://arxiv.org/abs/1603.05106}.

\bibitem[Rigotti et~al.(2013)Rigotti, Barak, Warden, Wang, Daw, Miller, and
  Fusi]{Rigotti2013}
Mattia Rigotti, Omri Barak, Melissa~R Warden, Xiao-Jing Wang, Nathaniel~D Daw,
  Earl~K Miller, and Stefano Fusi.
\newblock {The importance of mixed selectivity in complex cognitive tasks.}
\newblock \emph{Nature}, 497\penalty0 (7451):\penalty0 585--90, may 2013.
\newblock ISSN 1476-4687.
\newblock \doi{10.1038/nature12160}.

\bibitem[Robinson(1992)]{Robinson1992}
DA~Robinson.
\newblock {Implications of neural networks for how we think about brain
  function}.
\newblock \emph{Behavioral and brain sciences}, 1992.
\newblock URL
  \url{http://journals.cambridge.org/abstract{\_}S0140525X00072563}.

\bibitem[Roelfsema and van Ooyen(2005)]{Roelfsema2005}
Pieter~R Roelfsema and Arjen van Ooyen.
\newblock {Attention-gated reinforcement learning of internal representations
  for classification.}
\newblock \emph{Neural computation}, 17\penalty0 (10):\penalty0 2176--214, oct
  2005.
\newblock ISSN 0899-7667.
\newblock \doi{10.1162/0899766054615699}.
\newblock URL \url{http://www.ncbi.nlm.nih.gov/pubmed/16105222}.

\bibitem[Roelfsema et~al.(2010)Roelfsema, van Ooyen, and
  Watanabe]{Roelfsema2010}
PR~Roelfsema, A~van Ooyen, and T~Watanabe.
\newblock {Perceptual learning rules based on reinforcers and attention}.
\newblock \emph{Trends in cognitive sciences}, 2010.
\newblock URL
  \url{http://www.sciencedirect.com/science/article/pii/S1364661309002617}.

\bibitem[Rolls(2013)]{Rolls2013}
Edmund~T Rolls.
\newblock {The mechanisms for pattern completion and pattern separation in the
  hippocampus.}
\newblock \emph{Frontiers in systems neuroscience}, 7:\penalty0 74, jan 2013.
\newblock ISSN 1662-5137.
\newblock \doi{10.3389/fnsys.2013.00074}.
\newblock URL
  \url{http://journal.frontiersin.org/article/10.3389/fnsys.2013.00074/abstract}.

\bibitem[Rombouts et~al.(2015)Rombouts, Bohte, and Roelfsema]{Rombouts2015}
Jaldert~O Rombouts, Sander~M Bohte, and Pieter~R Roelfsema.
\newblock {How attention can create synaptic tags for the learning of working
  memories in sequential tasks.}
\newblock \emph{PLoS computational biology}, 11\penalty0 (3):\penalty0
  e1004060, mar 2015.
\newblock ISSN 1553-7358.
\newblock \doi{10.1371/journal.pcbi.1004060}.
\newblock URL
  \url{http://journals.plos.org/ploscompbiol/article?id=10.1371/journal.pcbi.1004060}.

\bibitem[Romero et~al.(2014)Romero, Ballas, Kahou, Chassang, Gatta, and
  Bengio]{romero2014fitnets}
Adriana Romero, Nicolas Ballas, Samira~Ebrahimi Kahou, Antoine Chassang, Carlo
  Gatta, and Yoshua Bengio.
\newblock Fitnets: Hints for thin deep nets.
\newblock \emph{arXiv preprint arXiv:1412.6550}, 2014.

\bibitem[Roudi and Taylor(2015)]{Roudi2015}
Y~Roudi and G~Taylor.
\newblock {Learning with hidden variables}.
\newblock \emph{Current opinion in neurobiology}, 2015.
\newblock URL
  \url{http://www.sciencedirect.com/science/article/pii/S0959438815001245}.

\bibitem[Rozell et~al.(2008)Rozell, Johnson, Baraniuk, and
  Olshausen]{Rozell2008}
Christopher~J Rozell, Don~H Johnson, Richard~G Baraniuk, and Bruno~A Olshausen.
\newblock {Sparse coding via thresholding and local competition in neural
  circuits.}
\newblock \emph{Neural computation}, 20\penalty0 (10):\penalty0 2526--63, oct
  2008.
\newblock ISSN 0899-7667.
\newblock \doi{10.1162/neco.2008.03-07-486}.
\newblock URL \url{http://www.ncbi.nlm.nih.gov/pubmed/18439138}.

\bibitem[Rubin et~al.(2015)Rubin, Geva, Sheintuch, and
  Ziv]{rubin2015hippocampal}
Alon Rubin, Nitzan Geva, Liron Sheintuch, and Yaniv Ziv.
\newblock Hippocampal ensemble dynamics timestamp events in long-term memory.
\newblock \emph{eLife}, page e12247, 2015.

\bibitem[Rumelhart et~al.(1986)Rumelhart, Hinton, and Williams]{Rumelhart1986}
David~E. Rumelhart, Geoffrey~E. Hinton, and Ronald~J. Williams.
\newblock {Learning representations by back-propagating errors}.
\newblock \emph{Nature}, 323\penalty0 (6088):\penalty0 533--536, oct 1986.
\newblock ISSN 0028-0836.
\newblock \doi{10.1038/323533a0}.
\newblock URL
  \url{http://www.nature.com/nature/journal/v323/n6088/pdf/323533a0.pdf}.

\bibitem[Sadtler et~al.(2014)Sadtler, Quick, Golub, Chase, Ryu, Tyler-Kabara,
  Yu, and Batista]{Sadtler2014}
Patrick~T Sadtler, Kristin~M Quick, Matthew~D Golub, Steven~M Chase, Stephen~I
  Ryu, Elizabeth~C Tyler-Kabara, Byron~M Yu, and Aaron~P Batista.
\newblock {Neural constraints on learning.}
\newblock \emph{Nature}, 512\penalty0 (7515):\penalty0 423--6, aug 2014.
\newblock ISSN 1476-4687.
\newblock \doi{10.1038/nature13665}.
\newblock URL \url{http://dx.doi.org/10.1038/nature13665}.

\bibitem[Sahani and Dayan(2003)]{sahani2003doubly}
Maneesh Sahani and Peter Dayan.
\newblock Doubly distributional population codes: simultaneous representation
  of uncertainty and multiplicity.
\newblock \emph{Neural Computation}, 15\penalty0 (10):\penalty0 2255--2279,
  2003.

\bibitem[Sandler et~al.(2016)Sandler, Shulman, and Schiller]{sandler2016novel}
Maya Sandler, Yoav Shulman, and Jackie Schiller.
\newblock A novel form of local plasticity in tuft dendrites of neocortical
  somatosensory layer 5 pyramidal neurons.
\newblock \emph{Neuron}, 2016.

\bibitem[Santoro et~al.(2016)Santoro, Bartunov, Botvinick, Wierstra, and
  Lillicrap]{Santoro2016}
Adam Santoro, Sergey Bartunov, Matthew Botvinick, Daan Wierstra, and Timothy
  Lillicrap.
\newblock {One-shot Learning with Memory-Augmented Neural Networks}.
\newblock page~13, may 2016.
\newblock URL \url{http://arxiv.org/abs/1605.06065}.

\bibitem[Saxe et~al.(2013)Saxe, McClelland, and Ganguli]{Saxe2013}
Andrew~M. Saxe, James~L. McClelland, and Surya Ganguli.
\newblock {Exact solutions to the nonlinear dynamics of learning in deep linear
  neural networks}.
\newblock dec 2013.
\newblock URL \url{http://arxiv.org/abs/1312.6120}.

\bibitem[Scellier and Bengio(2016)]{Scellier2016}
Benjamin Scellier and Yoshua Bengio.
\newblock {Towards a Biologically Plausible Backprop}.
\newblock feb 2016.
\newblock URL \url{http://arxiv.org/abs/1602.05179}.

\bibitem[Schmidhuber(2010)]{Schmidhuber2010}
J~Schmidhuber.
\newblock {Formal theory of creativity, fun, and intrinsic motivation
  (1990–2010)}.
\newblock \emph{Autonomous Mental Development, IEEE {\ldots}}, 2010.
\newblock URL
  \url{http://ieeexplore.ieee.org/xpls/abs{\_}all.jsp?arnumber=5508364}.

\bibitem[Schmidhuber(2015)]{schmidhuber2015deep}
J{\"u}rgen Schmidhuber.
\newblock Deep learning in neural networks: An overview.
\newblock \emph{Neural Networks}, 61:\penalty0 85--117, 2015.

\bibitem[Sejnowski and Poizner(2014)]{Sejnowski2014}
TJ~Sejnowski and H~Poizner.
\newblock {Prospective Optimization}.
\newblock \emph{Proceedings of the {\ldots}}, 2014.
\newblock URL
  \url{http://ieeexplore.ieee.org/xpls/abs{\_}all.jsp?arnumber=6803897}.

\bibitem[Sermanet and Kavukcuoglu(2013)]{Sermanet2013}
P~Sermanet and K~Kavukcuoglu.
\newblock {Pedestrian detection with unsupervised multi-stage feature
  learning}.
\newblock \emph{Proceedings of the {\ldots}}, 2013.

\bibitem[Serre et~al.(2007)Serre, Oliva, and Poggio]{Serre2007}
T.~Serre, A.~Oliva, and T.~Poggio.
\newblock {A feedforward architecture accounts for rapid categorization}.
\newblock \emph{Proceedings of the National Academy of Sciences}, 104\penalty0
  (15):\penalty0 6424--6429, apr 2007.
\newblock ISSN 0027-8424.
\newblock \doi{10.1073/pnas.0700622104}.
\newblock URL \url{http://www.pnas.org/content/104/15/6424.long}.

\bibitem[Servan-Schreiber and Anderson(1990)]{Servan-Schreiber1990}
E~Servan-Schreiber and JR~Anderson.
\newblock {Chunking as a mechanism of implicit learning}.
\newblock \emph{Journal of Experimental Psychology: Learning, {\ldots}}, 1990.

\bibitem[Seung(1998)]{SebastianSeung1998}
H.~Sebastian Seung.
\newblock {Continuous attractors and oculomotor control}.
\newblock \emph{Neural Networks}, 11\penalty0 (7-8):\penalty0 1253--1258, oct
  1998.
\newblock ISSN 08936080.
\newblock \doi{10.1016/S0893-6080(98)00064-1}.
\newblock URL
  \url{http://www.sciencedirect.com/science/article/pii/S0893608098000641}.

\bibitem[Seung(2003)]{Seung2003}
H.~Sebastian Seung.
\newblock {Learning in Spiking Neural Networks by Reinforcement of Stochastic
  Synaptic Transmission}.
\newblock \emph{Neuron}, 40\penalty0 (6):\penalty0 1063--1073, dec 2003.
\newblock ISSN 08966273.
\newblock \doi{10.1016/S0896-6273(03)00761-X}.
\newblock URL
  \url{http://www.sciencedirect.com/science/article/pii/S089662730300761X}.

\bibitem[Sherman(2005)]{Sherman2005}
S~Murray Sherman.
\newblock {Thalamic relays and cortical functioning.}
\newblock \emph{Progress in brain research}, 149:\penalty0 107--26, jan 2005.
\newblock ISSN 1875-7855.
\newblock \doi{10.1016/S0079-6123(05)49009-3}.
\newblock URL \url{http://www.ncbi.nlm.nih.gov/pubmed/16226580}.

\bibitem[Sherman(2007)]{Sherman2007}
S~Murray Sherman.
\newblock {The thalamus is more than just a relay.}
\newblock \emph{Current opinion in neurobiology}, 17\penalty0 (4):\penalty0
  417--22, aug 2007.
\newblock ISSN 0959-4388.
\newblock \doi{10.1016/j.conb.2007.07.003}.

\bibitem[Shimizu and Karten(2013)]{shimizu2013multiple}
Toru Shimizu and Harvey~J Karten.
\newblock Multiple origins of neocortex: Contributions of the dorsal.
\newblock \emph{The Neocortex: Ontogeny and Phylogeny}, 200:\penalty0 75, 2013.

\bibitem[Si(2004)]{si2004handbook}
Jennie Si.
\newblock \emph{Handbook of learning and approximate dynamic programming},
  volume~2.
\newblock John Wiley \& Sons, 2004.

\bibitem[Siegel et~al.(2009)Siegel, Warden, and Miller]{Siegel2009}
Markus Siegel, Melissa~R Warden, and Earl~K Miller.
\newblock {Phase-dependent neuronal coding of objects in short-term memory.}
\newblock \emph{Proceedings of the National Academy of Sciences of the United
  States of America}, 106\penalty0 (50):\penalty0 21341--6, dec 2009.
\newblock ISSN 1091-6490.
\newblock \doi{10.1073/pnas.0908193106}.
\newblock URL \url{http://www.pnas.org/content/106/50/21341.abstract}.

\bibitem[Singh and Eliasmith(2006)]{Singh2006}
Ray Singh and Chris Eliasmith.
\newblock {Higher-dimensional neurons explain the tuning and dynamics of
  working memory cells.}
\newblock \emph{The Journal of neuroscience : the official journal of the
  Society for Neuroscience}, 26\penalty0 (14):\penalty0 3667--78, apr 2006.
\newblock ISSN 1529-2401.
\newblock \doi{10.1523/JNEUROSCI.4864-05.2006}.
\newblock URL \url{http://www.ncbi.nlm.nih.gov/pubmed/16597721}.

\bibitem[Sj{\"{o}}str{\"{o}}m and Gerstner(2010)]{Sjostrom2010}
Jesper Sj{\"{o}}str{\"{o}}m and Wulfram Gerstner.
\newblock {Spike-timing dependent plasticity}.
\newblock \emph{Scholarpedia}, 5\penalty0 (2):\penalty0 1362, feb 2010.
\newblock ISSN 1941-6016.
\newblock \doi{10.4249/scholarpedia.1362}.
\newblock URL
  \url{http://www.scholarpedia.org/article/Spike-timing{\_}dependent{\_}plasticity}.

\bibitem[Skerry and Spelke(2014)]{Skerry2014}
Amy~E Skerry and Elizabeth~S Spelke.
\newblock {Preverbal infants identify emotional reactions that are incongruent
  with goal outcomes.}
\newblock \emph{Cognition}, 130\penalty0 (2):\penalty0 204--16, feb 2014.
\newblock ISSN 1873-7838.
\newblock \doi{10.1016/j.cognition.2013.11.002}.

\bibitem[Softky and Koch(1993)]{Softky1993}
WR~Softky and C~Koch.
\newblock {The highly irregular firing of cortical cells is inconsistent with
  temporal integration of random EPSPs}.
\newblock \emph{The Journal of Neuroscience}, 1993.
\newblock URL \url{http://www.jneurosci.org/content/13/1/334.short}.

\bibitem[Solari and Stoner(2011)]{Solari2011}
Soren Van~Hout Solari and Rich Stoner.
\newblock {Cognitive consilience: primate non-primary neuroanatomical circuits
  underlying cognition.}
\newblock \emph{Frontiers in neuroanatomy}, 5:\penalty0 65, jan 2011.
\newblock ISSN 1662-5129.
\newblock \doi{10.3389/fnana.2011.00065}.
\newblock URL
  \url{http://journal.frontiersin.org/article/10.3389/fnana.2011.00065/abstract}.

\bibitem[Sountsov and Miller(2015)]{Sountsov2015}
Pavel Sountsov and Paul Miller.
\newblock {Spiking neuron network Helmholtz machine.}
\newblock \emph{Frontiers in computational neuroscience}, 9:\penalty0 46, jan
  2015.
\newblock ISSN 1662-5188.
\newblock \doi{10.3389/fncom.2015.00046}.
\newblock URL
  \url{http://journal.frontiersin.org/article/10.3389/fncom.2015.00046/abstract}.

\bibitem[Squire(2004)]{squire2004memory}
Larry~R Squire.
\newblock Memory systems of the brain: a brief history and current perspective.
\newblock \emph{Neurobiology of learning and memory}, 82\penalty0 (3):\penalty0
  171--177, 2004.

\bibitem[Srivastava et~al.(2014)Srivastava, Hinton, Krizhevsky, Sutskever, and
  Salakhutdinov]{Srivastava2014}
Nitish Srivastava, Geoffrey Hinton, Alex Krizhevsky, Ilya Sutskever, and Ruslan
  Salakhutdinov.
\newblock {Dropout: a simple way to prevent neural networks from overfitting}.
\newblock \emph{The Journal of Machine Learning Research}, 15\penalty0
  (1):\penalty0 1929--1958, jan 2014.
\newblock ISSN 1532-4435.
\newblock URL \url{http://dl.acm.org/citation.cfm?id=2627435.2670313}.

\bibitem[Stachenfeld(2014)]{Stachenfeld2014}
KL~Stachenfeld.
\newblock {Design Principles of the Hippocampal Cognitive Map}.
\newblock \emph{Advances in Neural {\ldots}}, 2014.

\bibitem[Stewart and Eliasmith(2009)]{Stewart2009}
Terry Stewart and Chris Eliasmith.
\newblock {Compositionality and biologically plausible models}.
\newblock 2009.
\newblock URL \url{http://philpapers.org/rec/STECAB-2}.

\bibitem[Stocco et~al.(2010)Stocco, Lebiere, and Anderson]{Stocco2010}
Andrea Stocco, Christian Lebiere, and John~R Anderson.
\newblock {Conditional routing of information to the cortex: a model of the
  basal ganglia's role in cognitive coordination.}
\newblock \emph{Psychological review}, 117\penalty0 (2):\penalty0 541--74, apr
  2010.
\newblock ISSN 1939-1471.
\newblock \doi{10.1037/a0019077}.

\bibitem[Stork(1989)]{Stork1989}
D.~G. Stork.
\newblock {Is backpropagation biologically plausible?}
\newblock In \emph{International Joint Conference on Neural Networks}, pages
  241--246 vol.2. IEEE, 1989.
\newblock \doi{10.1109/IJCNN.1989.118705}.
\newblock URL
  \url{http://ieeexplore.ieee.org/articleDetails.jsp?arnumber=118705}.

\bibitem[Strausfeld and Hirth(2013)]{Strausfeld2013}
Nicholas~J Strausfeld and Frank Hirth.
\newblock {Deep homology of arthropod central complex and vertebrate basal
  ganglia.}
\newblock \emph{Science (New York, N.Y.)}, 340\penalty0 (6129):\penalty0
  157--61, apr 2013.
\newblock ISSN 1095-9203.
\newblock \doi{10.1126/science.1231828}.
\newblock URL \url{http://www.sciencemag.org/content/340/6129/157.short}.

\bibitem[Stănişor et~al.(2013)Stănişor, van~der Togt, Pennartz, and
  Roelfsema]{Stanisor2013}
Liviu Stănişor, Chris van~der Togt, Cyriel M~A Pennartz, and Pieter~R
  Roelfsema.
\newblock {A unified selection signal for attention and reward in primary
  visual cortex.}
\newblock \emph{Proceedings of the National Academy of Sciences of the United
  States of America}, 110\penalty0 (22):\penalty0 9136--41, may 2013.
\newblock ISSN 1091-6490.
\newblock \doi{10.1073/pnas.1300117110}.

\bibitem[Sukhbaatar et~al.(2014)Sukhbaatar, Bruna, Paluri, Bourdev, and
  Fergus]{sukhbaatar2014training}
Sainbayar Sukhbaatar, Joan Bruna, Manohar Paluri, Lubomir Bourdev, and Rob
  Fergus.
\newblock Training convolutional networks with noisy labels.
\newblock \emph{arXiv preprint arXiv:1406.2080}, 2014.

\bibitem[Sun et~al.(2011)Sun, Gomez, and Schmidhuber]{sun2011planning}
Yi~Sun, Faustino Gomez, and J{\"u}rgen Schmidhuber.
\newblock Planning to be surprised: Optimal bayesian exploration in dynamic
  environments.
\newblock In \emph{Artificial General Intelligence}, pages 41--51. Springer,
  2011.

\bibitem[Sussillo(2014)]{Sussillo2014}
D~Sussillo.
\newblock {Neural circuits as computational dynamical systems}.
\newblock \emph{Current opinion in neurobiology}, 2014.
\newblock URL
  \url{http://www.sciencedirect.com/science/article/pii/S0959438814000166}.

\bibitem[Sussillo and Abbott(2009)]{Sussillo2009}
D~Sussillo and LF~Abbott.
\newblock {Generating coherent patterns of activity from chaotic neural
  networks}.
\newblock \emph{Neuron}, 2009.
\newblock URL
  \url{http://www.sciencedirect.com/science/article/pii/S0896627309005479}.

\bibitem[Sussillo et~al.(2015)Sussillo, Churchland, Kaufman, and
  Shenoy]{Sussillo2015}
David Sussillo, Mark~M Churchland, Matthew~T Kaufman, and Krishna~V Shenoy.
\newblock {A neural network that finds a naturalistic solution for the
  production of muscle activity.}
\newblock \emph{Nature neuroscience}, 18\penalty0 (7):\penalty0 1025--33, jul
  2015.
\newblock ISSN 1546-1726.
\newblock \doi{10.1038/nn.4042}.
\newblock URL \url{http://dx.doi.org/10.1038/nn.4042}.

\bibitem[Sutskever and Martens(2013)]{Sutskever2013}
I~Sutskever and J~Martens.
\newblock {On the importance of initialization and momentum in deep learning}.
\newblock \emph{Proceedings of the {\ldots}}, 2013.
\newblock URL
  \url{http://machinelearning.wustl.edu/mlpapers/papers/icml2013{\_}sutskever13}.

\bibitem[Sutskever et~al.(2011)Sutskever, Martens, and
  Hinton]{sutskever2011generating}
Ilya Sutskever, James Martens, and Geoffrey~E Hinton.
\newblock Generating text with recurrent neural networks.
\newblock In \emph{Proceedings of the 28th International Conference on Machine
  Learning (ICML-11)}, pages 1017--1024, 2011.

\bibitem[Sutton and Barto(1998)]{sutton1998reinforcement}
Richard~S Sutton and Andrew~G Barto.
\newblock \emph{Reinforcement learning: An introduction}.
\newblock MIT press, 1998.

\bibitem[Takata et~al.(2011)Takata, Mishima, Hisatsune, Nagai, Ebisui,
  Mikoshiba, and Hirase]{Takata2011}
Norio Takata, Tsuneko Mishima, Chihiro Hisatsune, Terumi Nagai, Etsuko Ebisui,
  Katsuhiko Mikoshiba, and Hajime Hirase.
\newblock {Astrocyte calcium signaling transforms cholinergic modulation to
  cortical plasticity in vivo.}
\newblock \emph{The Journal of neuroscience : the official journal of the
  Society for Neuroscience}, 31\penalty0 (49):\penalty0 18155--65, dec 2011.
\newblock ISSN 1529-2401.
\newblock \doi{10.1523/JNEUROSCI.5289-11.2011}.
\newblock URL \url{http://www.ncbi.nlm.nih.gov/pubmed/22159127}.

\bibitem[Tamar et~al.(2016)Tamar, Levine, and Abbeel]{tamar2016value}
Aviv Tamar, Sergey Levine, and Pieter Abbeel.
\newblock Value iteration networks.
\newblock \emph{arXiv preprint arXiv:1602.02867}, 2016.

\bibitem[Tang et~al.(2013)Tang, Salakhutdinov, and Hinton]{Tang2013}
Y~Tang, R~Salakhutdinov, and G~Hinton.
\newblock {Tensor analyzers}.
\newblock \emph{Proceedings of The 30th International {\ldots}}, 2013.
\newblock URL \url{http://jmlr.org/proceedings/papers/v28/tang13.html}.

\bibitem[Tang et~al.(2012)Tang, Salakhutdinov, and Hinton]{Tang2012}
Yichuan Tang, Ruslan Salakhutdinov, and Geoffrey Hinton.
\newblock {Deep Mixtures of Factor Analysers}.
\newblock jun 2012.
\newblock URL \url{http://arxiv.org/abs/1206.4635}.

\bibitem[Tapson and van Schaik(2013)]{tapson2013learning}
Jonathan Tapson and Andr{\'e} van Schaik.
\newblock Learning the pseudoinverse solution to network weights.
\newblock \emph{Neural Networks}, 45:\penalty0 94--100, 2013.

\bibitem[Taylor and Faisal(2011)]{Taylor2011}
Scott~V Taylor and Aldo~A Faisal.
\newblock {Does the cost function of human motor control depend on the internal
  metabolic state?}
\newblock \emph{BMC Neuroscience}, 12\penalty0 (Suppl 1):\penalty0 P99, 2011.
\newblock ISSN 1471-2202.
\newblock \doi{10.1186/1471-2202-12-S1-P99}.
\newblock URL \url{http://www.ncbi.nlm.nih.gov/pmc/articles/PMC3240571/}.

\bibitem[{Terrence C. Stewart, Xuan Choo}(2010)]{TerrenceC.StewartXuanChoo2010}
Chris~Eliasmith {Terrence C. Stewart, Xuan Choo}.
\newblock {Symbolic reasoning in spiking neurons: A model of the cortex/basal
  ganglia/thalamus loop.}
\newblock \emph{32nd Annual Meeting of the Cognitive Science Society}, 2010.

\bibitem[Tervo et~al.(2016)Tervo, Tenenbaum, and Gershman]{Tervo2016}
D~Gowanlock~R Tervo, Joshua~B Tenenbaum, and Samuel~J Gershman.
\newblock {Toward the neural implementation of structure learning.}
\newblock \emph{Current opinion in neurobiology}, 37:\penalty0 99--105, feb
  2016.
\newblock ISSN 1873-6882.
\newblock \doi{10.1016/j.conb.2016.01.014}.
\newblock URL \url{http://www.ncbi.nlm.nih.gov/pubmed/26874471}.

\bibitem[Tesauro(1995)]{Tesauro1995}
G~Tesauro.
\newblock {Temporal difference learning and TD-Gammon}.
\newblock \emph{Communications of the ACM}, 1995.

\bibitem[Thalmeier et~al.(2015)Thalmeier, Uhlmann, Kappen, and
  Memmesheimer]{Thalmeier2015}
Dominik Thalmeier, Marvin Uhlmann, Hilbert~J. Kappen, and Raoul-Martin
  Memmesheimer.
\newblock {Learning universal computations with spikes}.
\newblock page~24, may 2015.
\newblock URL \url{http://arxiv.org/abs/1505.07866}.

\bibitem[Tinbergen(1965)]{tinbergen1965behavior}
N~Tinbergen.
\newblock Behavior and natural selection.
\newblock 1965.

\bibitem[Todorov(2002)]{todorov2002cosine}
Emanuel Todorov.
\newblock Cosine tuning minimizes motor errors.
\newblock \emph{Neural Computation}, 14\penalty0 (6):\penalty0 1233--1260,
  2002.

\bibitem[Todorov(2009)]{todorov2009efficient}
Emanuel Todorov.
\newblock Efficient computation of optimal actions.
\newblock \emph{Proceedings of the national academy of sciences}, 106\penalty0
  (28):\penalty0 11478--11483, 2009.

\bibitem[Todorov and Jordan(2002)]{todorov2002optimal}
Emanuel Todorov and Michael~I Jordan.
\newblock Optimal feedback control as a theory of motor coordination.
\newblock \emph{Nature neuroscience}, 5\penalty0 (11):\penalty0 1226--1235,
  2002.

\bibitem[Tripp and Eliasmith(2016)]{Tripp2016}
Bryan Tripp and Chris Eliasmith.
\newblock {Function approximation in inhibitory networks.}
\newblock \emph{Neural networks : the official journal of the International
  Neural Network Society}, 77:\penalty0 95--106, may 2016.
\newblock ISSN 1879-2782.
\newblock \doi{10.1016/j.neunet.2016.01.010}.
\newblock URL
  \url{http://www.sciencedirect.com/science/article/pii/S0893608016000113}.

\bibitem[Turrigiano(2012)]{Turrigiano2012}
Gina Turrigiano.
\newblock {Homeostatic synaptic plasticity: local and global mechanisms for
  stabilizing neuronal function.}
\newblock \emph{Cold Spring Harbor perspectives in biology}, 4\penalty0
  (1):\penalty0 a005736, jan 2012.
\newblock ISSN 1943-0264.
\newblock \doi{10.1101/cshperspect.a005736}.

\bibitem[Ullman et~al.(2012)Ullman, Harari, and Dorfman]{Ullman2012}
Shimon Ullman, Daniel Harari, and Nimrod Dorfman.
\newblock {From simple innate biases to complex visual concepts.}
\newblock \emph{Proceedings of the National Academy of Sciences of the United
  States of America}, 109\penalty0 (44):\penalty0 18215--20, oct 2012.
\newblock ISSN 1091-6490.
\newblock \doi{10.1073/pnas.1207690109}.
\newblock URL \url{http://www.pnas.org/content/109/44/18215.full}.

\bibitem[Urbanczik and Senn(2014)]{Urbanczik2014}
Robert Urbanczik and Walter Senn.
\newblock {Learning by the dendritic prediction of somatic spiking.}
\newblock \emph{Neuron}, 81\penalty0 (3):\penalty0 521--8, feb 2014.
\newblock ISSN 1097-4199.
\newblock \doi{10.1016/j.neuron.2013.11.030}.
\newblock URL \url{http://www.ncbi.nlm.nih.gov/pubmed/24507189}.

\bibitem[Valpola(2015)]{Valpola2015}
H~Valpola.
\newblock {From neural PCA to deep unsupervised learning}.
\newblock \emph{Adv. in Independent Component Analysis and {\ldots}}, 2015.

\bibitem[van~den Oord et~al.(2016)van~den Oord, Kalchbrenner, and
  Kavukcuoglu]{Oord2016}
Aaron van~den Oord, Nal Kalchbrenner, and Koray Kavukcuoglu.
\newblock {Pixel Recurrent Neural Networks}.
\newblock jan 2016.
\newblock URL \url{http://arxiv.org/abs/1601.06759}.

\bibitem[Van~Heijningen et~al.(2009)Van~Heijningen, De~Visser, Zuidema, and
  Ten~Cate]{van2009simple}
Caroline~AA Van~Heijningen, Jos De~Visser, Willem Zuidema, and Carel Ten~Cate.
\newblock Simple rules can explain discrimination of putative recursive
  syntactic structures by a songbird species.
\newblock \emph{Proceedings of the National Academy of Sciences}, 106\penalty0
  (48):\penalty0 20538--20543, 2009.

\bibitem[Veit et~al.(2016)Veit, Wilber, and Belongie]{Veit2016}
Andreas Veit, Michael Wilber, and Serge Belongie.
\newblock {Residual Networks are Exponential Ensembles of Relatively Shallow
  Networks}.
\newblock may 2016.
\newblock URL \url{http://arxiv.org/abs/1605.06431}.

\bibitem[Verney et~al.(1985)Verney, Baulac, Berger, Alvarez, Vigny, and
  Helle]{verney1985morphological}
C~Verney, M~Baulac, B~Berger, C~Alvarez, A~Vigny, and KB~Helle.
\newblock Morphological evidence for a dopaminergic terminal field in the
  hippocampal formation of young and adult rat.
\newblock \emph{Neuroscience}, 14\penalty0 (4):\penalty0 1039--1052, 1985.

\bibitem[Verwey(1996)]{verwey1996buffer}
Willem~B Verwey.
\newblock Buffer loading and chunking in sequential keypressing.
\newblock \emph{Journal of Experimental Psychology: Human Perception and
  Performance}, 22\penalty0 (3):\penalty0 544, 1996.

\bibitem[Wang et~al.(2015)Wang, Cohen, and Voss]{Wang2015}
Jane~X Wang, Neal~J Cohen, and Joel~L Voss.
\newblock {Covert rapid action-memory simulation (CRAMS): a hypothesis of
  hippocampal-prefrontal interactions for adaptive behavior.}
\newblock \emph{Neurobiology of learning and memory}, 117:\penalty0 22--33, jan
  2015.
\newblock ISSN 1095-9564.
\newblock \doi{10.1016/j.nlm.2014.04.003}.

\bibitem[Wang and Yuille(2014)]{Wang2014}
Jianyu Wang and Alan Yuille.
\newblock {Semantic Part Segmentation using Compositional Model combining Shape
  and Appearance}.
\newblock dec 2014.
\newblock URL \url{http://arxiv.org/abs/1412.6124}.

\bibitem[Wang(2012)]{Wang2012}
Xiao-Jing Wang.
\newblock {The Prefrontal Cortex as a Quintessential “Cognitive-Type”
  Neural Circuit : Principles of Frontal Lobe Function - oi}, 2012.
\newblock URL
  \url{http://oxfordindex.oup.com/view/10.1093/med/9780199837755.003.0018}.

\bibitem[Warden and Miller(2007)]{Warden2007}
Melissa~R Warden and Earl~K Miller.
\newblock {The representation of multiple objects in prefrontal neuronal delay
  activity.}
\newblock \emph{Cerebral cortex (New York, N.Y. : 1991)}, 17 Suppl 1:\penalty0
  i41--50, sep 2007.
\newblock ISSN 1047-3211.
\newblock \doi{10.1093/cercor/bhm070}.
\newblock URL \url{http://www.ncbi.nlm.nih.gov/pubmed/17726003}.

\bibitem[Warden and Miller(2010)]{Warden2010}
Melissa~R Warden and Earl~K Miller.
\newblock {Task-dependent changes in short-term memory in the prefrontal
  cortex.}
\newblock \emph{The Journal of neuroscience : the official journal of the
  Society for Neuroscience}, 30\penalty0 (47):\penalty0 15801--10, nov 2010.
\newblock ISSN 1529-2401.
\newblock \doi{10.1523/JNEUROSCI.1569-10.2010}.
\newblock URL \url{http://www.jneurosci.org/content/30/47/15801.short}.

\bibitem[Watter et~al.(2015)Watter, Springenberg, Boedecker, and
  Riedmiller]{watter2015embed}
Manuel Watter, Jost Springenberg, Joschka Boedecker, and Martin Riedmiller.
\newblock Embed to control: A locally linear latent dynamics model for control
  from raw images.
\newblock In \emph{Advances in Neural Information Processing Systems}, pages
  2728--2736, 2015.

\bibitem[Wayne and Abbott(2014)]{Wayne2014}
Greg Wayne and L~F Abbott.
\newblock {Hierarchical control using networks trained with higher-level
  forward models.}
\newblock \emph{Neural computation}, 26\penalty0 (10):\penalty0 2163--93, oct
  2014.
\newblock ISSN 1530-888X.

\bibitem[Werbos(1974)]{werbos1974beyond}
Paul Werbos.
\newblock Beyond regression: New tools for prediction and analysis in the
  behavioral sciences.
\newblock 1974.

\bibitem[Werbos(1982)]{Werbos1982}
PJ~Werbos.
\newblock {Applications of advances in nonlinear sensitivity analysis}.
\newblock \emph{System modeling and optimization}, 1982.
\newblock URL \url{http://link.springer.com/chapter/10.1007/BFb0006203}.

\bibitem[Werbos(1990)]{Werbos1990}
PJ~Werbos.
\newblock {Backpropagation through time: what it does and how to do it}.
\newblock \emph{Proceedings of the IEEE}, 1990.
\newblock URL
  \url{http://ieeexplore.ieee.org/xpls/abs{\_}all.jsp?arnumber=58337}.

\bibitem[Werfel et~al.(2005)Werfel, Xie, and Seung]{Werfel2005}
Justin Werfel, Xiaohui Xie, and H~Sebastian Seung.
\newblock {Learning curves for stochastic gradient descent in linear
  feedforward networks.}
\newblock \emph{Neural computation}, 17\penalty0 (12):\penalty0 2699--718, dec
  2005.
\newblock ISSN 0899-7667.
\newblock \doi{10.1162/089976605774320539}.
\newblock URL
  \url{http://ieeexplore.ieee.org/articleDetails.jsp?arnumber=6790355}.

\bibitem[Weston et~al.(2014)Weston, Chopra, and Bordes]{Weston2014}
Jason Weston, Sumit Chopra, and Antoine Bordes.
\newblock {Memory Networks}.
\newblock oct 2014.
\newblock URL \url{http://arxiv.org/abs/1410.3916}.

\bibitem[Whitney et~al.(2016)Whitney, Chang, Kulkarni, and
  Tenenbaum]{Whitney2016}
William~F. Whitney, Michael Chang, Tejas Kulkarni, and Joshua~B. Tenenbaum.
\newblock {Understanding Visual Concepts with Continuation Learning}.
\newblock feb 2016.
\newblock URL \url{http://arxiv.org/abs/1602.06822}.

\bibitem[Williams(1992)]{Williams1992}
Ronald~J. Williams.
\newblock {Simple statistical gradient-following algorithms for connectionist
  reinforcement learning}.
\newblock \emph{Machine Learning}, 8\penalty0 (3-4):\penalty0 229--256, may
  1992.
\newblock ISSN 0885-6125.
\newblock \doi{10.1007/BF00992696}.
\newblock URL \url{http://link.springer.com/10.1007/BF00992696}.

\bibitem[Williams and Baird(1993)]{williams1993tight}
Ronald~J Williams and Leemon~C Baird.
\newblock Tight performance bounds on greedy policies based on imperfect value
  functions.
\newblock Technical report, Citeseer, 1993.

\bibitem[Williams and Stuart(2000)]{Williams2000}
S~R Williams and G~J Stuart.
\newblock {Backpropagation of physiological spike trains in neocortical
  pyramidal neurons: implications for temporal coding in dendrites.}
\newblock \emph{The Journal of neuroscience : the official journal of the
  Society for Neuroscience}, 20\penalty0 (22):\penalty0 8238--46, nov 2000.
\newblock ISSN 1529-2401.
\newblock URL \url{http://www.ncbi.nlm.nih.gov/pubmed/11069929}.

\bibitem[Wilson and Nicoll(2001)]{Wilson2001}
R~I Wilson and R~A Nicoll.
\newblock {Endogenous cannabinoids mediate retrograde signalling at hippocampal
  synapses.}
\newblock \emph{Nature}, 410\penalty0 (6828):\penalty0 588--92, mar 2001.
\newblock ISSN 0028-0836.
\newblock \doi{10.1038/35069076}.
\newblock URL \url{http://www.ncbi.nlm.nih.gov/pubmed/11279497}.

\bibitem[Winston(2011)]{winston2011strong}
Patrick~Henry Winston.
\newblock The strong story hypothesis and the directed perception hypothesis.
\newblock 2011.

\bibitem[Wiskott and Sejnowski(2002)]{Wiskott2002}
Laurenz Wiskott and Terrence~J Sejnowski.
\newblock {Slow feature analysis: unsupervised learning of invariances.}
\newblock \emph{Neural computation}, 14\penalty0 (4):\penalty0 715--70, apr
  2002.
\newblock ISSN 0899-7667.
\newblock \doi{10.1162/089976602317318938}.
\newblock URL \url{http://www.ncbi.nlm.nih.gov/pubmed/11936959}.

\bibitem[Wyss et~al.(2006)Wyss, K{\"{o}}nig, and Verschure]{Wyss2006}
Reto Wyss, Peter K{\"{o}}nig, and Paul F M~J Verschure.
\newblock {A model of the ventral visual system based on temporal stability and
  local memory.}
\newblock \emph{PLoS biology}, 4\penalty0 (5):\penalty0 e120, may 2006.
\newblock ISSN 1545-7885.
\newblock \doi{10.1371/journal.pbio.0040120}.
\newblock URL
  \url{http://journals.plos.org/plosbiology/article?id=10.1371/journal.pbio.0040120}.

\bibitem[Xie and Seung(2000)]{Xie2000}
X~Xie and HS~Seung.
\newblock {Spike-based learning rules and stabilization of persistent neural
  activity}.
\newblock \emph{Advances in neural information processing {\ldots}}, 2000.

\bibitem[Xie and Seung(2003)]{Xie2003}
Xiaohui Xie and H~Sebastian Seung.
\newblock {Equivalence of backpropagation and contrastive Hebbian learning in a
  layered network.}
\newblock \emph{Neural computation}, 15\penalty0 (2):\penalty0 441--54, feb
  2003.
\newblock ISSN 0899-7667.
\newblock \doi{10.1162/089976603762552988}.
\newblock URL \url{http://www.ncbi.nlm.nih.gov/pubmed/12590814}.

\bibitem[Xiong et~al.(2016)Xiong, Merity, and Socher]{xiong2016dynamic}
Caiming Xiong, Stephen Merity, and Richard Socher.
\newblock Dynamic memory networks for visual and textual question answering.
\newblock \emph{arXiv preprint arXiv:1603.01417}, 2016.

\bibitem[Yamins and DiCarlo(2016{\natexlab{a}})]{yamins2016eight}
Daniel~LK Yamins and James~J DiCarlo.
\newblock Eight open questions in the computational modeling of higher sensory
  cortex.
\newblock \emph{Current opinion in neurobiology}, 37:\penalty0 114--120,
  2016{\natexlab{a}}.

\bibitem[Yamins and DiCarlo(2016{\natexlab{b}})]{Yamins2016}
DLK Yamins and JJ~DiCarlo.
\newblock {Using goal-driven deep learning models to understand sensory
  cortex}.
\newblock \emph{Nature neuroscience}, 2016{\natexlab{b}}.
\newblock URL
  \url{http://www.nature.com/neuro/journal/v19/n3/abs/nn.4244.html}.

\bibitem[Yosinski et~al.(2014)Yosinski, Clune, Bengio, and
  Lipson]{yosinski2014transferable}
Jason Yosinski, Jeff Clune, Yoshua Bengio, and Hod Lipson.
\newblock How transferable are features in deep neural networks?
\newblock In \emph{Advances in Neural Information Processing Systems}, pages
  3320--3328, 2014.

\bibitem[Yttri and Dudman(2016)]{yttri2016opponent}
Eric~A Yttri and Joshua~T Dudman.
\newblock Opponent and bidirectional control of movement velocity in the basal
  ganglia.
\newblock \emph{Nature}, 533\penalty0 (7603):\penalty0 402--406, 2016.

\bibitem[Yu and Smith(2013)]{yu2013joint}
Chen Yu and Linda~B Smith.
\newblock Joint attention without gaze following: Human infants and their
  parents coordinate visual attention to objects through eye-hand coordination.
\newblock \emph{PloS one}, 8\penalty0 (11):\penalty0 e79659, 2013.

\bibitem[Yuste et~al.(2005)Yuste, MacLean, Smith, and Lansner]{Yuste2005}
Rafael Yuste, Jason~N MacLean, Jeffrey Smith, and Anders Lansner.
\newblock {The cortex as a central pattern generator.}
\newblock \emph{Nature reviews. Neuroscience}, 6\penalty0 (6):\penalty0
  477--83, jun 2005.
\newblock ISSN 1471-003X.
\newblock \doi{10.1038/nrn1686}.
\newblock URL \url{http://www.ncbi.nlm.nih.gov/pubmed/15928717}.

\bibitem[Zaremba and Sutskever(2015)]{zaremba2015reinforcement}
Wojciech Zaremba and Ilya Sutskever.
\newblock Reinforcement learning neural turing machines.
\newblock \emph{arXiv preprint arXiv:1505.00521}, 2015.

\bibitem[Zeisel et~al.(2015)Zeisel, Manchado, Codeluppi, Lonnerberg, {La
  Manno}, Jureus, Marques, Munguba, He, Betsholtz, Rolny, Castelo-Branco,
  Hjerling-Leffler, and Linnarsson]{Zeisel2015}
A.~Zeisel, A.~B.~M. Manchado, S.~Codeluppi, P.~Lonnerberg, G.~{La Manno},
  A.~Jureus, S.~Marques, H.~Munguba, L.~He, C.~Betsholtz, C.~Rolny,
  G.~Castelo-Branco, J.~Hjerling-Leffler, and S.~Linnarsson.
\newblock {Cell types in the mouse cortex and hippocampus revealed by
  single-cell RNA-seq}.
\newblock \emph{Science}, 347\penalty0 (6226):\penalty0 1138--42, feb 2015.
\newblock ISSN 0036-8075.
\newblock \doi{10.1126/science.aaa1934}.
\newblock URL
  \url{http://science.sciencemag.org/content/347/6226/1138.abstract}.

\bibitem[Zemel and Dayan(1997)]{zemel1997combining}
Richard~S Zemel and Peter Dayan.
\newblock Combining probabilistic population codes.
\newblock In \emph{IJCAI}, pages 1114--1119, 1997.

\bibitem[Zipser and Andersen(1988)]{Zipser1988}
D~Zipser and RA~Andersen.
\newblock {A back-propagation programmed network that simulates response
  properties of a subset of posterior parietal neurons}.
\newblock \emph{Nature}, 1988.
\newblock URL \url{https://cortex.vis.caltech.edu/Papers/PDFs of journal
  articles/Nature/V331{\_}88.pdf}.
\end{thebibliography}
\sloppy

} 
\end{multicols} 

\end{document}